\documentclass[aip,preprint]{revtex4-1}
\usepackage[utf8]{inputenc}
\usepackage{graphicx}
\usepackage{float}
\usepackage{latexsym}
\usepackage{amstext,amsfonts,amssymb,amsmath}
\usepackage{color}
\usepackage{bm} 	
\usepackage{caption}
\usepackage{subcaption}

\usepackage{tikz}
\usepackage{amsmath}
\usepackage{amssymb}
\usepackage{etoolbox}

\newrobustcmd*{\myVtriangle}[2]{\tikz{\filldraw[draw=#1,fill=#2] (0cm,0.2cm) --
(0.2cm,0.2cm) -- (0.1cm,0cm) -- (0cm,0.2cm);}}
\newrobustcmd*{\mythickVtriangle}[2]{\tikz{\filldraw[line width=0.3mm,draw=#1,fill=#2] (0cm,0.2cm) --
(0.2cm,0.2cm) -- (0.1cm,0cm) -- (0cm,0.2cm);}}
\newrobustcmd*{\mytriangle}[2]{\tikz{\filldraw[draw=#1,fill=#2] (0.0cm,0.0cm) --
(0.2cm,0cm) -- (0.1cm,0.2cm) -- (0cm,0cm);}}
\newrobustcmd*{\mysquare}[2]{\tikz{\draw[draw=#1,fill=#2] (0cm,0cm)
rectangle (0.2cm,0.2cm)}}
\newrobustcmd*{\mythicktriangle}[2]{\tikz{\filldraw[line width=0.3mm,draw=#1,fill=#2] (0.0cm,0cm) --
(0.2cm,0cm) -- (0.1cm,0.2cm) -- (0.0cm,0cm);}}
\newrobustcmd*{\mythicksquare}[2]{\tikz{\draw[line width=0.3mm,draw=#1,fill=#2] (0cm,0cm)
rectangle (0.2cm,0.2cm)}}
\newrobustcmd*{\mybarredtriangle}[2]{\tikz{\draw[draw=#1,fill=#2] (0,0) --
(0.2cm,0) -- (0.1cm,0.2cm) -- (0cm,0cm); \draw[draw=#1] (-0.1cm, 0.07cm) -- (0.3cm, 0.07cm)}}
\newrobustcmd*{\mythickbarredtriangle}[2]{\tikz{\draw[line width=0.3mm,draw=#1,fill=#2] (0,0) --
(0.2cm,0) -- (0.1cm,0.2cm) -- (0cm,0cm); \draw[draw=#1] (-0.1cm, 0.07cm) -- (0.3cm, 0.07cm)}}
\newrobustcmd*{\mybarredsquare}[2]{\tikz{\draw[draw=#1,fill=#2] (0,0)
rectangle (0.2cm,0.2cm); \draw[draw=#1] (-0.1cm, 0.1cm) -- (0.3cm, 0.1cm)}}
\newrobustcmd*{\mythickbarredsquare}[2]{\tikz{\draw[line width=0.3mm,draw=#1,fill=#2] (0,0)
rectangle (0.2cm,0.2cm); \draw[draw=#1] (-0.1cm, 0.1cm) -- (0.3cm, 0.1cm)}}
\newrobustcmd*{\mybarredcircle}[2]{\tikz{\draw[draw=#1,fill=#2] (0,0)
circle (0.1cm); \draw[draw=#1] (-0.2cm, 0.0cm) -- (0.2cm, 0.0cm)}}
\newrobustcmd*{\mythickbarredcircle}[2]{\tikz{\draw[line width=0.3mm,draw=#1,fill=#2] (0,0)
circle (0.1cm); \draw[draw=#1] (-0.2cm, 0.0cm) -- (0.2cm, 0.0cm)}}
\newrobustcmd*{\mydashedline}[1]{\tikz{\draw[draw=#1] (-0.2cm, 0.2cm) -- (-0.1cm, 0.2cm); \draw[draw=#1] (-0.0cm, 0.2cm) -- (0.1cm, 0.2cm)}}
\newrobustcmd*{\mythickcross}[1]{\tikz{\draw[line width=0.3mm,draw=#1] (0,0) --
(0.2cm,0); \draw[line width=0.3mm,draw=#1] (0.1cm,-0.1cm) -- (0.1cm,0.1cm);}}
\newrobustcmd*{\mybarredcross}[1]{\tikz{\draw[line width=0.3mm,draw=#1] (0,0) --
(0.2cm,0); \draw[line width=0.3mm,draw=#1] (0.1cm,-0.1cm) -- (0.1cm,0.1cm); \draw[draw=#1] (-0.1cm,0) -- (0.3cm,0);}}
\newrobustcmd*{\myline}[1]{\tikz{\draw[draw=#1] (-0.15cm, 0.1cm) -- (0.15cm, 0.1cm);\draw[line width=0.3mm,draw=#1] (-0.0cm, 0.0cm);}}
\newrobustcmd*{\mythickline}[1]{\tikz{\draw[line width=0.3mm,draw=#1] (-0.15cm, 0.1cm) -- (0.15cm, 0.1cm);\draw[line width=0.3mm,draw=#1] (-0.0cm, 0.0cm);}}
\newrobustcmd*{\mythickdashedline}[1]{\tikz{\draw[line width=0.3mm,draw=#1] (-0.2, 0.1cm) -- (-0.1cm, 0.1cm); \draw[line width=0.3mm,draw=#1] (-0.0cm, 0.1cm) -- (0.1cm, 0.1cm); \draw[line width=0.3mm,draw=#1] (-0.0cm, 0.0cm);}}
\newrobustcmd*{\mythickdasheddottedline}[1]{\tikz{\draw[line width=0.3mm,draw=#1] (-0.22, 0.1cm) -- (-0.13cm, 0.1cm); \draw[line width=0.3mm,draw=#1] (-0.085cm, 0.1cm) -- (-0.055cm, 0.1cm); \draw[line width=0.3mm,draw=#1] (-0.01cm, 0.1cm) -- (0.08cm, 0.1cm); \draw[line width=0.3mm,draw=#1] (-0.0cm, 0.0cm);}}

\newrobustcmd*{\mycircle}[2]{\tikz{\draw[draw=#1,fill=#2] (0,0)
circle (0.1cm);}}
\newrobustcmd*{\mythickcircle}[2]{\tikz{\draw[line width=0.3mm,draw=#1,fill=#2] (0,0)
circle (0.1cm);}}
\newrobustcmd*{\mydot}[1]{\tikz{\draw[line width=0.3mm,draw=#1] (0,0)
circle (0.025cm);}}

\definecolor{blue-violet}{rgb}{0.54, 0.17, 0.89}
\definecolor{bostonuniversityred}{rgb}{0.8, 0.0, 0.0}
\definecolor{blue(ryb)}{rgb}{0.01, 0.28, 1.0}
\definecolor{ao(english)}{rgb}{0.0, 0.5, 0.0}

\draft 

\begin{document}
\raggedbottom
\title{Using approximate inertial manifold approach to model turbulent non-premixed combustion}

\author{Maryam Akram}
\email[akramrym@umich.edu]
\author{Venkat Raman}
\affiliation{Department of Aerospace Engineering, University of Michigan, Ann Arbor, MI 48109, United States}

\begin{abstract}
The theory of inertial manifolds (IM) is used to develop reduced-order models of turbulent combustion. In this approach, the dynamics of the system are tracked in a low-dimensional manifold determined in-situ without invoking laminar flame structures or statistical assumptions about the underlying turbulent flow. The primary concept in approximate IM (AIM) is that slow dominant dynamical behavior of the system is confined to a low-dimension manifold, and fast dynamics respond to the dynamics on the IM instantaneously. Decomposition of slow/fast dynamics and formulation of an AIM is accomplished by only exploiting the governing equations. Direct numerical simulations (DNS) of initially non-premixed fuel-air mixtures developing in forced isotropic turbulence have been carried out to investigate the proposed model. Reaction rate parameters are varied to allow for varying levels of extinction and reignition. The AIM performance in capturing different flame behaviors is assessed both {\it a priori} and {\it a posteriori}. It is shown that AIM captures the dynamics of the flames including extinction and reignition. Moreover, AIM provides scalar dissipation rate, mixing time for reactive scalars, and closures for nonlinear terms without any additional modeling. The AIM formulation is found promising, and provides a new approach to modeling turbulent combustion.
\end{abstract}


\maketitle 

\section{Introduction}

Large eddy simulation (LES) has become the preeminent tool for modeling complex and turbulent reacting flows \cite{pitsch2006large,menon2000subgrid,raman2019emerging}. In LES, the large energy-containing scales are resolved directly, while the small-scales contributions are modeled. Combustion is controlled by molecular diffusion of reactants at scales competing with the smallest scales of turbulence, and needs to be exclusively modeled in all LES formulations. While this may seem to contradict the premise of LES, combustion in canonical \cite{fiorina2014,rieth2014comparison,PITSCH2008466,MUELLER20122166,kaul2013analysis} and practical \cite{koo2017large,chen2013large,sanjose2011fuel,boudier2009thermo} applications have been successfully modeled. In general, LES is far more accurate compared to the Reynolds-averaged Navier-Stokes equations when the large scale mixing controls the combustion process, for instance in flames far away from extinction \cite{pitsch2006large}. In particular, the use of manifold-based methods, such as flame-generated manifold \cite{van2002flamelet}, flamelet-progress variable \cite{pierce2004progress}, or unsteady flamelets \cite{pitsch2005unsteady} reduces the computational cost while providing excellent predictive capabilities for such stable flames \cite{raman2019emerging}. 

In practical systems, the development of such manifold methods faces some inherent hurdles. The primary issue is that manifolds are pre-generated by using an auxiliary system, such as counterflow or burner-stabilized laminar flames. It has been shown that the choice of the auxiliary system has a direct impact on the results \cite{mueller_2d,tang2019comprehensive, yihao_pci,chrigui2012partially,yihao_counterflow}. Essentially, it is assumed that the flame structure is similar to that of the auxiliary system, which is difficult to ensure in complex flows. A second issue concerns the necessary models once a manifold is selected. In most LES applications, closures for sub-filter variance of mixture fraction and progress variable \cite{cook1994subgrid,kaul2013analysis}, as well as the one-point one-time distribution function \cite{raman2005large, givi2006filtered} are needed. These models can also impact simulation predictions, especially when local extinction and reignition become important \cite{knudsen2012modeling,kaul2013large}. Alternative approaches, such as the filtered density function (FDF) method \cite{colucci1998filtered,raman2007consistent}, face related issues in terms of the modeling of small-scale mixing. 

In this regard, different approaches to LES modeling have been considered. The deconvolution based modeling procedure \cite{stolz1999approximate,mathew2016explicit,germano2009new,wang2017regularized} assumes that given the large scale field, the unresolved scales can be reconstructed, albeit only to some accuracy, to directly obtain sub-filter models. In particular, such methods have been linked to an explicit filtering procedure, where different filter shapes are explicitly assumed and the unresolved fields are recovered using an inversion procedure \cite{mathew2016explicit}. Another approach is the variational multiscale method \cite{hughes1998variational,gravemeier2006scale}, where transport equations are written for groups of length scales instead of partitioning the entire flow into two groups (i.e., resolved and unresolved). The deconvolution approach is particularly interesting since it extracts unresolved fields directly from the resolved fields, implying that a single unresolved field is linked to each resolved field. This is a simplification of the ideal LES formulation of Langford and Moser \cite{langford1999optimal}, in the sense that the optimal evolution is obtained by assuming that the distribution of sub-filter fields for a given filtered field is a delta function.

In the field of turbulence, there has been a long history of treating the flow as a chaotic dynamical system \cite{temam1989induced,keefe,vastano1991short}, and deriving closures based on the notion of a manifold \cite{temam1989inertial}. While nominally related to the flamelet-type methods, which also generate a manifold, these approaches treat the discretized form of the full set of governing equations (i.e., transport equations for mass, momentum, energy and scalars) as a finite-dimensional system. In statistically stationary flows, it is hypothesized that the trajectory followed by the flow in this high-dimensional state space is confined to a low-dimensional subspace. There have been many attempts to characterize the size of this subspace \cite{temam1989induced,keefe,vastano1991short,hassanaly2019numerical,hassanaly2019ensemble}. While theoretical scaling suggests that the manifold dimension will increase as $Re^n$, where $n > 2$, in statistically stationary turbulence \cite{temam1989inertial}, more recent numerical studies have concluded that the manifold may be lower dimensional. For instance, application to the Sandia flame series \cite{hassanaly2019ensemble} showed that the dimension is much smaller than the degrees of freedom generated by discretization. For context, the intrinsic low-dimensional manifold (ILDM) approach \cite{maas1992simplifying} can be considered as the manifold obtained from a spatially zero-dimensional system.

The manifold-based approach has also been used for model reduction in combustion. For instance, computational singular perturbation (CSP, \cite{lam1994csp}) separates the state space into fast and slow dimensions, with the variables corresponding to the slow dimensions explicitly solved while the fast dimensions obtained through closure. The G-scheme \cite{valorani2009g} extends this approach to spatially extended systems. In these approaches, the fast and slow sub-spaces are identified using eigenvectors of the Jacobian of the evolution equations \cite{valorani2009g}. As a result, the projection operators change as the system moves through the state-space. From a practical viewpoint, if the computational domain is discretized using $N = n_g \times n_s$ degrees of freedom, where $n_g$ is the number of control volumes and $n_s$ is the number of variables used to describe the thermochemical state (typically multiple species and energy), then the Jacobian will be of size $N^2$, which can become computationally intractable, especially for LES-type discretizations. 

In this work, a different manifold-based approach is used, which utilizes the notion of inertial manifolds (IM). In certain dissipative dynamical systems, trajectories of the system are exponentially attracted to a low-dimensional subspace of the state space known as the inertial manifold (IM) \cite{foias1988inertial}. All the trajectories of the system are attracted to the IM exponentially, which contains the attractor of the system. The existence of IM has been proven for many dissipative systems \cite{foias1988inertial,chung2015unified,foias1986inertial,constantin1988integral}, but its existence is under investigation for three-dimensional Navier-Stokes equations \cite{guermond2008fully,chueshov1995approximate,temam1989induced,foias1988inertial}. The main advantage of the IM-based approach is that the manifold can be identified using the linear part of the governing equations, namely the viscous or diffusion operators, rather than by the eigendecomposition of the Jacobian matrix.

Based on this concept, an analysis of the use of the IM theory for modeling turbulent reacting flows is conducted. Here, an approximate inertial manifold (AIM) is utilized \cite{temam1989induced,temam1989inertial,marion1989approximate,foias1988modelling,foias1986inertial,foias1988algebraic}, where the flow is decomposed into resolved and unresolved scales similar to conventional LES filtering. However, unlike in conventional LES, the small scales are directly reconstructed through a steady-state assumption. Previously, this approach has been tested for non-reacting turbulent flows \cite{AIM-AIAA,akram_jcp}. The proposed methodology has been investigated for chaotic systems, such as the one-dimensional Kuramoto-Sivashinsky equation and homogeneous isotropic turbulence (HIT), and promising results have been achieved for both systems. The AIM approach combines adaptive deconvolution principles with state space decomposition such as ILDM or CSP. In this work, AIM is extended to turbulent combustion. In other words, unlike the flamelet-type approaches where the manifold is constructed {\it a priori}, the AIM approach will extract this manifold {\it in-situ}. As a result, additional sub-filter closures for scalar dissipation or mixing time scales are not needed. 

The paper proceeds as follows. Section~\ref{sec:math} describes the inertial manifold theory and AIM development. In Sec.~\ref{sec:dns}, problem configuration and direct numerical simulation results are discussed followed by an investigation of the AIM model in Sec.~\ref{sec:AIM}. Concluding remarks are discussed in Sec.~\ref{sec:conclusion}.

\section{Approximate inertial manifold approach}
\label{sec:math}
Evolution of reacting flows can be fully described by transporting a set of variables such as momentum, energy, conserved and reacting scalars depending on the problem configuration. 
The variables of interest, ${\bm \xi} = \{\xi_1, \xi_2,\cdots,\xi_{n_v}\}$, are governed by a set of partial differential equations: 
\begin{equation}
\frac{\partial \bm \xi}{\partial t } + \nabla \cdot {\cal N}{\bm \xi}  + \nabla \cdot {\cal G}{\bm \xi} + S({\bm \xi}) = 0,
\end{equation}
where ${\cal G}$ is a linear operator, ${\cal N}$ is a nonlinear operator, and $S$ represents volumetric source term. The nonlinear term creates a spectrum of length-scales, which introduces the inherent computational complexity in solving these equations. For reacting scalars, including species mass fractions and temperature, the volumetric chemical source term involves contributions from many fast reactions, which introduces an additional source of nonlinearity. Consequently, the above set of equations governing practical reacting flows cannot be directly solved due to computational cost \cite{vervisch1998direct}, and a reduced description is necessary. 

The modeling approach based on the inertial manifold theory is developed for dissipative dynamical systems. Here, the reacting flow can be cast in a dynamical system framework after proper spatial discretization. In this discretized form, the governing equations can be written as
\begin{equation}
    \frac{d {\bm v}}{dt} + {\cal A}{\bm v} + {\cal F}({\bm v}) = 0,~~ {\bm v}(t=0) = {\bm v}_0,
    \label{eq:dyn}
\end{equation}
where ${\bm v}$ is the discrete vector of variables such that ${\cal D}: {\bm \xi} \rightarrow {\bm v}$, with ${\cal D}$ being the discretization operation using $n_g$ grid points. The inertial manifold theory requires ${\cal A}$ to be a linear, positive and self-adjoint operator such that the set of eigenvectors of ${\cal A}$ forms an orthonormal eigenbasis for the Hilbert space ${\cal H}$, where dynamics of the system reside \cite{foias1988inertial}. Thus, ${\cal F}$ in this formulation contains the discretized nonlinear terms including the source term. 

Here the objective is to develop a reduced order model predicting the dynamical behavior of the system in a low-dimensional manifold described by dynamics of a subset of the variables of interest. Unlike the predominant approach in turbulence modeling which uses a spatial filtering operator for separation of variables, here the governing equations are leveraged to define an orthogonal projection operator $P$ which decomposes the vector of variables (${\bm v}$) into the resolved ${\bm u}$ and the unresolved ${\bm w}$ subsets. The projection $P$ is performed on the space spanned by the first $m$ eigenfunctions of the linear operator ${\cal A}$. The choice of $m$ depends on the spectral properties of the linear operator \cite{foias1986inertial,temam1989induced,akram_jcp}.
By applying the projection operator to Eq.~\ref{eq:dyn}, the governing equations for the resolved and unresolved variables can be obtained as
\begin{equation}
    \frac{d {\bm u}}{dt} + {\cal A}{\bm u} + P {\cal F}({\bm v}) = 0,~~ {\bm u}(t=0) = P {\bm v}_0,
    \label{u-gov}
\end{equation}
and
\begin{equation}
    \frac{d {\bm w}}{dt} + {\cal A}{\bm w} + Q {\cal F}({\bm v}) = 0,~~ {\bm w}(t=0) = Q {\bm v}_0,
    \label{w-gov}
\end{equation}
where $Q = I - P$ is the complement operator of the projection $P$, and it maps its operand to the null-space of the projection operator. As $P$ and $Q$ are orthogonal projections in the Hilbert space, ${\cal H}$, they commute with the linear operator, ${\cal A}$, and its powers \cite{temam1989induced}. 

For tracking the dynamics of the approximate inertial manifold by only the resolved variables, the projected nonlinear term $P {\cal F} ({\bm v})$ should be modeled as it cannot be described using only ${\bm u}$. 
The goal is to reconstruct ${\bm w}$ given only information of ${\bm u}$ and compute the nonlinear terms with the recovered full-dimensional vector of variables. Using the IM theory, it is assumed that the dynamics of ${\bm w}$ respond instantaneously to the dynamics of ${\bm u}$. With the approximation $d{\bm w}/dt = 0$ \cite{foias1988modelling}, Eq.~\ref{w-gov} results in:
\begin{equation}
    {\bm w} = - {\cal A}^{-1} Q{\cal F}({\bm u},{\bm w}).
    \label{eq:AIM-app}
\end{equation}
The above nonlinear equation can be iterated starting from an initial guess to obtain a converged solution for ${\bm w}$. It has been shown that this approximation satisfies continuity \cite{akram_jcp}. Mass conservation and boundedness of scalars is achieved by clipping the solution to the physically bounded values. With this approximation of the unresolved dynamics, the nonlinear term $P {\cal F} ({\bm v})$, and hence the governing equations of the resolved modes ${\bm u}$, are closed. 

This modeling ansatz has been tested on canonical turbulent flows such as the one-dimensional Kuramoto-Sivashinsky equation and homogeneous isotropic turbulence \cite{akram_jcp}. The focus of this work is in investigating the validity of this approximation for turbulent combustion of canonical diffusion flames. Compared to the previous analysis on chaotic flows \cite{akram_jcp}, this problem challenges the AIM approximation in different ways: 1) The non-linear chemical source term lacks the scale separation property of the convection term, 2) the thermochemical properties of the chemistry field affect the level of local flame strength causing extinction or reignition, which imply strong interaction between turbulence and chemistry. 

To assess the IM assumptions for reacting flows, an {\it a priori} analysis has been conducted here such that DNS data is used to study the accuracy of AIM approximation in the reconstruction of the unresolved dynamics. This analysis is followed by an {\it a posteriori} study, and the proposed closure for the resolved variables is used to forecast the dynamics of the system in a low-dimensional approximate IM rather than the full-dimensional system. With this in mind, the following section describes the problem configuration.  

\section{Flow Configuration}
\label{sec:dns}
Homogeneous isotropic turbulence is used to assess the AIM approach. One-step chemical global reaction is used to emulate combustion, while neglecting density variations. The reaction rate parameters are obtained from prior work \cite{kops1999numerical}, which has been previously used to study unsteady flamelet modeling \cite{pitsch2003flamelet}. This flow configuration is solved using a pseudo-spectral approach, with the full resolution of the length and time scales providing the DNS solution. Additionally, the AIM-based simulations with reduced degrees of freedom, as well as a truncated simulation with the same degrees of freedom as the AIM but with no additional modeling, are conducted. The governing equations include transport equations for the three velocity components, mixture fraction and progress variable. The velocity field is considered to be incompressible, leading to the following set of equations: 

\begin{equation} 
\begin{aligned}
&\frac{\partial \xi_i}{\partial t} + \xi_j\frac{\partial \xi_i}{\partial x_j} = -\frac{1}{\rho}\frac{\partial p}{\partial x_i} + \nu \frac{\partial}{\partial x_j} (\frac{\partial \xi_i}{\partial x_j}) + B\xi_i, \\
&\frac{\partial \xi_i}{\partial x_i} = 0,
\end{aligned}
\label{eq:HIT}
\end{equation} 
where $\xi_i$ is the velocity component in the $i^{th}$ direction, $p$ is the hydrodynamic pressure, $\nu$ is the kinematic viscosity and $\rho$ is the density. In order to obtain statistical stationarity of the turbulence features, large-scale forcing is employed. Here, a constant linear forcing term is used with the coefficient $B$ tuned to compensate for the viscous dissipation \cite{rosales2005linear,lundgren_linearforcing}. 

The global one-step chemistry is the following reversible reaction
\begin{equation}
    F+rO \rightleftharpoons (r+1)P,
    \label{eq:reaction}
\end{equation}
where $r$ is the stoichiometric ratio, which is the mass of oxidizer consumed with unit mass of fuel. For modeling purposes, mixture fraction is defined as
\begin{equation}
    Z = \frac{rY_F-Y_O+Y_{O\infty}}{rY_{F\infty}+Y_{O\infty}},
\end{equation}
where $Y_F$ and $Y_O$ are mass fractions of fuel and oxidizer respectively. The unmixed values of fuel and oxidizer $(Y_{F\infty}, \ Y_{O\infty})$ are unity which results in $Z_s = 1/(r+1)$ for the stoichiometric value of the mixture fraction. In this work, $r = 1$ has been chosen resulting in $Z_s = 0.5$. The transport equations for mixture fraction and progress variable are
\begin{equation}
    \frac{\partial Y_i}{\partial t} = -{\bm \xi} \cdot \nabla Y_i +D\nabla^2Y_i + \omega_i.
    \label{eq:species}
\end{equation}
Here, $D$ is the coefficient of molecular diffusion of species, and the reaction rate is zero in the governing equation of mixture fraction. In the current formulation, the product mass fraction ($Y_P$) is equivalent to the normalized temperature $\theta$,  and the reacting field is determined by solving for $Z$ and $\theta$, with the Schmidt number ($Sc = \nu/D$) equal to $0.7$ for both scalars. For the reversible chemistry (Eq.~\ref{eq:reaction}) with an equilibrium constant $K$, the production rate of the products can be expressed as \cite{lee1995nonpremixed}
\begin{equation}
    \omega_P = (r+1)A \exp{\left(\frac{-\beta}{\alpha}\right)} \exp{\left(\frac{-\beta(1-Y_P)}{1-\alpha(1-Y_P)}\right)}\left(Y_FY_O-\frac{1}{K}Y_P^{r+1}\right),
    \label{eq:wp}
\end{equation}
where $A$ is a pre-exponential factor, $\beta$ is the Zeldovich number, and $\alpha$ is the dimensionless heat release parameter.  These parameters are defined such that the reaction rate is strongly temperature dependent, which can lead to local extinction and reignition controlled by the interaction of turbulent mixing and reaction \cite{kops1999numerical}.

The relative impact of turbulence on flame stability can be controlled using the reaction rate parameters. For this purpose, a Damkohler number is defined as $Da = \tau_{\chi}/\tau_c$, which is the ratio of turbulent mixing and reaction time scales. The local mixing time scale is characterized by the inverse of the scalar dissipation rate, $\chi = 2D \nabla Z \cdot \nabla Z$. Considering the chemistry of the flame, the Damkohler number can be expressed as \cite{cuenot1996asymptotic}
\begin{equation}
    Da = 16 r^r (1-Z_{st})^2 l_r^{2+r} A \exp\bigg(\frac{-\beta}{\alpha}\bigg)/\chi_{s},
    \label{eq:Da}
\end{equation}
where $l_r$ characterizes the width of the reaction zone, and $\chi_{s}$ is the dissipation rate of the mixture fraction at the stoichimetric mixture. Reaction thickness $l_r$ is defined as \cite{vervisch1998direct}: $l_r = l_d Da^{-1/3}$, where $l_d = (D/\chi)^{1/2}$ is the diffusive thickness. The statistics conditioned on the stoichiometric mixture fraction are obtained by averaging over this reaction thickness. Here, the pre-exponent $A$ is chosen as the control parameter that is used to modify Damkohler number (Table~\ref{table:t2}). Five different flame behaviors ranging from stable combustion to localized extinction and reignition and finally global extinction are observed for different values of $Da$ considered. Table~\ref{table:t1} provides characteristics of the flow field and thermochemistry parameters of the reacting field.

\begin{table}[H]
\begin{center} 
\begin{tabular}{ ||c|c|c|c|c|| } 
 \hline
 Flame $I$ & Flame $II$ & Flame $III$ & Flame $IV$ & Flame $V$  \\ [0.5ex] 
 \hline
 $8 \times 10^3$ & $8 \times 10^4$ & $10^5$ & $8 \times 10^5$ & $8 \times 10^6$ \\ 
 \hline
\end{tabular}
\caption{Pre-exponent factor $A$ for different flames}
\label{table:t2}
\end{center}
\end{table}

\begin{table}[H]
\begin{center} 
\begin{tabular}{ ||c|c|c|c|c|c|c|c|| } 
 \hline
 $n_g$ & $Re_{\lambda}$ & $k_{max}\eta$ & B & Sc & $\alpha$ & $\beta$ & K \\ [0.5ex] 
 \hline
 $256^3$ & 83.25 & 1.72 & 5 & 0.7 &  0.87 & 4.0 & 100 \\ 
 \hline
\end{tabular}
\caption{Characteristics of turbulence and thermochemistry parameters}
\label{table:t1}
\end{center}
\end{table}

For the DNS calculations, the set of governing equations is solved in a $2\pi$ length cube with periodic boundary conditions using a pseudo-spectral code \cite{moin1998direct,orszag1972numerical}. A two-thirds dealiasing rule is used for the non-linear terms \cite{orszag1971elimination}. Exact time integration is used for the linear viscous term and RK2 is used for the other terms \cite{hochbruck2005explicit,schulze2009exponential}. Since small incompressibility errors can  grow fast in a spectral formulation, it is necessary to remove the divergence error at every time step \cite{malik_thesis,eswaran1988examination}. At each time step, the velocity field is projected on the divergence-free space \cite{malik_thesis}. Turbulent field statistics such as the Taylor microscale Reynolds number $Re_{\lambda} = \frac{u^{\prime} \lambda_g}{\nu}$ and the Kolmogorov length scale $\eta = (\frac{\nu^3}{\epsilon})^{1/4}$ are monitored over the initialization time to make sure the turbulent field is fully developed, where $\lambda_g$ is computed as  $\sqrt{15\frac{\nu}{\epsilon}} u^{\prime}$ \cite{popebook}, $u^{\prime}$ is the fluctuating velocity and $\epsilon$ is the dissipation of turbulent kinetic energy. To resolve the velocity and conserved scalar field $k_{max} \eta > 1$ is required, where $k_{max} = 15/32 N$ is the cut-off wavenumber and $N$ is the number of grid points in one direction of the computational domain \cite{eswaran1988direct}. The flow statistics are monitored for approximately $200$ eddy turnover times ($\tau$) to ensure a fully developed forced statistically stationary flow field. The long transient time is chosen to make sure that the forcing does not lead to instability and energy pile-up at small scales, and the forcing energy is balanced with dissipation. The combustion and AIM investigation period starts when the flow becomes statistically stationary. 

After the flow reaches statistical stationarity, the computational domain is filled homogeneously with blobs of fuel and oxidizer, corresponding to $Z$ equal to one and zero respectively, such that the volume average of mixture fraction is equal to its stoichiometric value ($\langle Z_0 \rangle = Z_s$). The progress variable (normalized temperature) is initialized with solutions of the steady flamelet model, which is obtained from the steady version of the following time-dependent flamelet model: 
\begin{equation}
    \frac{\partial Y_i}{\partial t} = \frac{\chi}{2} \frac{\partial^2Y_i}{\partial Z^2}+\omega_i.
    \label{eq:flamelet}
\end{equation}
For initializing the temperature field, the scalar dissipation rate is modeled as \cite{sripakagorn2004extinction}: $\chi(Z) = \chi_s \exp{ \left(-2 \text{erf}^{-1} \left[ (2Z-1)^2 \right] \right)}$.

\begin{figure}[H]
\begin{subfigure}{0.33\textwidth}
\centering
\includegraphics[width=\linewidth]{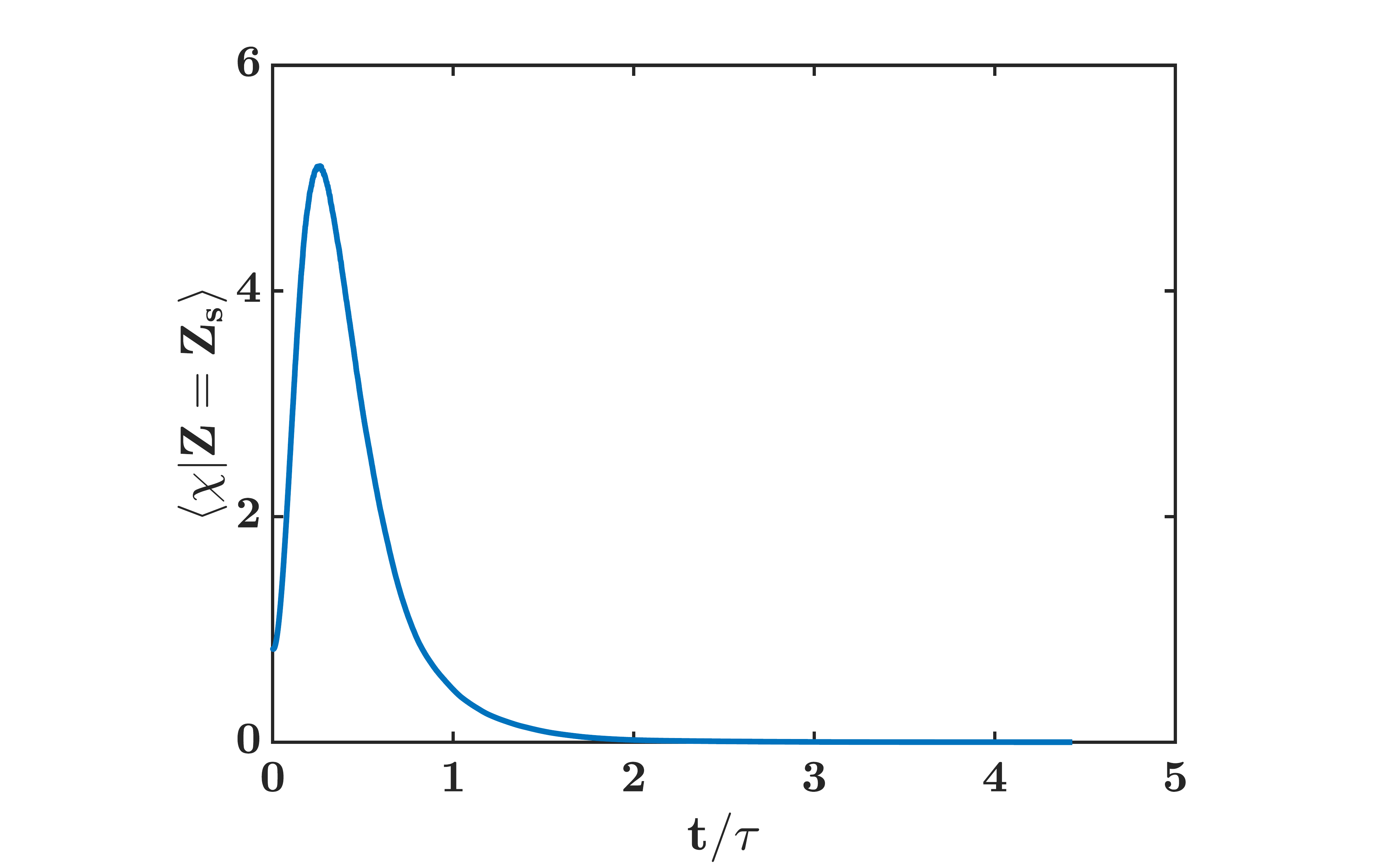}
\end{subfigure}%
\begin{subfigure}{0.33\textwidth}
\centering
\includegraphics[width=\linewidth]{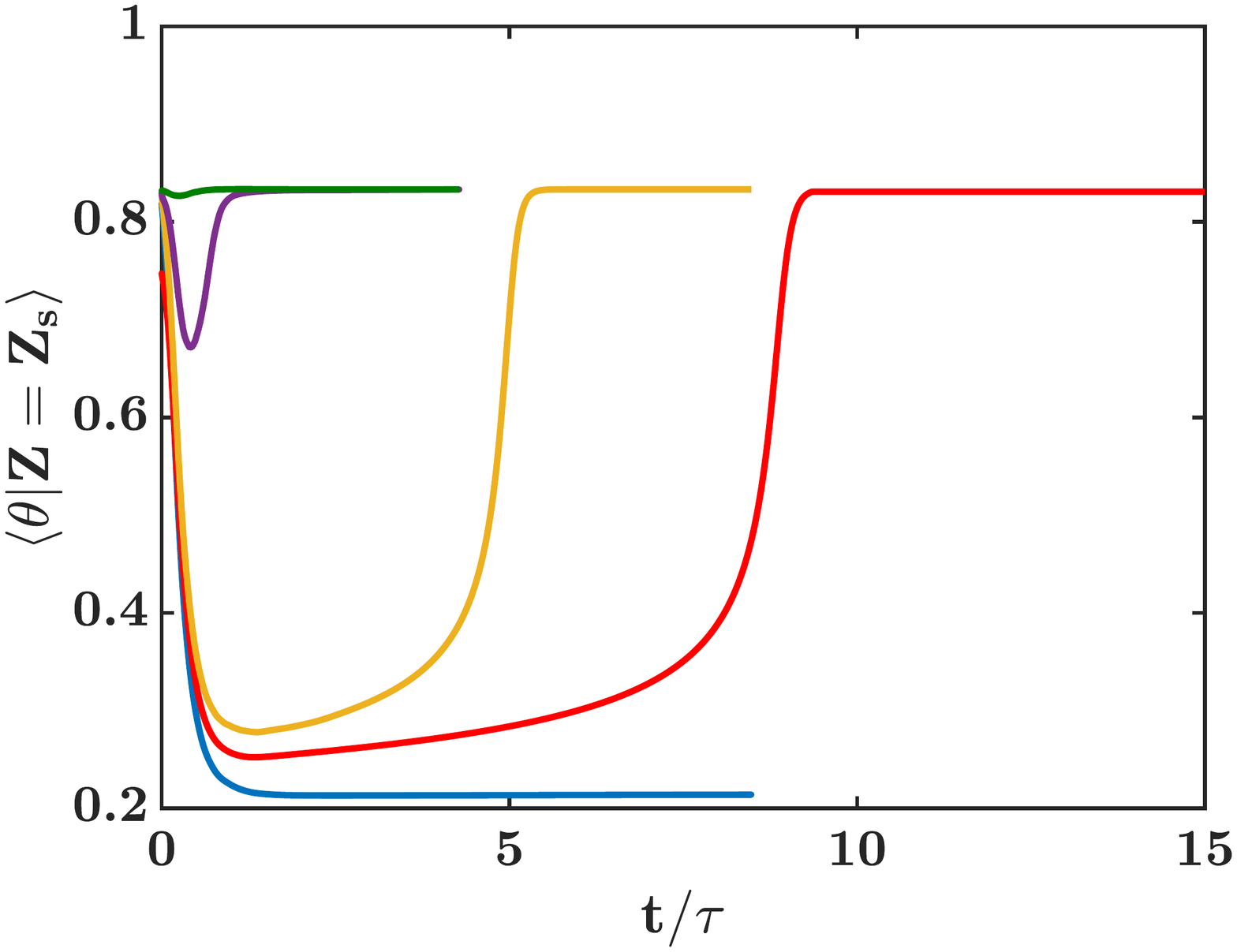}
\end{subfigure}%
\begin{subfigure}{0.33\textwidth}
\centering
\includegraphics[width=\linewidth]{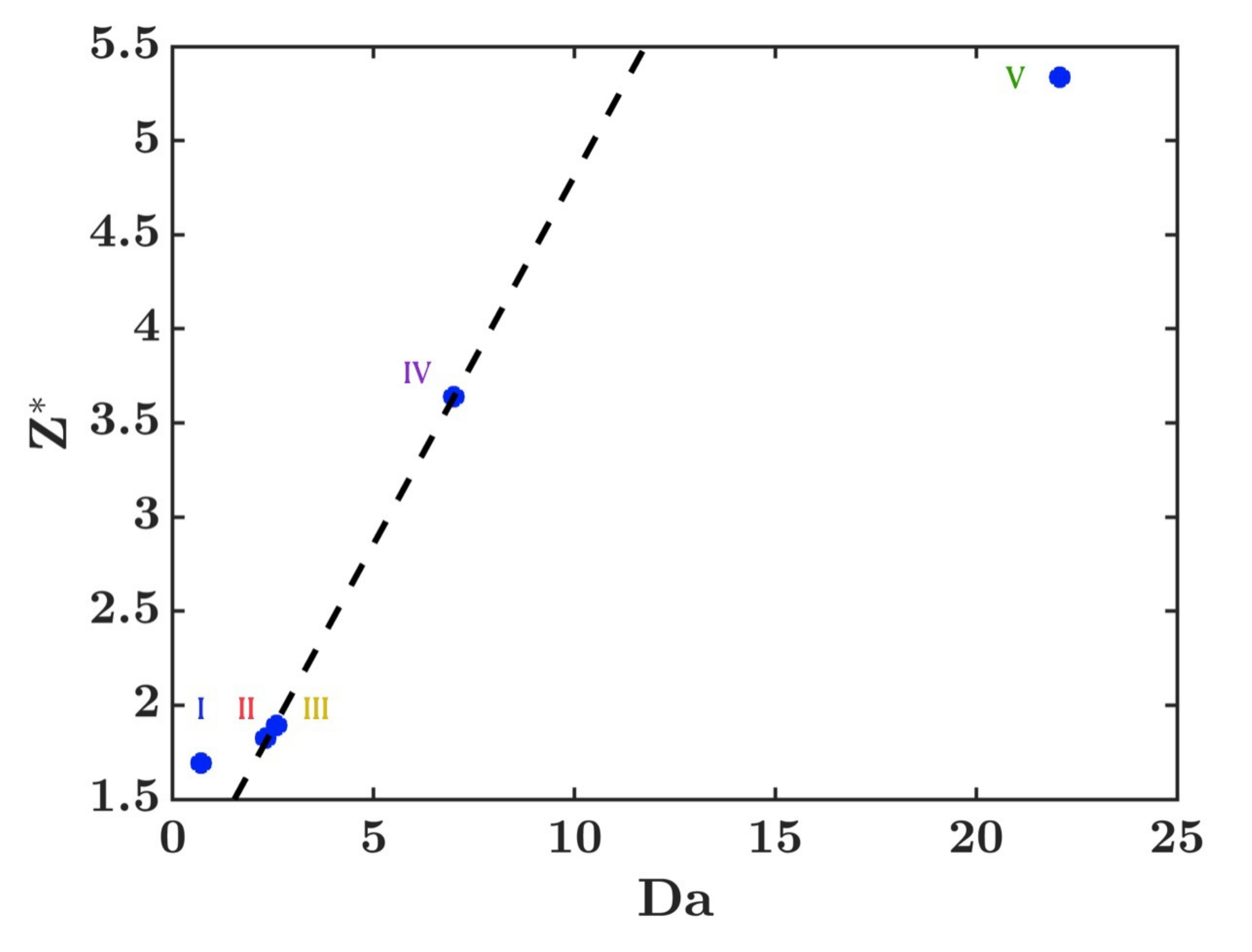}
\end{subfigure}%
\caption{Left: time evolution of conditionally averaged mixture fraction dissipation rate, $\langle \chi|Z=Z_s \rangle$; middle: time evolution of conditionally averaged temperature, $\langle \theta|Z=Z_s \rangle$, for different flames with different pre-exponential factors, $A$. Flame $I$: \mythickline{blue(ryb)}, flame $II$: \mythickline{red}, flame $III$: \mythickline{yellow}, flame $IV$: \mythickline{violet}, flame $V$: \mythickline{ao(english)}. Right: Sketch of $Da - Z^*$ relation for different flame behaviors. Dashed line shows the critical $Da$ number. All values are
computed at transient moment when $\langle \chi|Z=Z_s \rangle$ has its maximum value ($t/\tau \approx 0.25$).}
\label{fig:DNS-flame-info}
\end{figure}

Figure~\ref{fig:DNS-flame-info} shows time evolution of mixture fraction dissipation rate (left) and temperature (middle) at the stoichiometric values of mixture fraction. Due to straining, the scalar dissipation rate increases initially and its fluctuations disrupt the stoichiometric surface. Strong fluctuations of turbulent mixing cause excessive heat losses at the flame surface leading to extinguished regions. When reaction is faster than mixing, combustion heat release compensates for this heat loss leading to sustained high temperature. By increasing the pre-exponential factor $A$, the flame behavior changes from global extinction to various degrees of localized extinction and reignition, and finally to the stably burning flame at equilibrium conditions. It is clear that local extinction is due to the fluctuations of the scalar dissipation rate. When the maximum fluctuation of scalar occurs, the reacting surface deviates from the equilibrium condition. At this transient moment, $Da$ is computed from Eq.~\ref{eq:Da} for flames $I-V$ corresponding to $A \in (8 \times 10^3, 8 \times 10^6)$. Figure~\ref{fig:DNS-flame-info} (right) shows the $Da$ number versus the flame thickness parameter $Z^* = Z_{rms}/l_r$ where $Z_{rms}$ is the root mean square mixture fraction, and $l_r$ is the width of the reaction zone in mixture fraction space. Relation between $Da$ and $Z^*$ shows the competition between turbulence and chemistry time and length scales. When reaction is on the average faster than mixing with the smaller reaction layer thickness (larger $Z^*$), stable combustion occurs (flame $V$). As $Da$ is decreased, for given $Z^*$, departure from equilibrium increases, and beyond a critical $Da$ reaction can no longer balance heat loss of turbulent mixing and global extinction occurs (flame $I$). At this critical $Da$, when localized extinction occurs, molecular mixing eventually overcomes the influence of turbulent straining and leads to the decrease of the scalar gradients. This process is accompanied by gradual reignition; ultimately the reacting scalar field returns to the stable burning solution. In the case of diffusion flames, there is no exact definition of critical $Da$, and it is determined based on the DNS cases considered \cite{lee1995nonpremixed} (flames $II$, $III$ and $IV$ in this study).

\section{Numerical study of the AIM}
\label{sec:AIM}
In this section, the proposed approximate inertial manifold (Sec.~\ref{sec:math}) is implemented and studied for the non-premixed combustion cases discussed in Sec.~\ref{sec:dns}. To develop the AIM, governing equations of the flow field and combustion scalars can be rearranged as Eq.~\ref{eq:dyn} using the Stokes operator ($\nu \nabla^2 \xi_i$ in Eq.~\ref{eq:HIT}) as the linear operator of the Navier-Stokes equations and the diffusion operator as the linear operator of the conserved and reacting scalars. The projection operator that defines the resolved scales is parameterized using a three-dimensional wavenumber $k_m$ such that all the modes with wave numbers $\sqrt{k_x^2 + k_y^2 + k_z^2} \le k_m$ are included in the resolved space. This projection is equivalent to the sharp spectral filter in spatial filtering used in LES. The number of modes satisfying this requirement is the dimension of AIM, $m$. In this study, different values of $k_m$ have been used to examine the performance and convergence of AIM approximation. The goal is to assess AIM performance in prediction of different flame behaviors.

\subsection{A priori analysis of AIM}
\label{sec:apriori}
In this section, the performance of AIM approximation is assessed {\it a priori} such that the unresolved dynamics are approximated by the information of the exact resolved dynamics. For each AIM resolution ($m$), the resolved modes (${\bm u}$) are obtained by projection of the DNS data, then the unresolved variables (${\bm w}$) are approximated using Eq.~\ref{eq:AIM-app}. The full-dimensional vector of variables reconstructed by AIM (${\bm u}_{DNS}+ {\bm w}_{AIM}$) is compared against the DNS data. In this regard, this analysis is different from conventional LES analyses. Note that LES models approximate the effect of the unresolved features in the resolved scale equation. As a result, the unresolved features are not directly reconstructed in conventional LES. Hence {\it a priori} analysis of sub-filter terms typically involve comparison of the modeled term to the true sub-filter term. In the AIM approach, similar to deconvolution-based models \cite{stolz1999approximate,germano2009new,wang2017regularized,adams2001deconvolution,adams1999approximate}, a representation of the unresolved scales is obtained. Hence, the full reconstructed field can be directly compared to the DNS field. 

\begin{figure}[H]
\begin{subfigure}{0.33\textwidth}
\centering
\includegraphics[width=\linewidth]{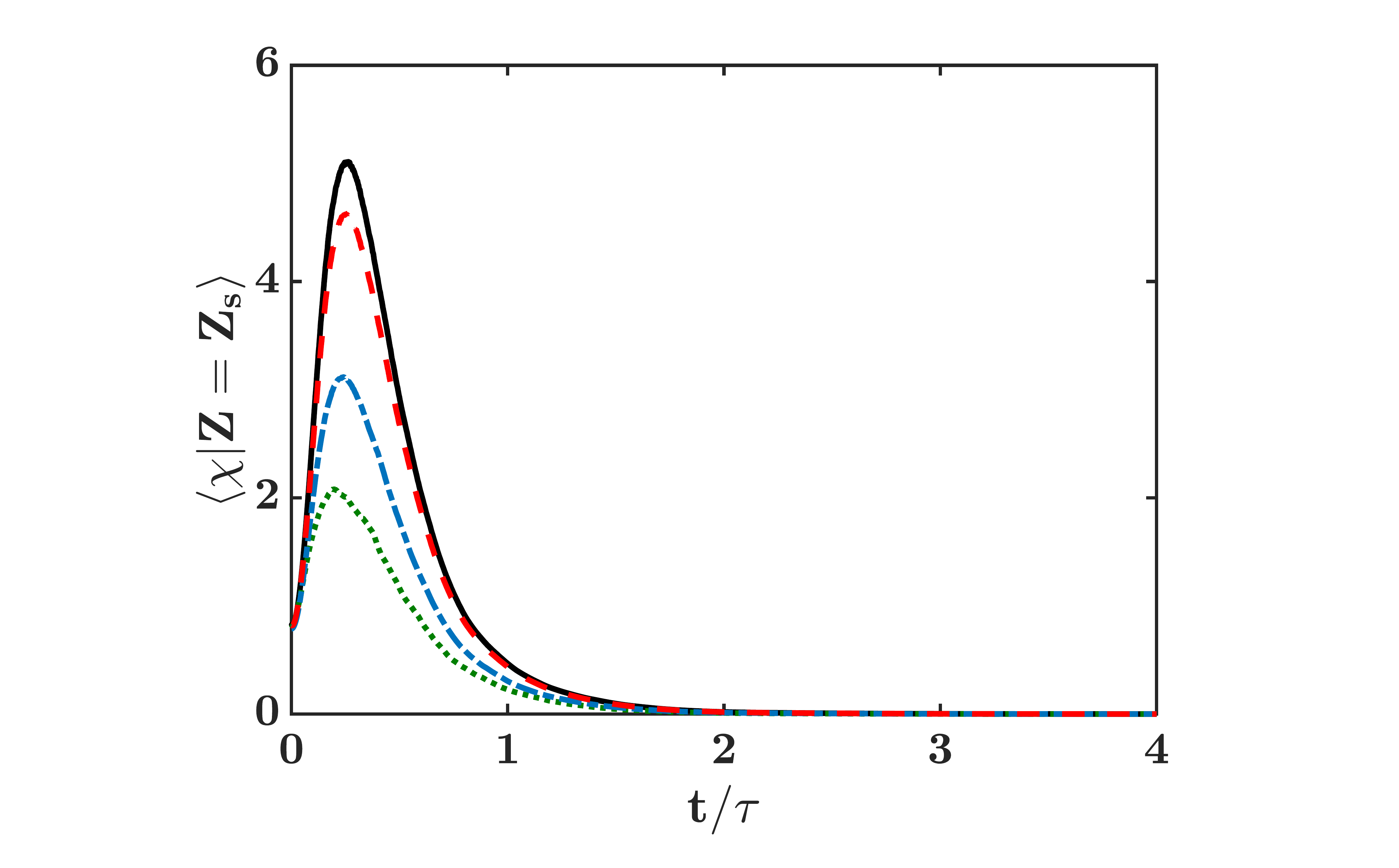}
\end{subfigure}%
\begin{subfigure}{0.33\textwidth}
\centering
\includegraphics[width=\linewidth]{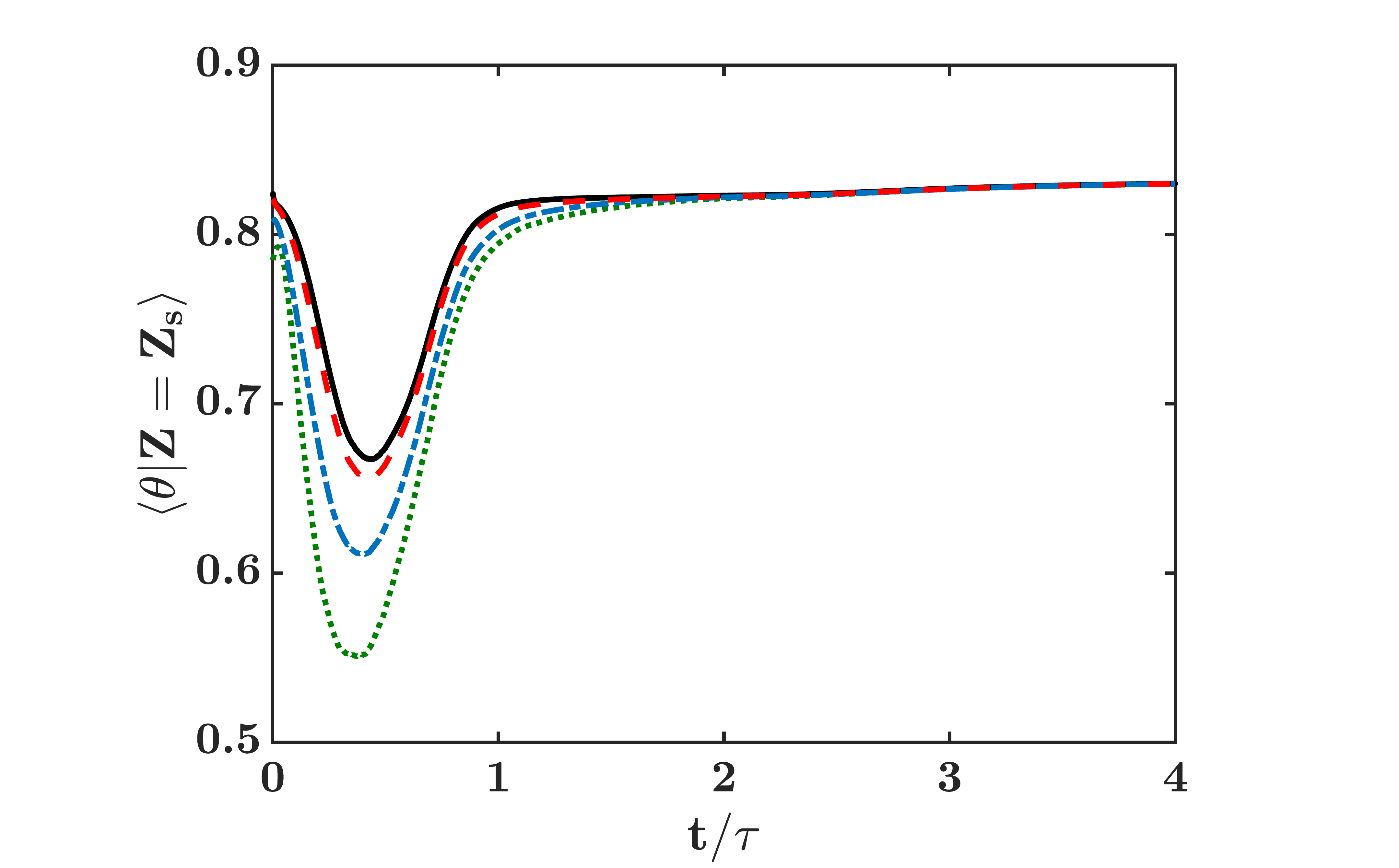}
\end{subfigure}%
\begin{subfigure}{0.33\textwidth}
\centering
\includegraphics[width=\linewidth]{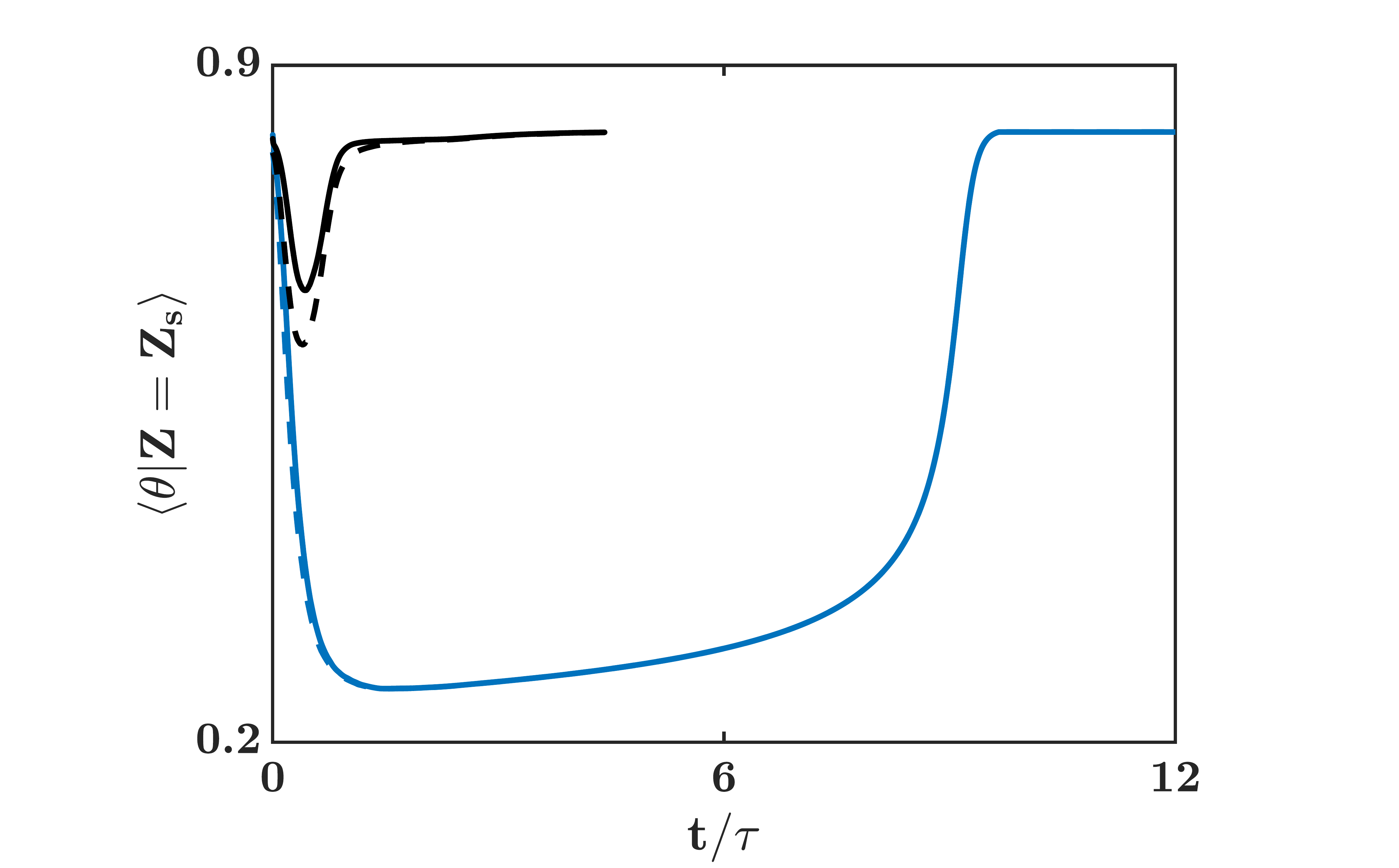}
\end{subfigure}%
\caption{Left: time evolution of conditionally-averaged mixture fraction dissipation rate, $\langle \chi| Z=Z_s \rangle$; middle: time evolution of conditionally averaged temperature, $\langle \theta|Z=Z_s \rangle$ for flame $IV$; DNS: {\mythickline{black}}, AIM with $k_m = 16$: {\mydot{ao(english)}}\mydot{{ao(english)}}{\mydot{ao(english)}}, AIM with $k_m = 32$: 
{\mythickdasheddottedline{blue(ryb)}}, AIM with $k_m = 64$: {\mythickdashedline{bostonuniversityred}}. Right: time evolution of conditionally averaged temperature, $\langle \theta|Z=Z_s \rangle$. DNS of flame $II$: {\mythickline{blue(ryb)}}, and flame $IV$: {\mythickline{black}}, AIM with $k_m = 32$ for flame $II$: {\mythickdashedline{blue(ryb)}} and flame $IV$: {\mythickdashedline{black}}.}
\label{fig:Stoi-AIM-conv}
\end{figure}

Global behavior of different flames is studied for various AIM resolutions in Fig.~\ref{fig:Stoi-AIM-conv}. Here, time evolution of mixture fraction dissipation rate and temperature at the stoichiometric surface are compared. As discussed in Sec.~\ref{sec:dns}, enhanced gradients imposed by the underlying turbulent field disrupts the flame surface and induces localized extinction. Competition between chemistry and mixing time scales either result in reignition or higher levels of extinction that causes global quenching of the flame. Figure~\ref{fig:Stoi-AIM-conv} (left) compares time evolution of the mixture fraction dissipation rate modeled by AIM against the DNS results. Since AIM reconstructs the full-dimensional vector of variables, the dissipation rate can be computed directly without further modeling of the small scales. By increasing the AIM resolution, $m$, the approximation becomes closer to the exact field, however, the maximum dissipation rate is underestimated by AIM. Here the highest resolution AIM is obtained with $k_m = 64$ and is spanned by only three percent of the DNS modes.

Approximations of the conditionally averaged temperature, $\langle \theta |Z = Z_s \rangle$, by different AIM resolutions ($m$) is compared against the DNS data for flame $IV$ (Fig.~\ref{fig:Stoi-AIM-conv}, middle). In this case, initially the flame experiences nominal temperature drop at the stoichiometric surface which shows localized extinction due to straining. Since the stoichiometric surface is still engulfed by relatively high temperature fuel and oxidizer mixture, reaction rate increases and the flame re-ignites and reaches equilibrium state quickly. It is shown that flame-front temperature reconstructed by AIM follows similar behavior, but experiences more severe extinction and slower return to stably burning flame. The reason behind this discrepancy between AIM and DNS can be the fast moving trajectory of the system on the manifold such that the scale-separation between resolved and unresolved dynamics is not present. 

The reasoning above can be verified by comparing AIM prediction of flames $II$ and $IV$. Figure~\ref{fig:Stoi-AIM-conv} (right) compares conditionally averaged temperature at the flame surface reconstructed by AIM with $k_m=32$ against the DNS data for fast and slow extinction and reignition events. At the same resolution, AIM approximation is significantly more accurate in the extent of the extinction and pace of the reignition for flame $II$. In this flame, chemical reaction is slower, on the average, than turbulence mixing. Initially, scalar fluctuations caused by turbulent mixing leads to extinction, and since chemical reaction is slower, it cannot overcome the influence of turbulent straining. After a while, scalar gradients are decreased due to molecular mixing. This process is accompanied by gradual reignition; ultimately the reacting scalar field returns to the steady flamelet solution.
 
\begin{figure}[H]
\begin{subfigure}{0.33\textwidth}
\centering
\includegraphics[width=\linewidth]{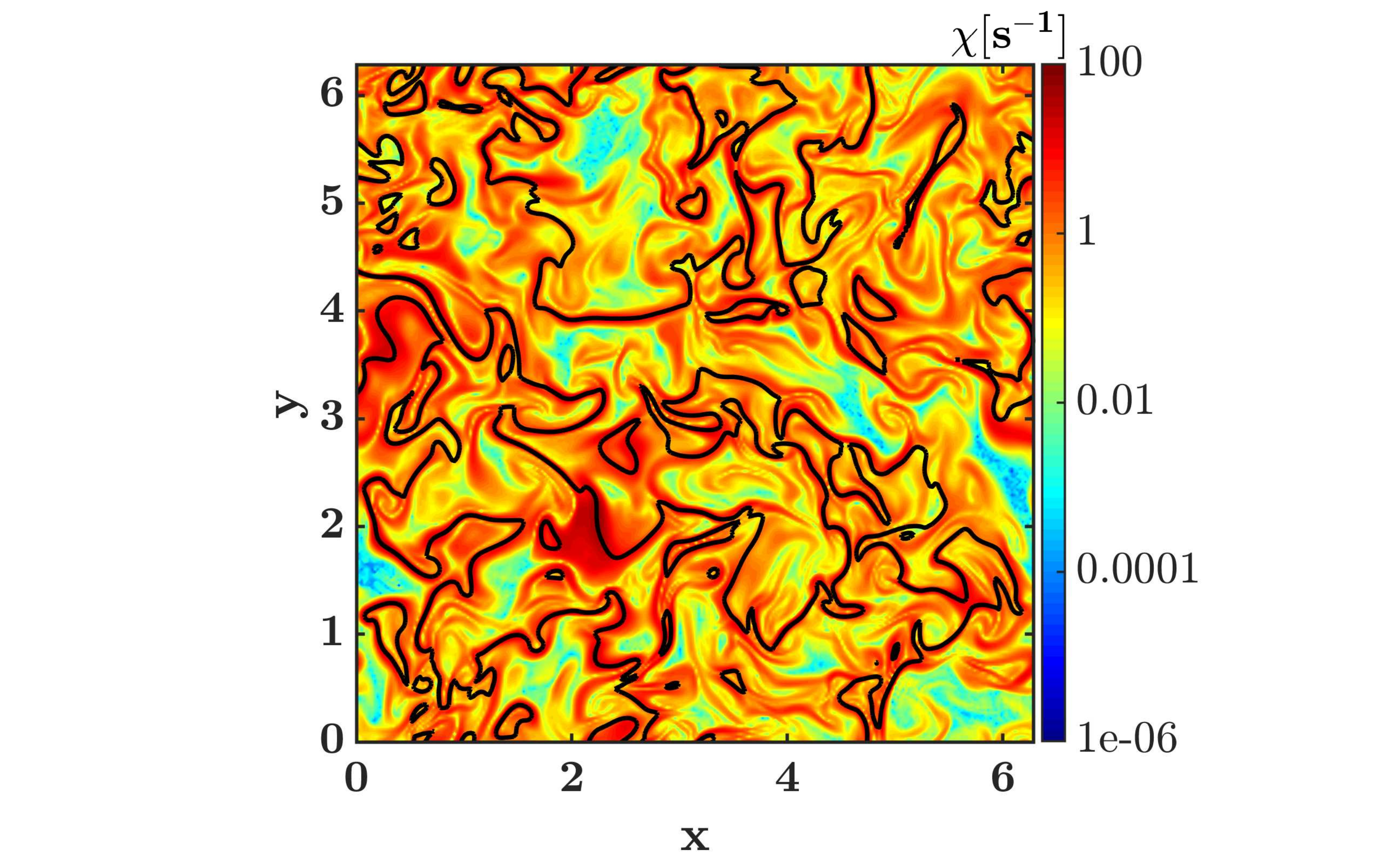}
\end{subfigure}%
\begin{subfigure}{0.33\textwidth}
\centering
\includegraphics[width=\linewidth]{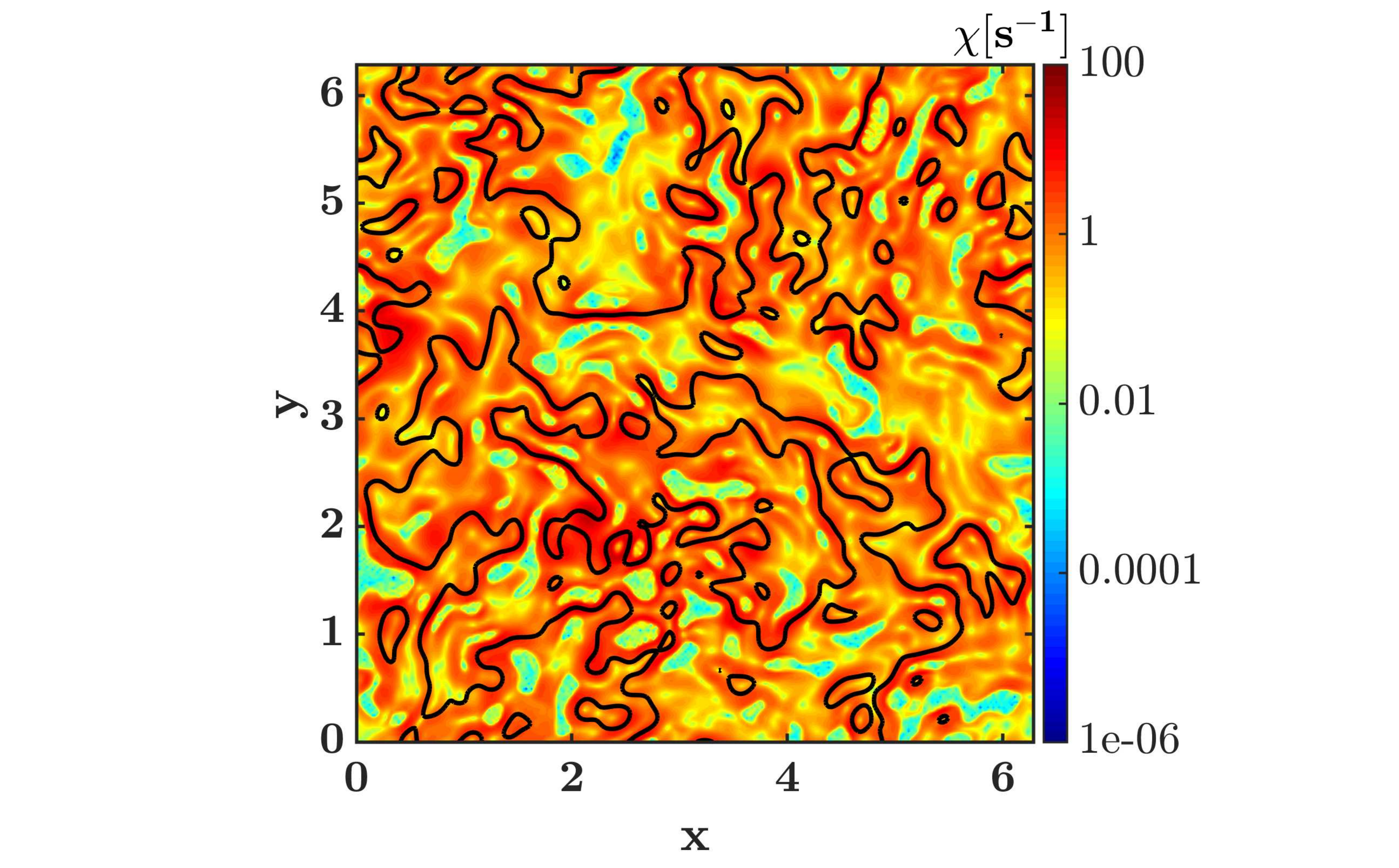}
\end{subfigure}%
\begin{subfigure}{0.33\textwidth}
\centering
\includegraphics[width=\linewidth]{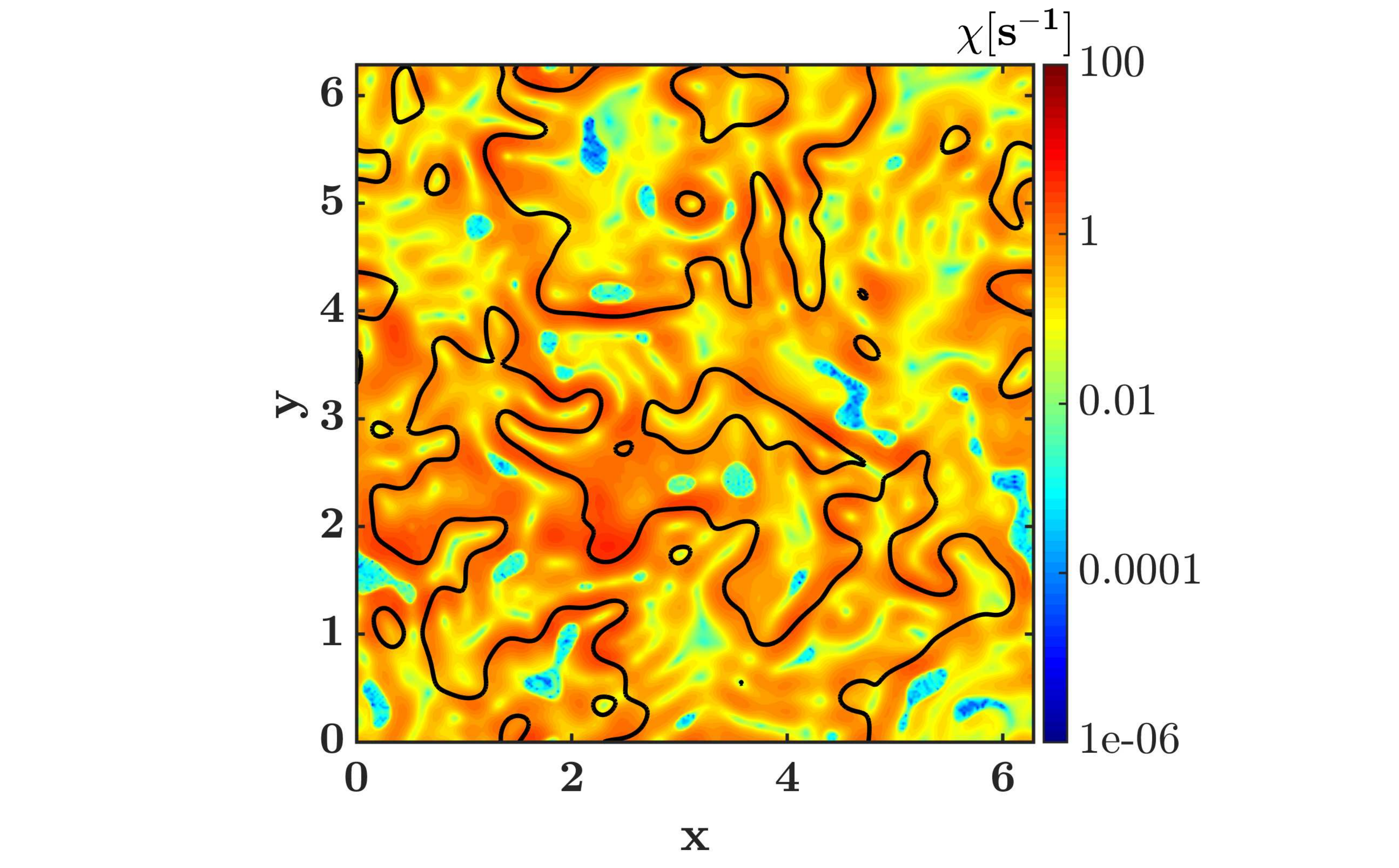}
\end{subfigure}%
\caption{Mixture fraction dissipation rate ($\chi = 2D (\nabla Z)^2$) in a plane of the computational domain at $t/\tau \approx 0.5$ when flame $IV$ is locally extinguished. Left: DNS, middle: AIM, right: projected DNS field (only resolved scales). Projection cut-off wavenumber is $k_m = 32$. Black lines represent stoichiometric mixture ($Z =  Z_s$).}
\label{fig:AIM1-Xai-Z-contours}
\end{figure}

In diffusion flames, chemical reaction is controlled by strain and is proportional to dissipation rate of reactants. By reconstruction of the small-scale quantities with AIM, scalar dissipation and chemical reaction rates can be directly computed from reconstructed temperature and mixture fraction fields without explicit modeling. Figure~\ref{fig:AIM1-Xai-Z-contours} shows the dissipation rate contours plotted when the mean temperature is the lowest (see Fig.~\ref{fig:Stoi-AIM-conv}). The presence of high dissipation rate values at the stoichiometric surface leads to local extinction. It is seen that the AIM reconstruction overall captures such structures, but differences can be found in top right and top left corners of the domain where only small scale reaction structures are present. At this cut-off wavenumber projection ($k_m = 16$), AIM dimension ($m$) is $0.05$ percent of the DNS degrees of freedom. The disipation rate computed from the projected DNS field, which includes only the resolved scales, is shown in Fig.~\ref{fig:AIM1-Xai-Z-contours} (right). At this projection cut-off wavenumber ($k_m = 16$), large features of the field are preserved but small scales and maximum dissipation rates are not captured. AIM approximation is able to recover many of the small features. 

An important component of turbulent combustion models is the prescription of the scalar dissipation rate, both for mixture fraction \cite{kaul2013large,jimenez2001subgrid} and the reacting scalars \cite{pope2013small,givi2006filtered,raman2007consistent}. More commonly, a time-scale is prescribed for the dissipation of scalar variance \cite{dopazo1994pdf,raman2007consistent}. The characteristic time scale $T_j$, for dissipation of scalar energy can be defined as \cite{cha2003model}
\begin{equation}
    T_j = \langle Y^2_j \rangle / \langle 2D (\nabla Y_j)^2 \rangle.
\end{equation}
Models for this time scale typically assume that scalar dissipation is directly proportional to turbulent energy dissipation \cite{popebook}. It has been widely observed that the modeling of these terms have first order impact on the simulation predictions, but models even for non-reacting scalars such as mixture fraction can introduce large errors \cite{kaul2013analysis}. Since turbulent mixing enhances the gradients of species, which is nonlinearly impacted by Arrhenius chemistry, reacting scalar gradients can be locally steepened due to small-scale reaction zones. As a result, the length and time scales associated with reacting species can be vastly different from the predominantly turbulent mixing controlled non-reacting scalar properties.  In addition, there is a time-lag in the response of chemistry to turbulent straining, which can be seen in Figure~\ref{fig:DNS-flame-info} (left). It shows that the scalar dissipation rate is initially increasing due to the straining effects of the turbulent field, and maximum of scalar variations at the flame surface occurs at $t/\tau \approx 0.25$. Interplay of mixing and reaction leads to extinction of flames $I - IV$, but this extinction happens after a finite period of time depending on the chemistry time scale. For instance, flames $II$ and $III$ are quenched by $t/\tau \approx 1.37$ when almost all of the scalar variations are dampened. On the other hand, flame $IV$ responds much quicker to the turbulent mixing effects and experiences localized extinction at $t/\tau \approx 0.5$. So, even if the time scale for the dissipation of conserved scalar energy can be assumed to be proportional to the turbulent energy dissipation rate, the characteristic time scale for the reacting scalars need not follow this relation. 

\begin{figure}[H]
\begin{subfigure}{0.5\textwidth}
\centering
\includegraphics[width=\linewidth]{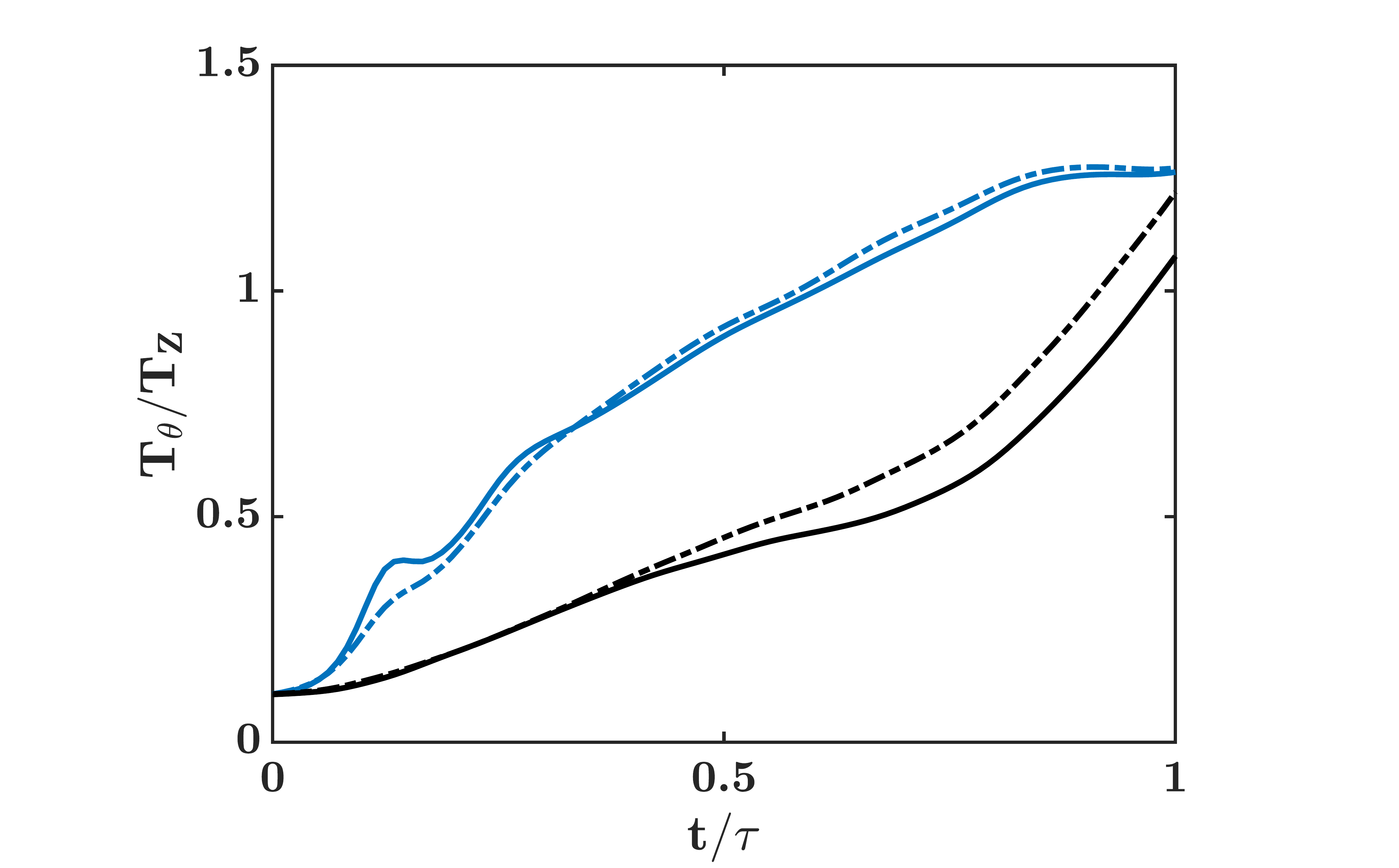}
\end{subfigure}%
\begin{subfigure}{0.5\textwidth}
\centering
\includegraphics[width=\linewidth]{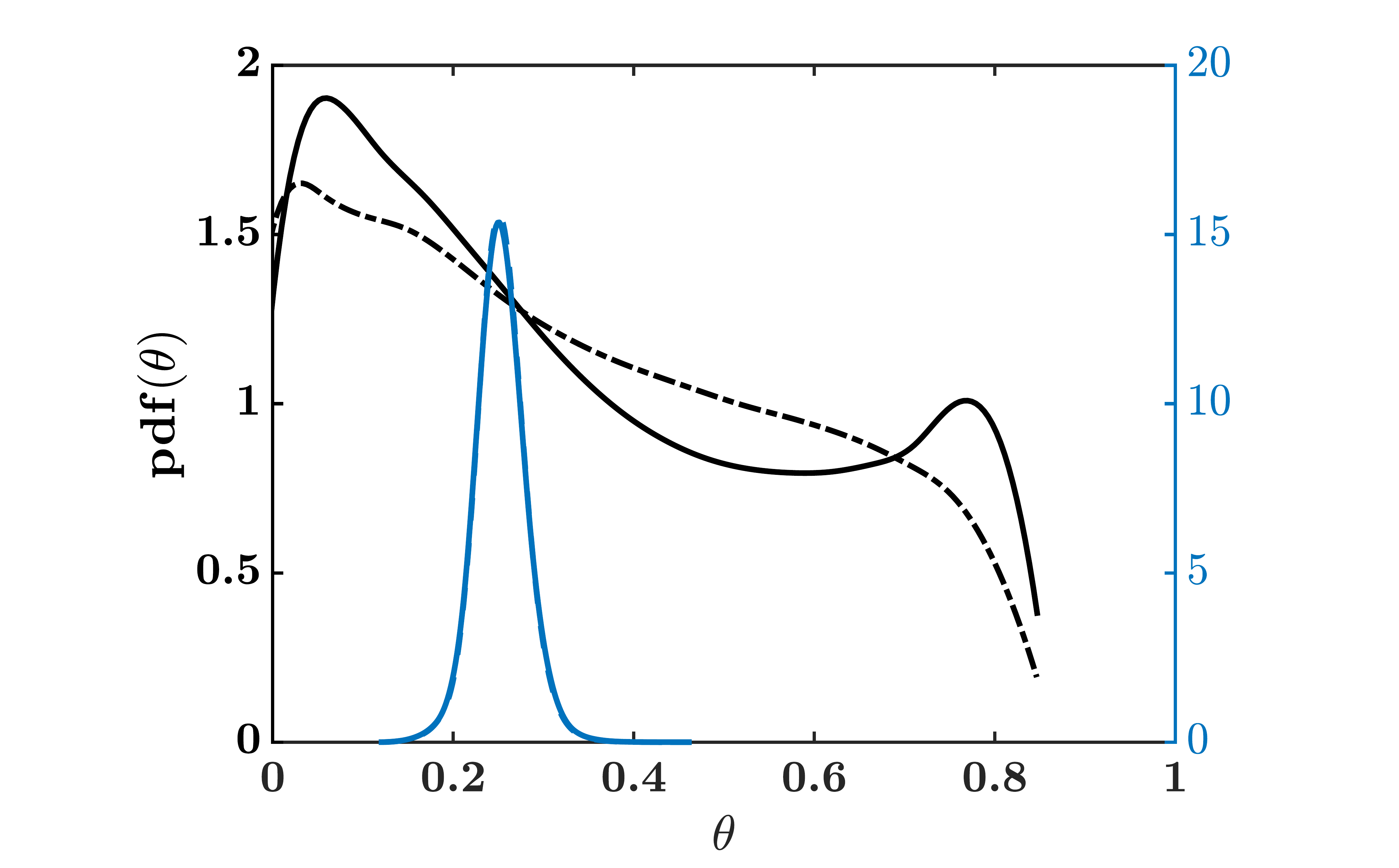}
\end{subfigure}%

\begin{subfigure}{0.5\textwidth}
\centering
\includegraphics[width=\linewidth]{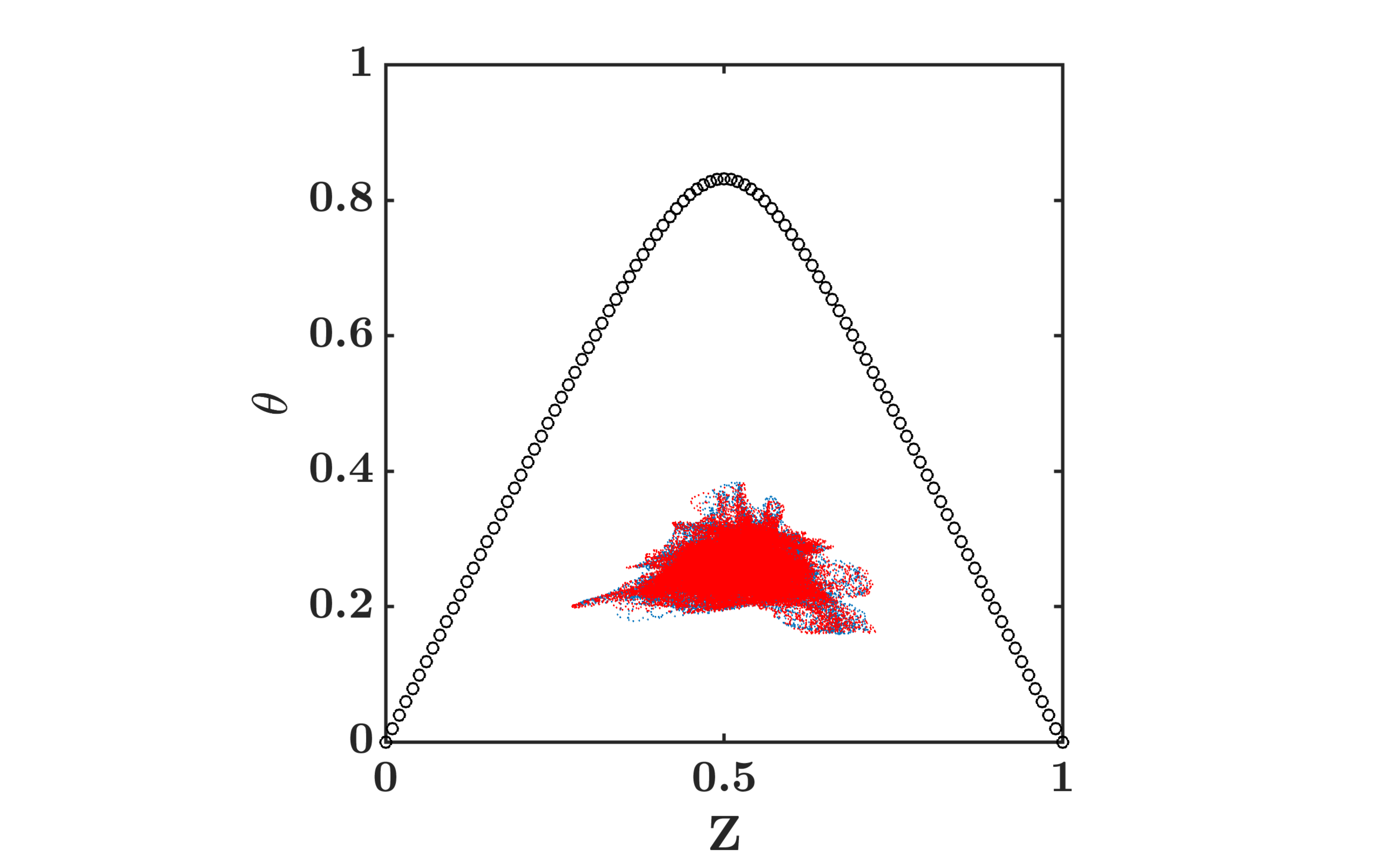}
\end{subfigure}%
\begin{subfigure}{0.5\textwidth}
\centering
\includegraphics[width=\linewidth]{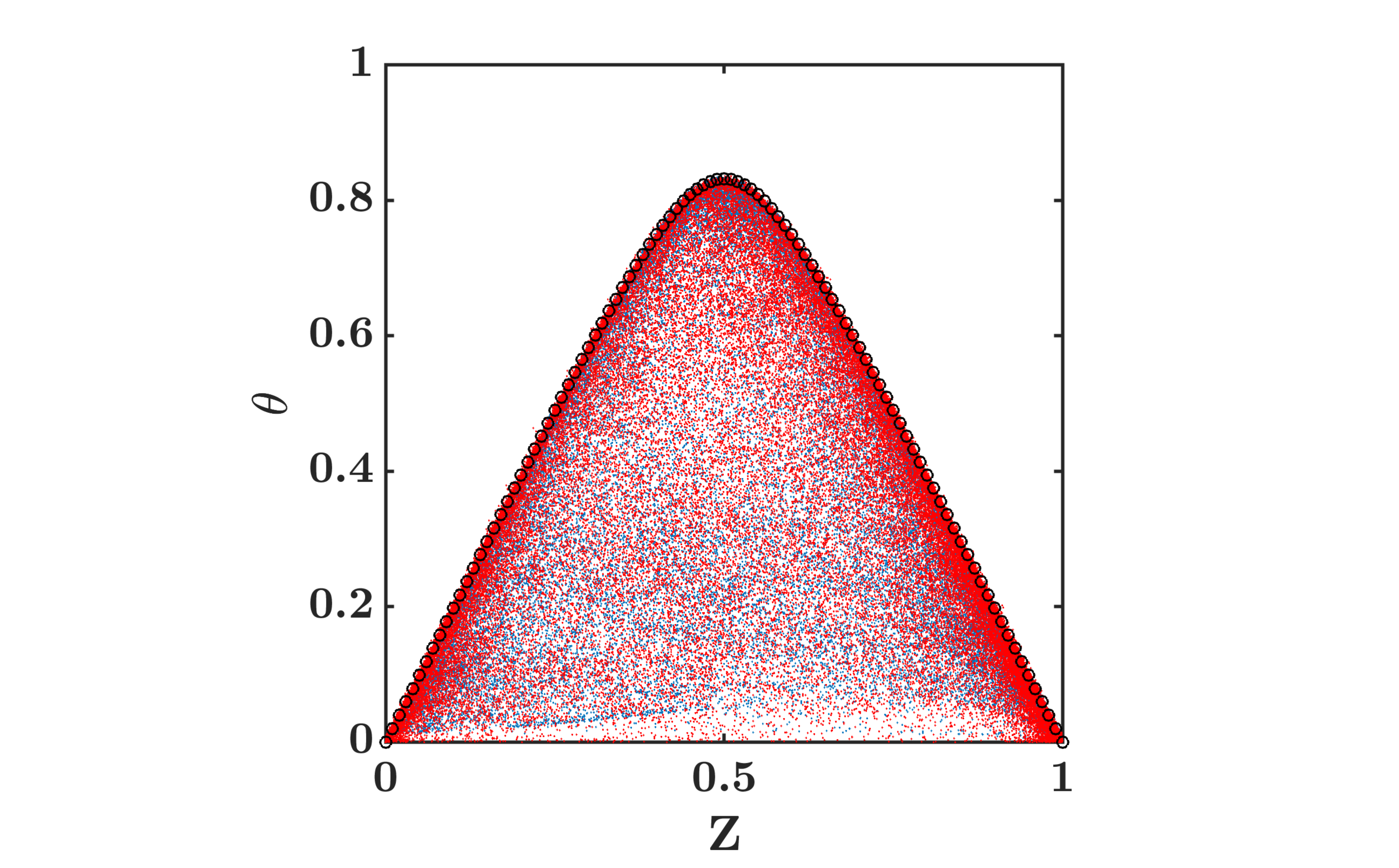}
\end{subfigure}%
\caption{Top: time evolution of mixing time scale ratio of reactive and conserved scalars ($T_{\theta}/T_Z$) (left) and probability distribution of temperature at $t/\tau = 0.5$ (right) for flames $II$ (blue) and $IV$ (black). DNS of flame $II$: {\mythickline{blue(ryb)}}, and flame $IV$: {\mythickline{black}}, AIM with $k_m = 32$ for flame $II$: {\mythickdashedline{blue(ryb)}} and flame $IV$: {\mythickdashedline{black}}. Bottom: conditional distribution of temperature at $t/\tau = 0.5$ for flame $II$ (left) and flame $IV$ (right); DNS: blue dots, AIM with $k_m = 32$: red dots and steady flamelet solutions: black circles.}
\label{fig:AIM1-TimeScaleRatio}
\end{figure}

Only for infinitely fast irreversible chemistry or slow chemistry, where the assumption of dissipation balancing production is sufficiently accurate, is the characteristic time scale of all reacting scalars nearly identical and can be assumed to be proportional to the conserved scalar time scale \cite{petersturbulent}. Here, the ratio of progress variable and mixture fraction time scales is compared for two flames experiencing minimal and maximal extinction under the straining effects of the turbulent field. Figure~\ref{fig:AIM1-TimeScaleRatio} compares some properties of flame $II$ and flame $IV$, and the capability of AIM to reproduce them. Time scale ratio of reacting and conserved scalars ($T_{\theta}/T_Z$), show that on average, reaction is faster than mixing. However, in flame $IV$ with minimal extinction, the difference between time scales is more prominent, since this flame reignites quickly when the micro-scale mixing overcomes straining effects (Fig.~\ref{fig:Stoi-AIM-conv}, right). 

Properties of temperature (progress variable) as these flames experience their extinction are compared further. Conditional distribution of temperature depicts very different behaviors. Flame $II$ is globally extinguished, and there is little variation in the temperature field. On the contrary, flame $IV$ experiences local extinction due to breakage of flamelets and since micromixing of reactants is fast enough to enhance gradients, the flame re-ignites across the entire range of mixture fraction values. This is shown by the large variation of temperature (Fig.~\ref{fig:AIM1-TimeScaleRatio}, bottom). Comparison of probability distribution of temperature for these two flames illustrates their different responses to the straining turbulent field. AIM succeeds in representing these characteristics quite accurately, but the AIM approach is marginally better in flame $II$. This is expected since the trajectory is varying quickly in flame $IV$ as compared to flame $II$.

\begin{figure}[H]
\begin{subfigure}{0.5\textwidth}
\centering
\includegraphics[width=\linewidth]{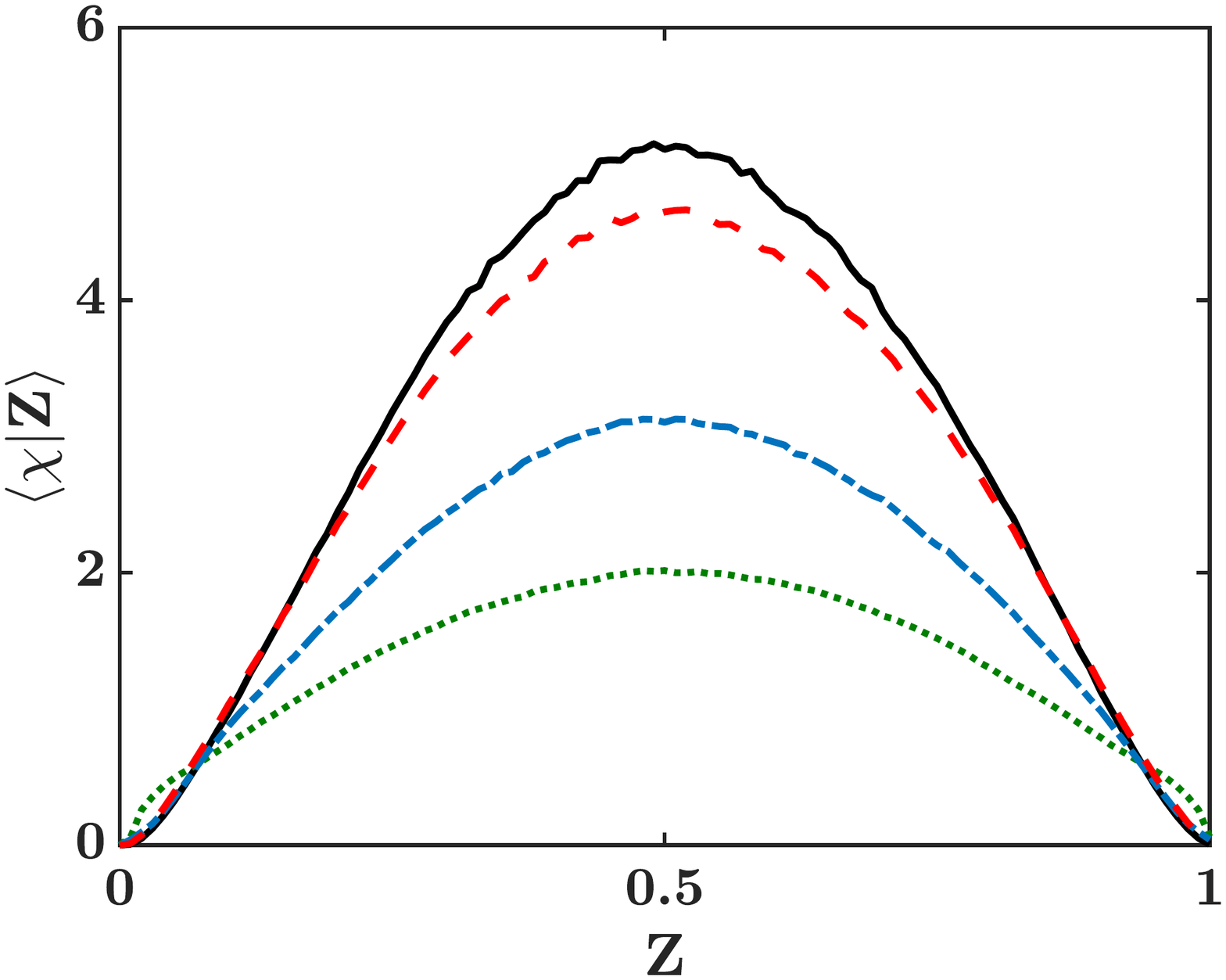}
\end{subfigure}%
\begin{subfigure}{0.5\textwidth}
\centering
\includegraphics[width=\linewidth]{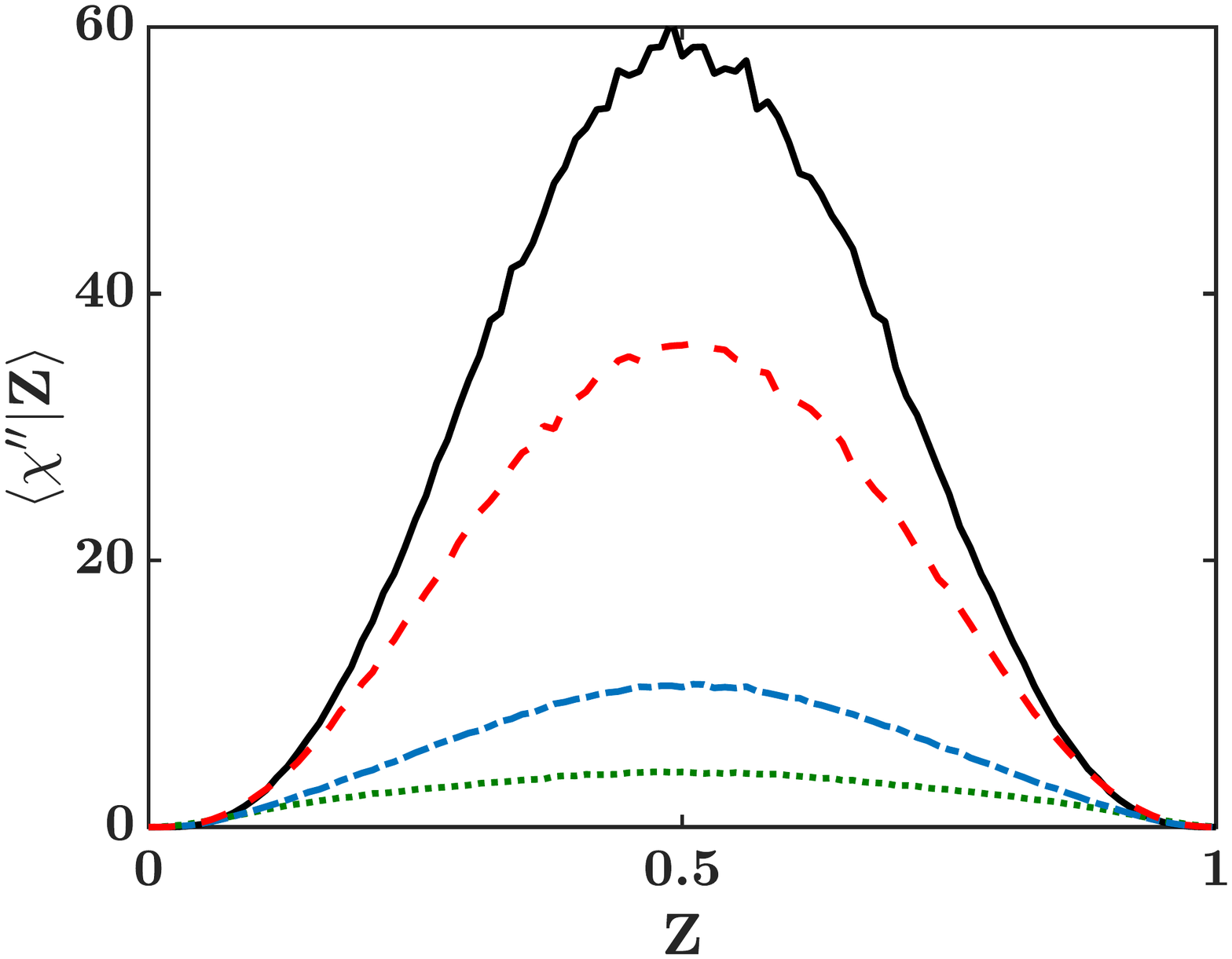}
\end{subfigure}%

\begin{subfigure}{0.5\textwidth}
\centering
\includegraphics[width=\linewidth]{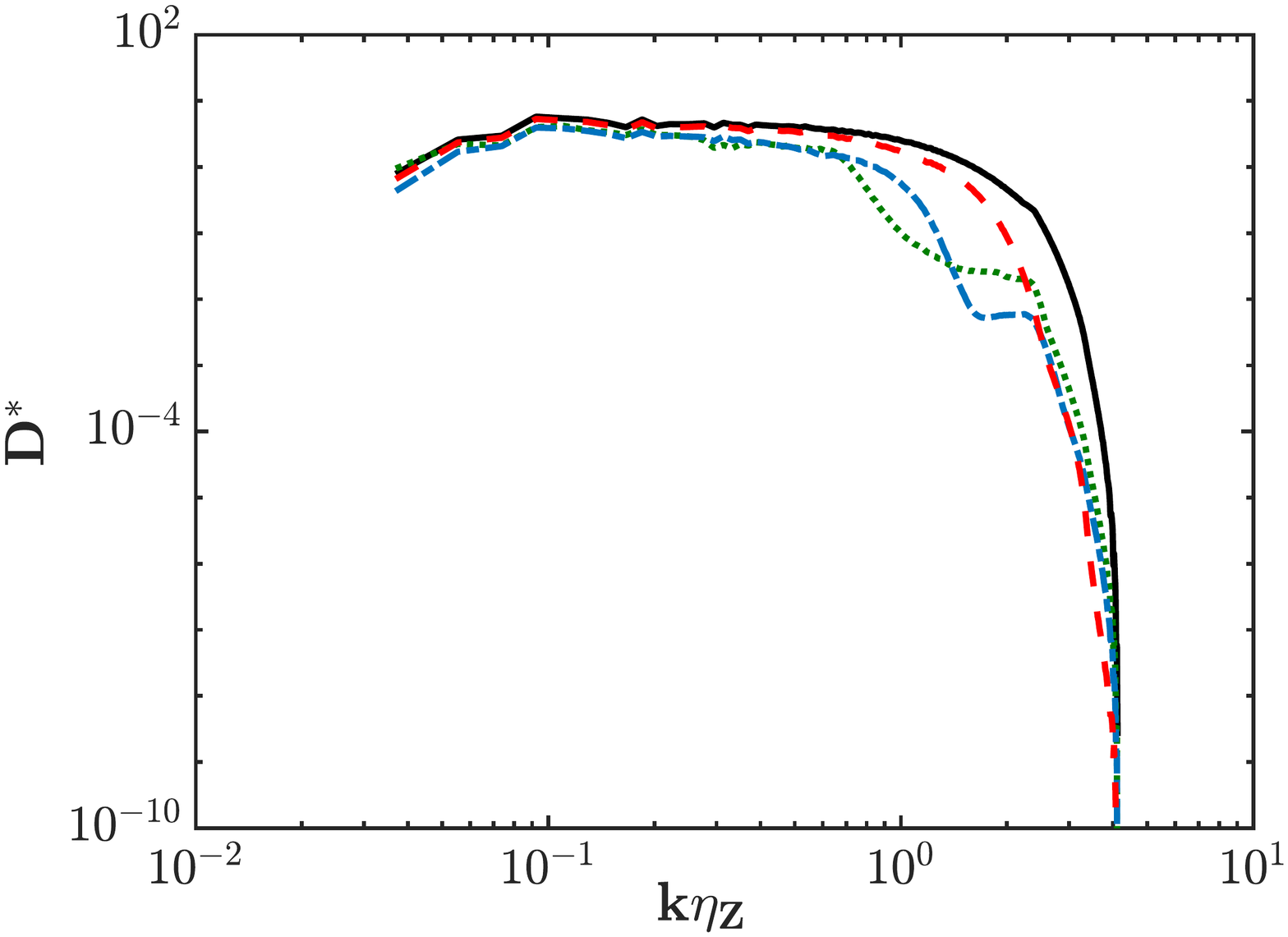}
\end{subfigure}%
\begin{subfigure}{0.5\textwidth}
\centering
\includegraphics[width=\linewidth]{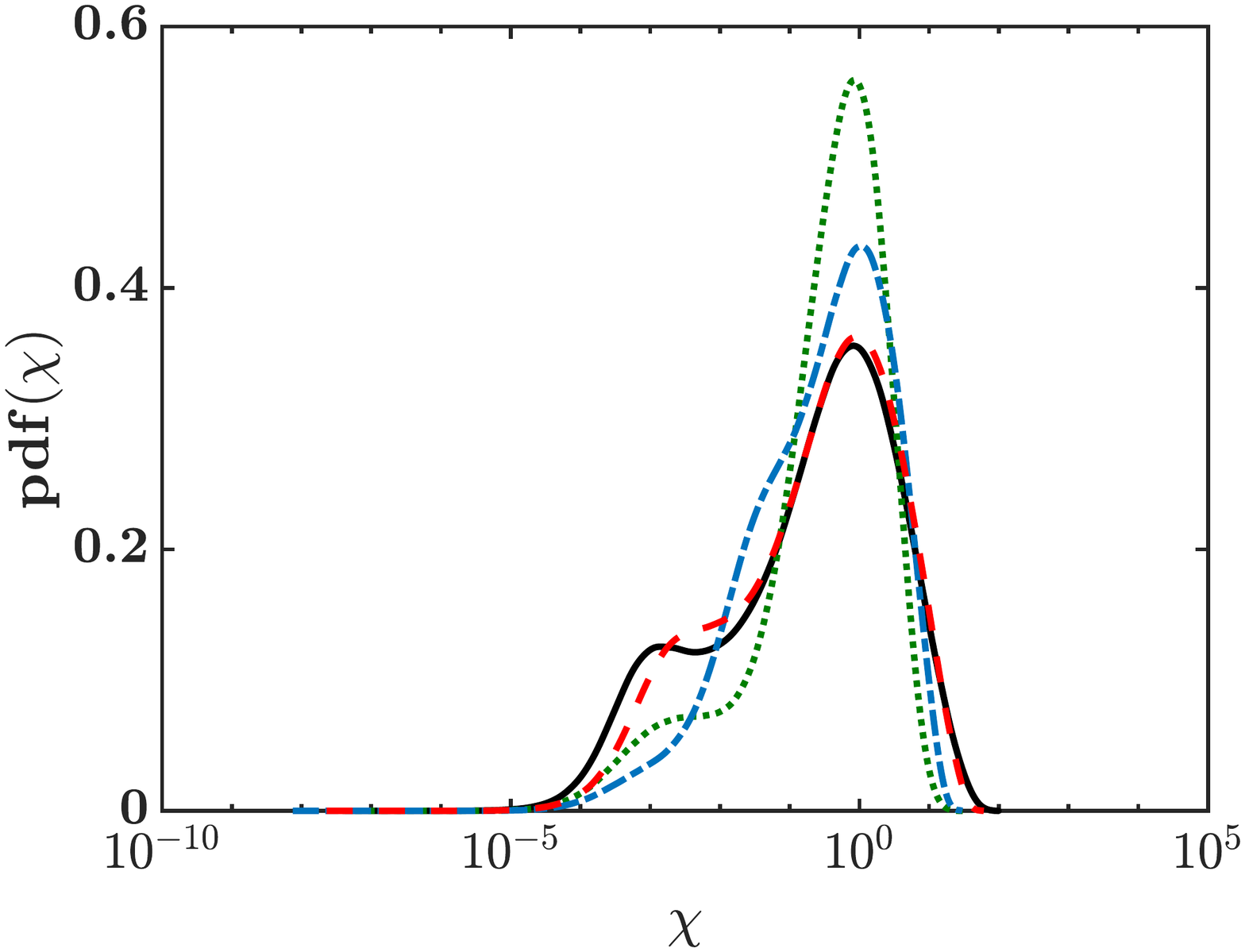}
\end{subfigure}%
\caption{Statistical properties of mixture fraction dissipation rate, $\chi$, at $t/\tau \approx 0.25$. Top: conditional average (left) and conditional variance (right) of $\chi$ in mixture fraction space. Bottom: spectrum of the scalar dissipation rate normalized by $D (\epsilon \eta_Z)^{-1/3} \langle \chi \rangle $, where $\epsilon$ is the total dissipation, and $\eta_Z = \eta Sc^{-3/4}$ (left); and probability distribution function of $\chi$ (right). DNS: {\mythickline{black}}, AIM with $k_m = 16$: {\mydot{ao(english)}}{\mydot{ao(english)}}{\mydot{ao(english)}}, AIM with $k_m = 32$: {\mythickdasheddottedline{blue}}, AIM with $k_m = 64$ \mythickdashedline{bostonuniversityred}.}
\label{fig:Case3-apriori-X}
\end{figure}

Additional statistics of the scalar dissipation rate computed from the AIM reconstructed mixture fraction field is provided in Fig.~\ref{fig:Case3-apriori-X}, where three approximate inertial manifolds are compared against the DNS data at a time instance where the mixture fraction fluctuations (standard deviation of the field) is maximum. Over all, by increasing the AIM resolution ($m$), the approximation improves and converges to the exact solution. Figure~\ref{fig:Case3-apriori-X} (top) shows that the conditional average of $\chi$ is captured quite accurately with the higher dimensional AIMs, but conditionally averaged fluctuations of mixture fraction are underestimated closer to the stoichiometric mixture fraction value even at the highest resolution AIM. Distribution of the range of scales in the mixture fraction field fluctuations is compared between DNS and AIM reconstructed fields with the normalized spectrum of the mixture fraction dissipation rate (Fig.~\ref{fig:Case3-apriori-X}, bottom left). The wavenumber space is scaled with the smallest scale of the scalar field $\eta_Z = \eta Sc^{-3/4}$, where $\eta$ is the Kolmogorov scale and $\eta_Z$ is the Batchelor's scale. AIM approximation is able to model the dissipation rate of mixture fraction at large scales, where there is a plateau in the spectrum corresponding to the inertial range of the turbulent scales, but dissipation rate of the mixture fraction is underestimated at the small scales. The approximation improves by increasing the AIM resolution, but it cannot recover the small-scale dissipation rate entirely. Finally, probability distribution of the dissipation rate (Fig.~\ref{fig:Case3-apriori-X} bottom right) shows that in the lower-dimensional AIM, the maximum of $\chi$ is underestimated, and the range of scales is not captured accurately. However, the approximation is considerably improved by increasing $m$.  

\begin{figure}[H]
\begin{subfigure}{0.5\textwidth}
\centering
\includegraphics[width=\linewidth]{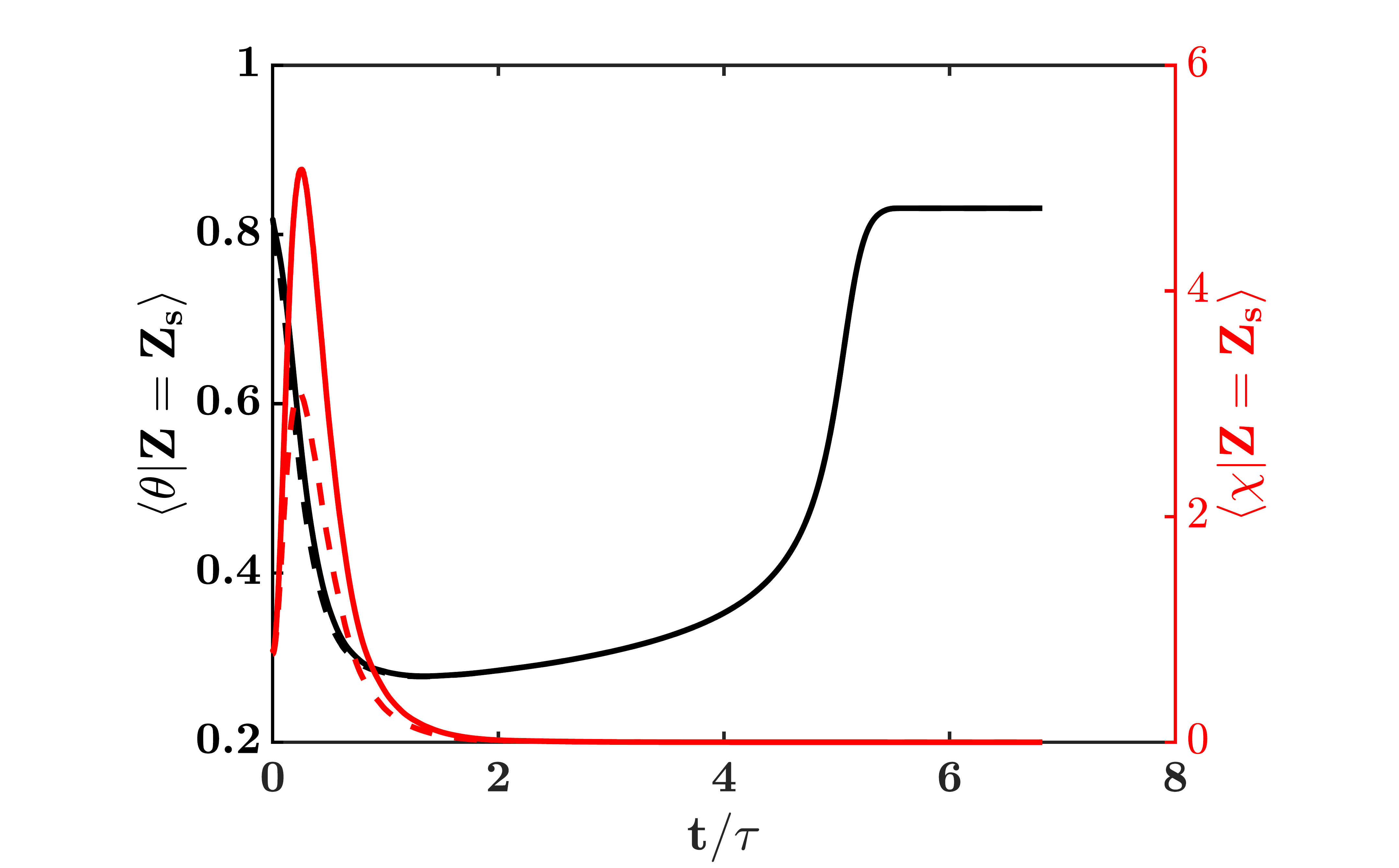}
\end{subfigure}%
\begin{subfigure}{0.5\textwidth}
\centering
\includegraphics[width=\linewidth]{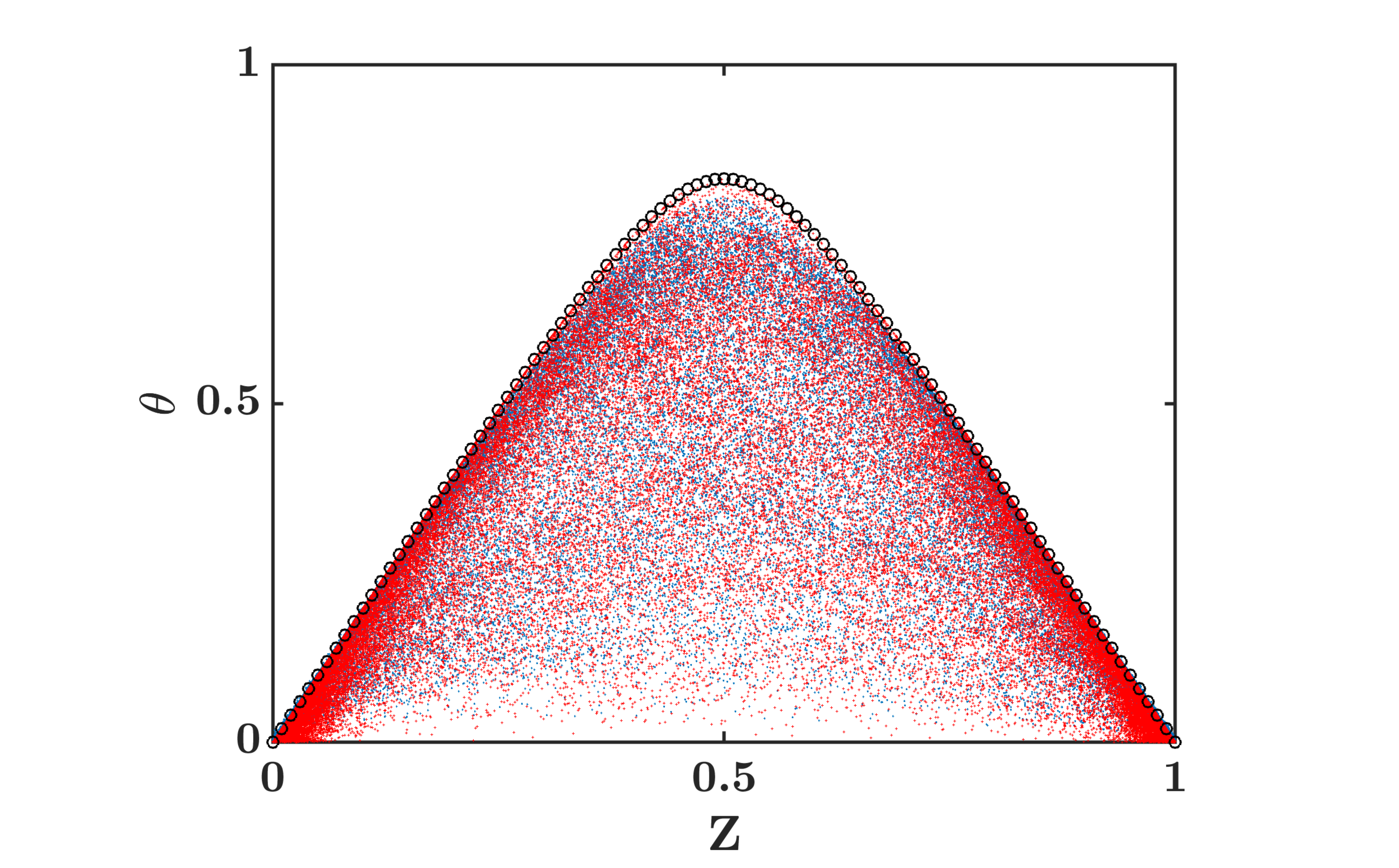}
\end{subfigure}%

\begin{subfigure}{0.5\textwidth}
\centering
\includegraphics[width=\linewidth]{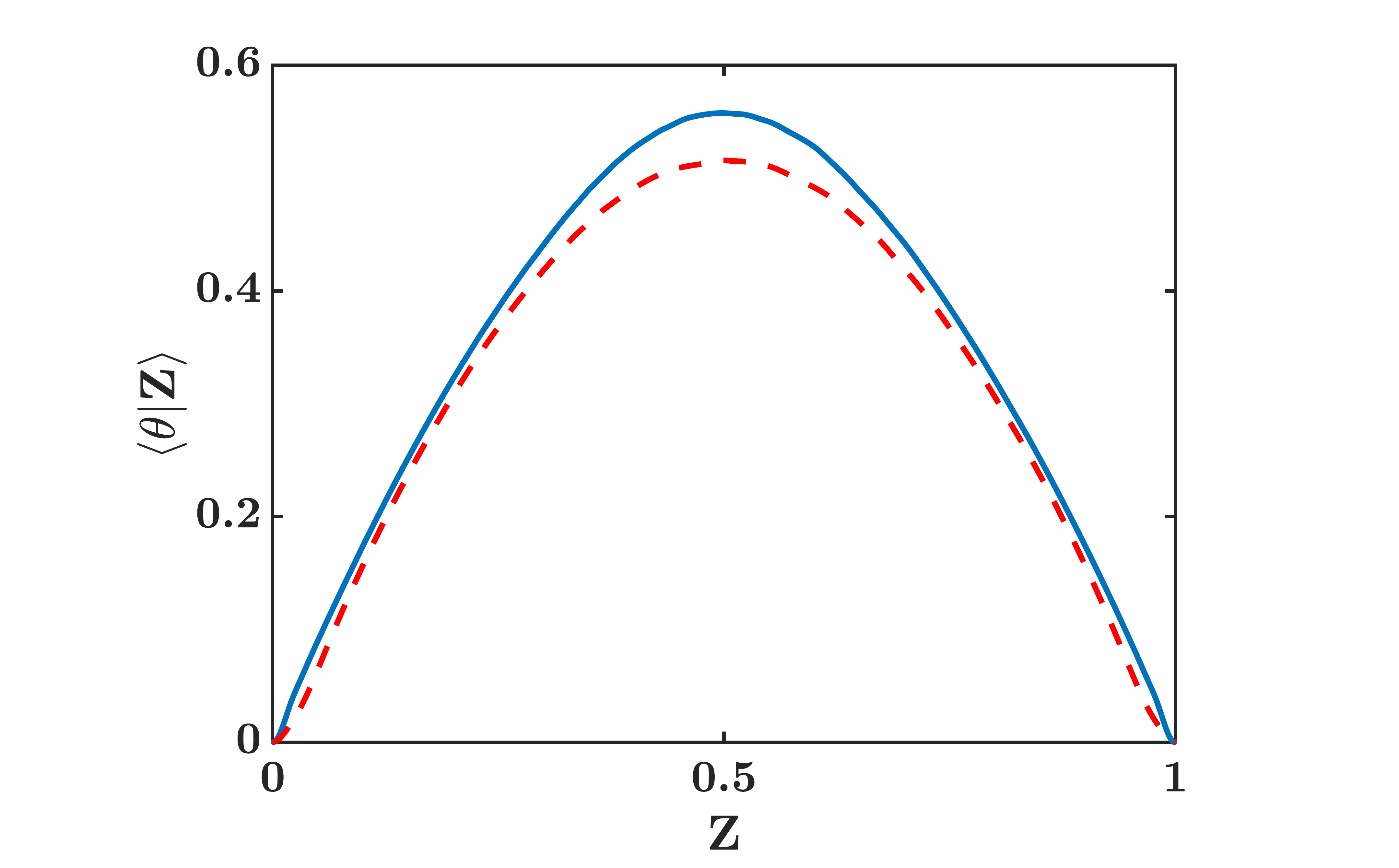}
\end{subfigure}%
\begin{subfigure}{0.5\textwidth}
\centering
\includegraphics[width=\linewidth]{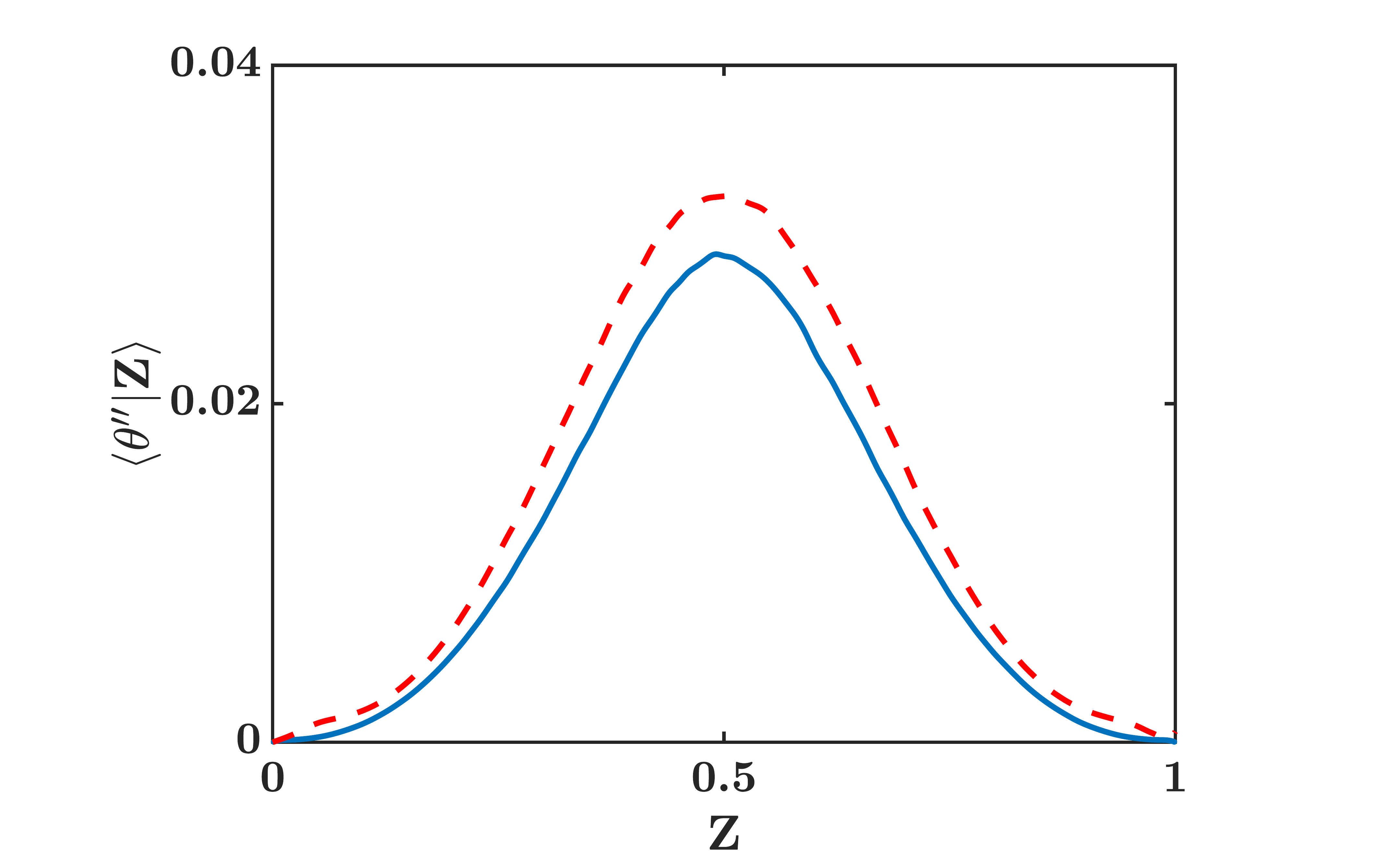}
\end{subfigure}%

\begin{subfigure}{0.5\textwidth}
\centering
\includegraphics[width=\linewidth]{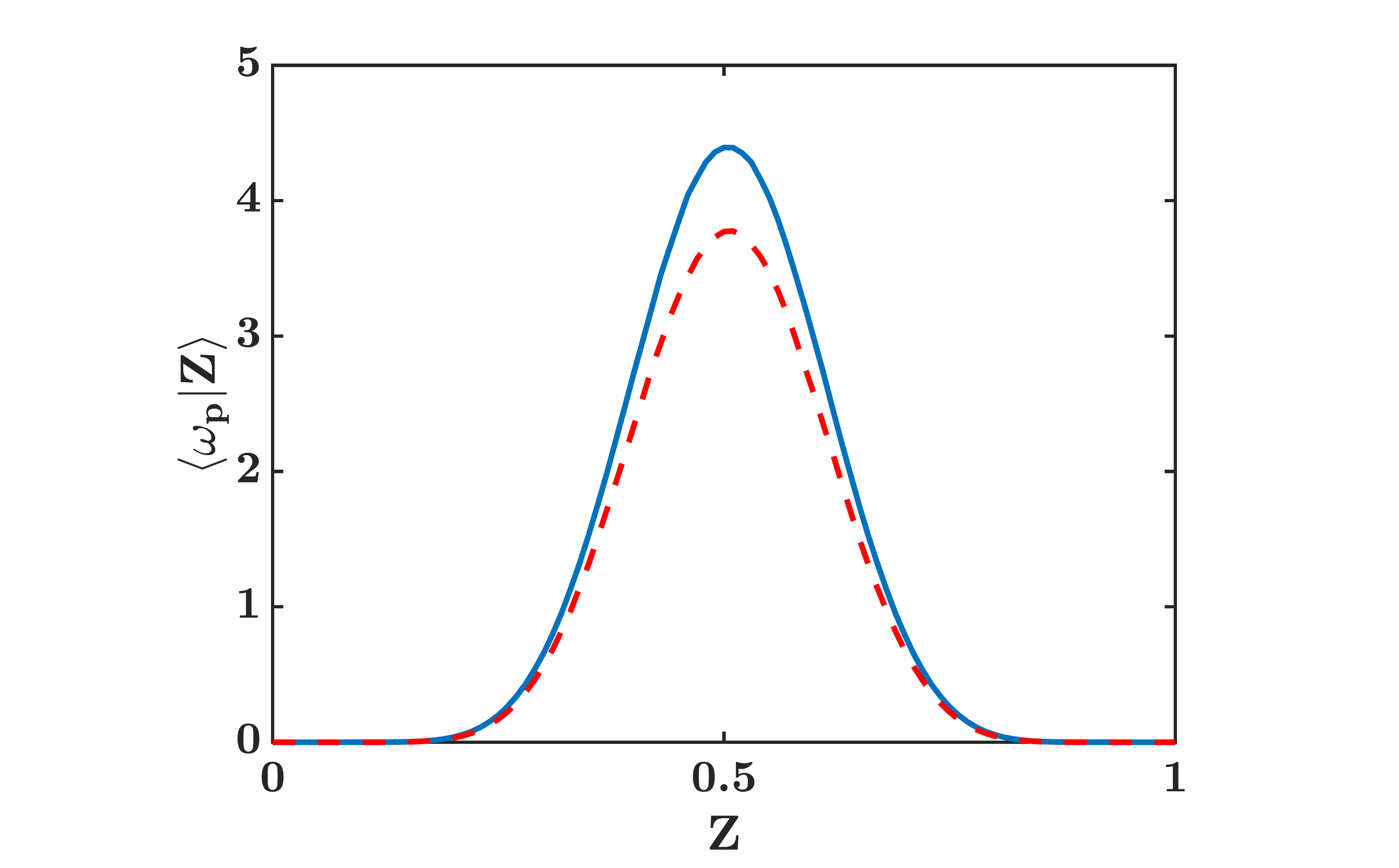}
\end{subfigure}%
\begin{subfigure}{0.5\textwidth}
\centering
\includegraphics[width=\linewidth]{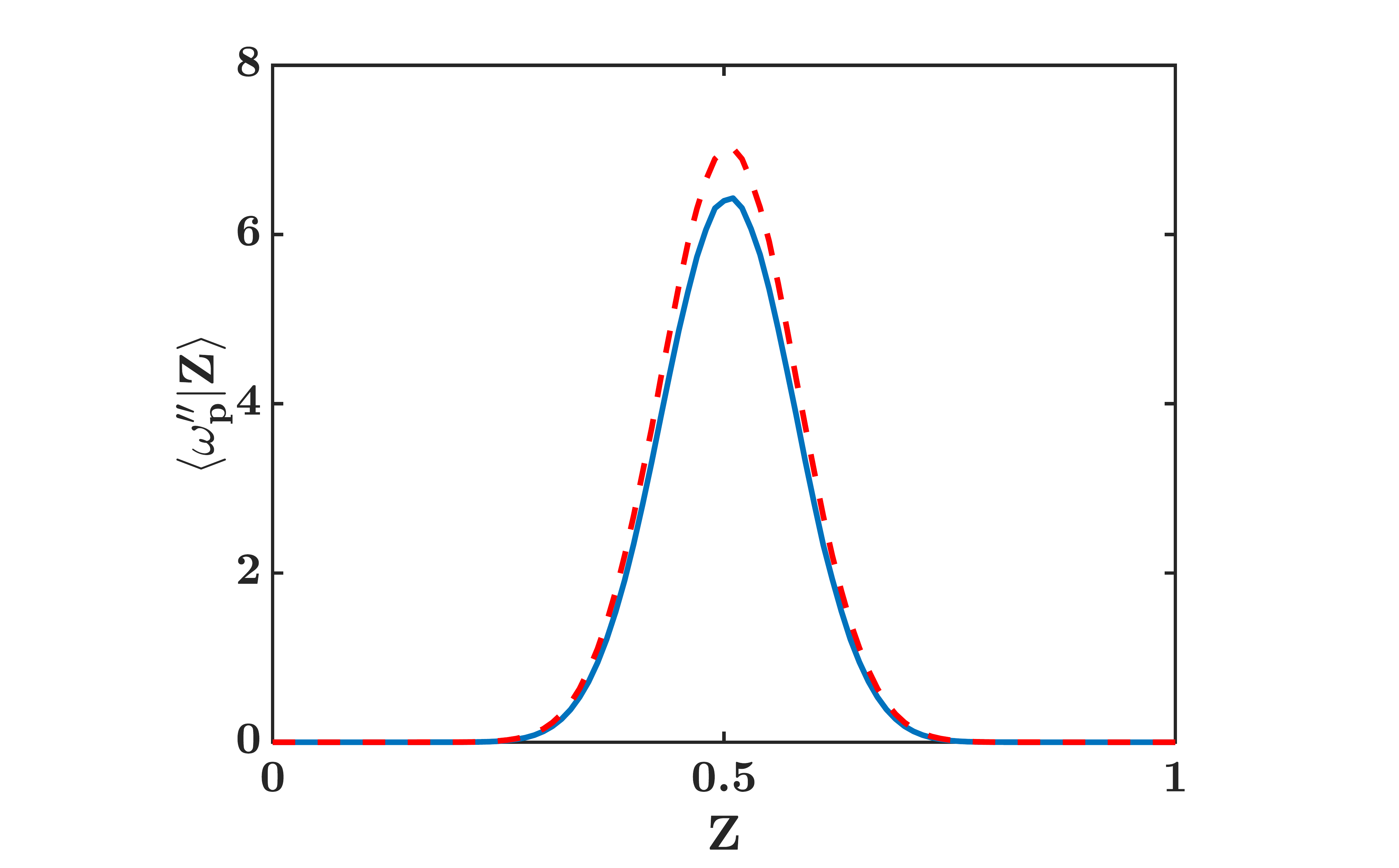}
\end{subfigure}%
\caption{Statistical behavior of flame $III$; top left: time evolution of conditionally averaged temperature and mixture fraction dissipation rate at $Z = Z_s$; DNS: solid lines, AIM: dashed lines. Each color represents values for the matching vertical axis. Top right: Conditional distribution of temperature in mixture fraction space (same legend as Fig.~\ref{fig:AIM1-TimeScaleRatio}, bottom). Middle: conditional mean (left) and variance (right) of temperature. Bottom: conditional mean (left) and variance (right) of chemical reaction source term ($w_p$). DNS: {\mythickline{blue(ryb)}}, AIM: {\mythickdashedline{red}}. Cut-off wavenumber for AIM projection is $k_m = 32$.}
\label{fig:Case3-apriori-Y-Wp}
\end{figure}

Statistics of the temperature field and chemical reaction source term for flame $III$ are presented in Fig.~\ref{fig:Case3-apriori-Y-Wp}, showing that flame $III$ experiences a mild extinction followed by reignition (top left). The conditional distribution of temperature in mixture fraction space shows large scatter, indicating regions of local extinction as well as fully burning pockets. AIM approximation with $k_m = 32$ is able to capture this variation, both in terms of the conditional mean and variance of temperature. While there is a slight undeprediction of chemical source term conditional mean of reaction source term (bottom left), this effect is minimal in the prediction of the conditional mean of temperature. AIM reconstruction overestimates variance of chemical reaction source term, indicating that overall the predictions show higher degree of extinction. 

\begin{figure}[H]
\begin{subfigure}{0.5\textwidth}
\centering
\includegraphics[width=\linewidth]{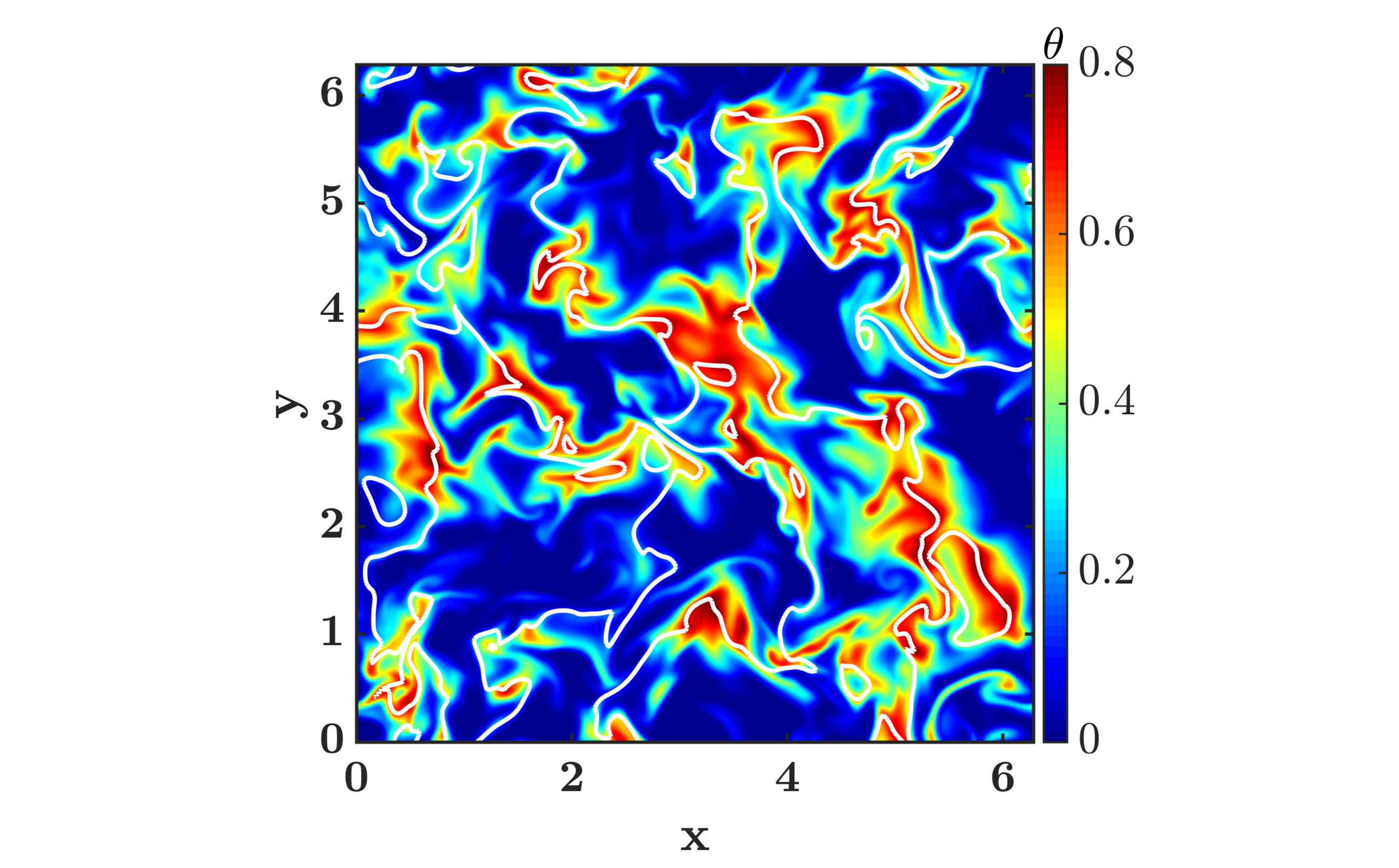}
\end{subfigure}%
\begin{subfigure}{0.5\textwidth}
\centering
\includegraphics[width=\linewidth]{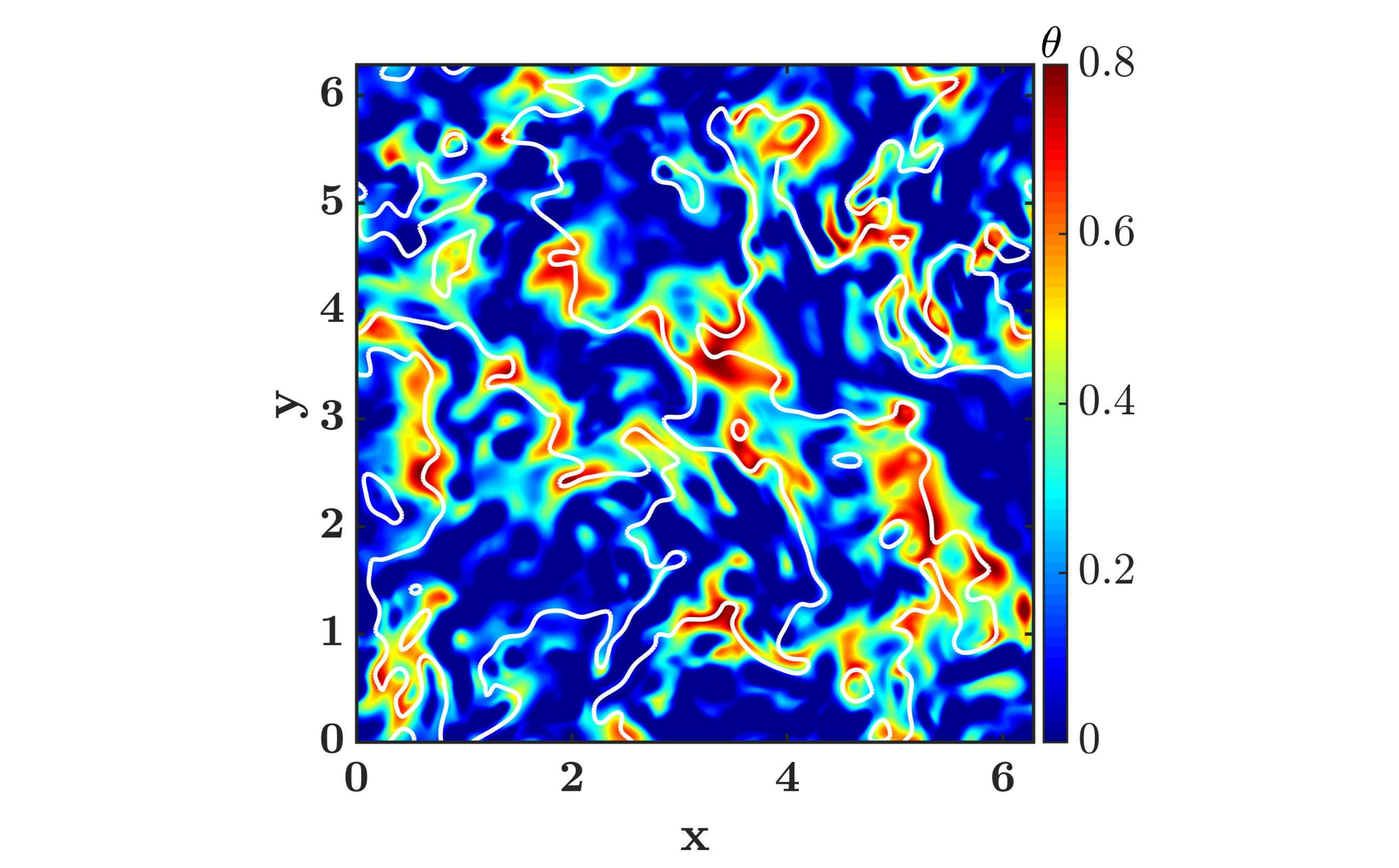}
\end{subfigure}%
\caption{Temperature field for flame $III$ at $t/\tau = 0.25$, when $\langle \chi | Z = Z_s \rangle$ is maximum. Left: DNS, right: AIM with $k_m = 16$. White contour lines show $Z = Z_s$.}
\label{fig:Case3-apriori-contour-Y}
\end{figure}

Finally, temperature field of flame $III$ reconstructed by a lower resolution AIM is compared against the DNS field in Fig.~\ref{fig:Case3-apriori-contour-Y}. Snapshots are taken when the mean dissipation rate at the flame surface is the highest. Details of the flame distribution is captured in the AIM reconstructed field, but the flame temperature is underestimated in some regions. 

\subsection{A posteriori analysis of AIM}
\label{sec:aposteriori}
In this section, the performance of AIM approximation is assessed {\it a posteriori} such that only the resolved dynamics are evolved to the next time step (Eq.~\ref{u-gov}), while the unresolved variables are approximated using the information of the resolved variables (Eq.~\ref{eq:AIM-app}), and the nonlinear term ($P{\cal F}({\bm u}+ {\bm w})$), is computed from the AIM-reconstructed full-dimensional vector of variables to close the governing equation of resolved variables. For each AIM resolution ($m$), the full-dimensional field modeled by AIM is compared against the DNS field. Here, three different resolutions ($k_m = 16, 32, 64$) are analyzed for different Damkohler numbers. The velocity and scalar fields are evolved using the AIM, but the scalar field properties are discussed below.

\begin{figure}[H]
\begin{subfigure}{0.33\textwidth}
\centering
\includegraphics[width=\linewidth]{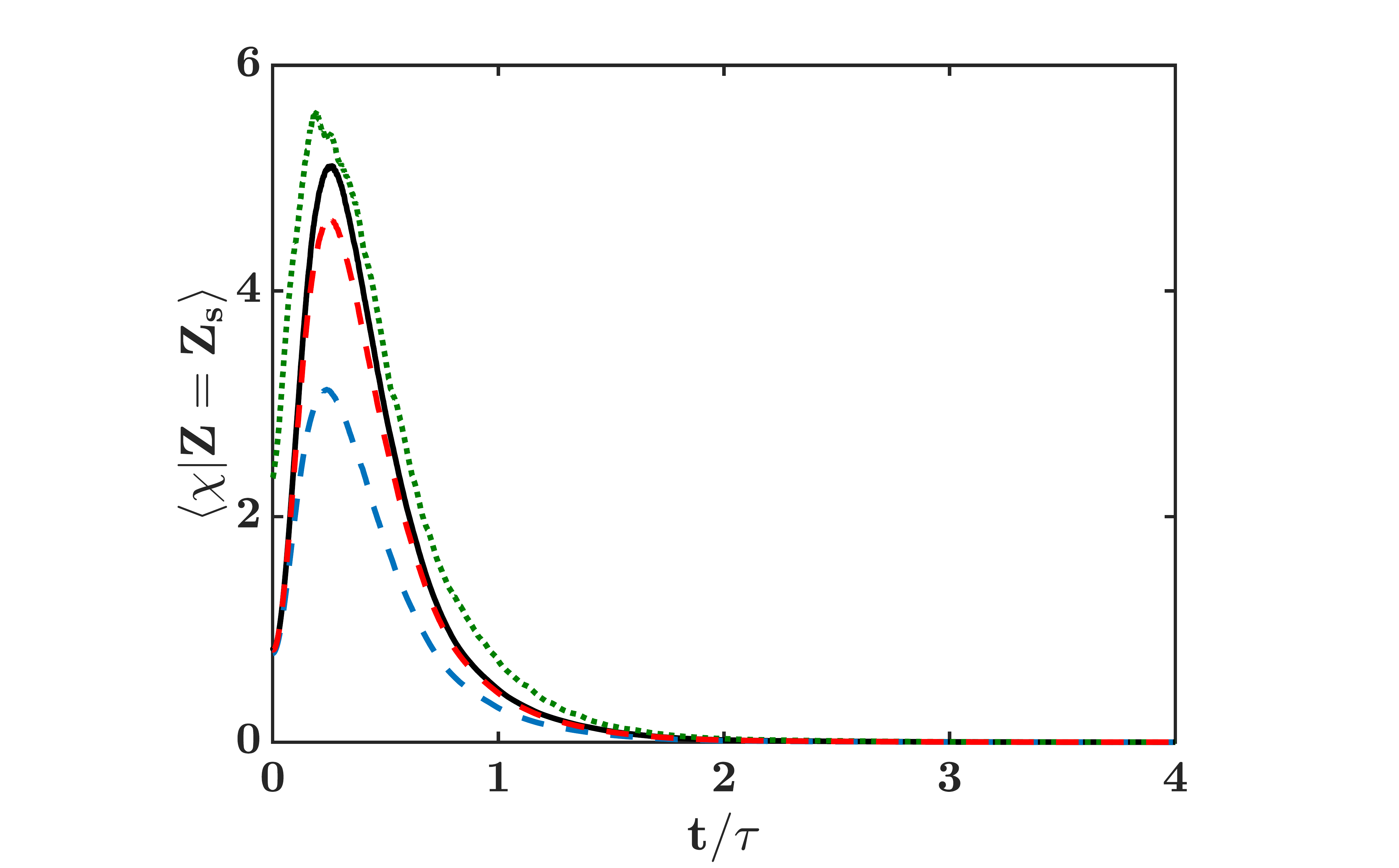}
\end{subfigure}%
\begin{subfigure}{0.33\textwidth}
\centering
\includegraphics[width=\linewidth]{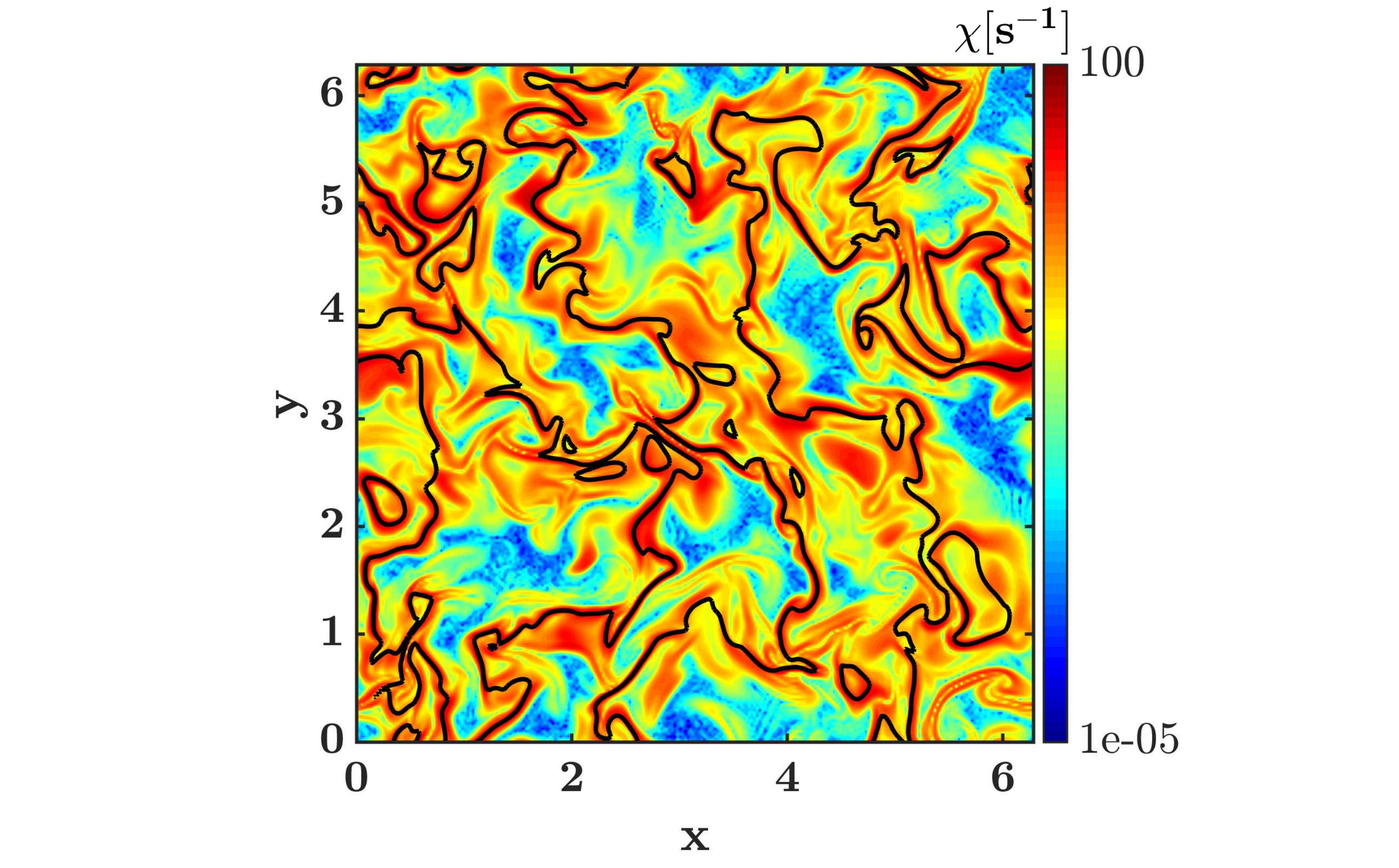}
\end{subfigure}%
\begin{subfigure}{0.33\textwidth}
\centering
\includegraphics[width=\linewidth]{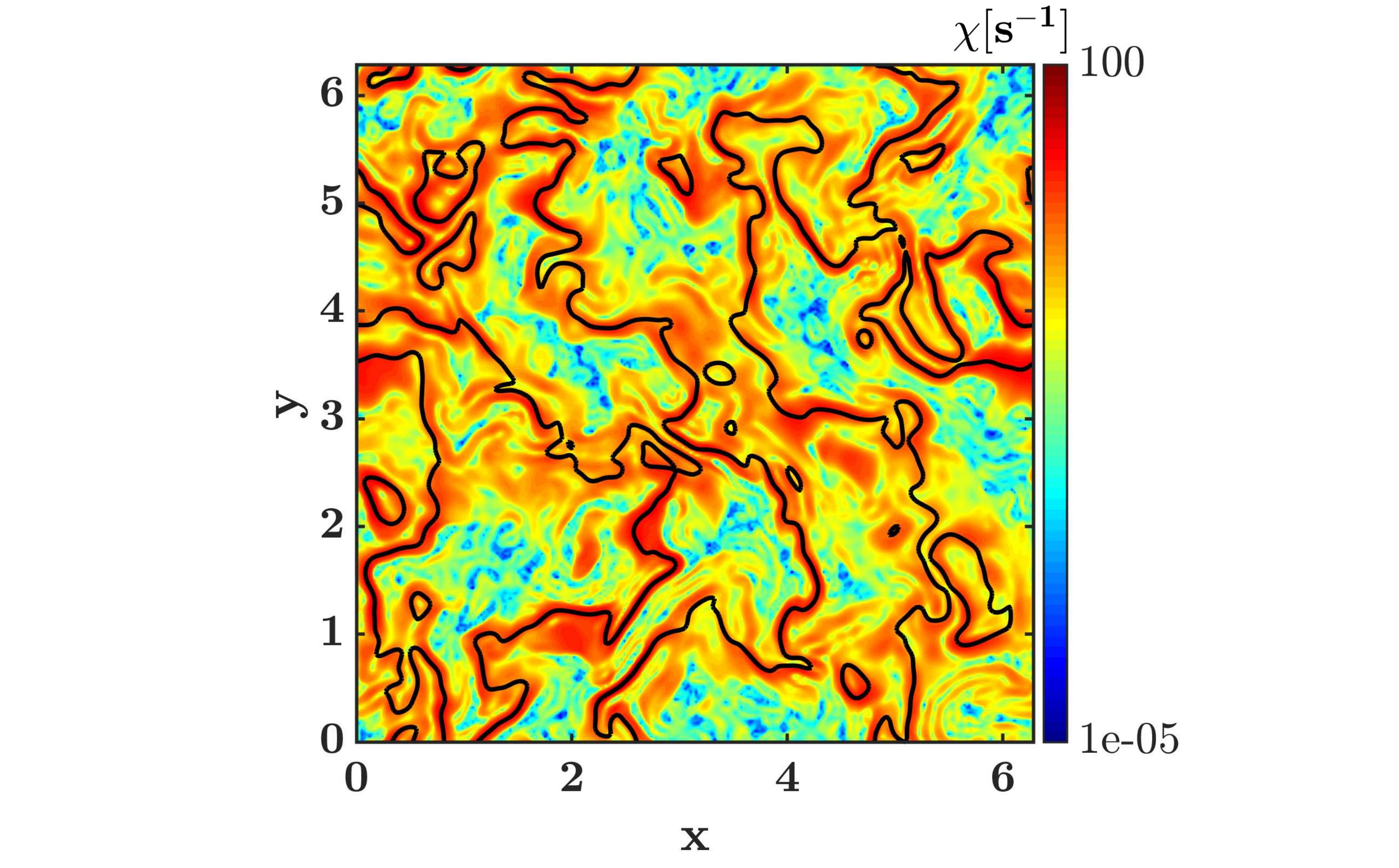}
\end{subfigure}%
\caption{Left: time evolution of conditionally averaged $\chi$ at stoichiometric mixture; DNS:{\mythickline{black}}, AIM with $k_m = 16$: {\mydot{ao(english)}}{\mydot{ao(english)}}{\mydot{ao(english)}}, AIM with $k_m = 32$: {\mythickdasheddottedline{blue(ryb)}}, AIM with $k_m = 64$ \mythickdashedline{bostonuniversityred}. Middle: contour of $\chi$ in a plane of domain obtained from the DNS data. Right: contour of $\chi$ in a plane of domain obtained from AIM model with $k_m = 32$. Black lines in contour plots mark stoichiometric mixture ($Z = Z_s$), and fields are extracted at $t/\tau \approx 0.25 $. }
\label{fig:AIM2-X-Prop-Convergence}
\end{figure}

Time evolution of dissipation rate of mixture fraction at stoichiometric mixture modeled by AIM is shown in Fig.~\ref{fig:AIM2-X-Prop-Convergence} (left). Convergence of AIM prediction to the exact solution is different from the {\it a priori} analysis, and the lowest dimensional AIM ($k_m = 16$) overpredicts the dissipation rate. The reason for this behavior is that the approximate inertial manifold at this resolution ($k_m = 16, m = 8937$) is lower dimensional than the estimated attractor dimension for the underlying turbulent field \cite{hassanaly2019lyapunov}. The insufficient resolution cannot reconstruct the dissipative range of the turbulent motions which manifests as enhanced mixing and higher degree of turbulence. Spatial variation of the mixture fraction dissipation rate under the effect of turbulent straining is compared in Fig.~\ref{fig:AIM2-X-Prop-Convergence} (middle and right). When turbulent mixing is dominant ($t/\tau \approx 0.25$), the stoichiometric mixture, represented with black lines, experiences the steepest gradients and becomes corrugated. It can be seen that a moderate resolution AIM ($k_m = 32$) is able to capture details of spatial structures quite accurately but it overestimates the scalar dissipation rate at less dissipative regions. 

\begin{figure}[H]
\begin{subfigure}{0.33\textwidth}
\centering
\includegraphics[width=\linewidth]{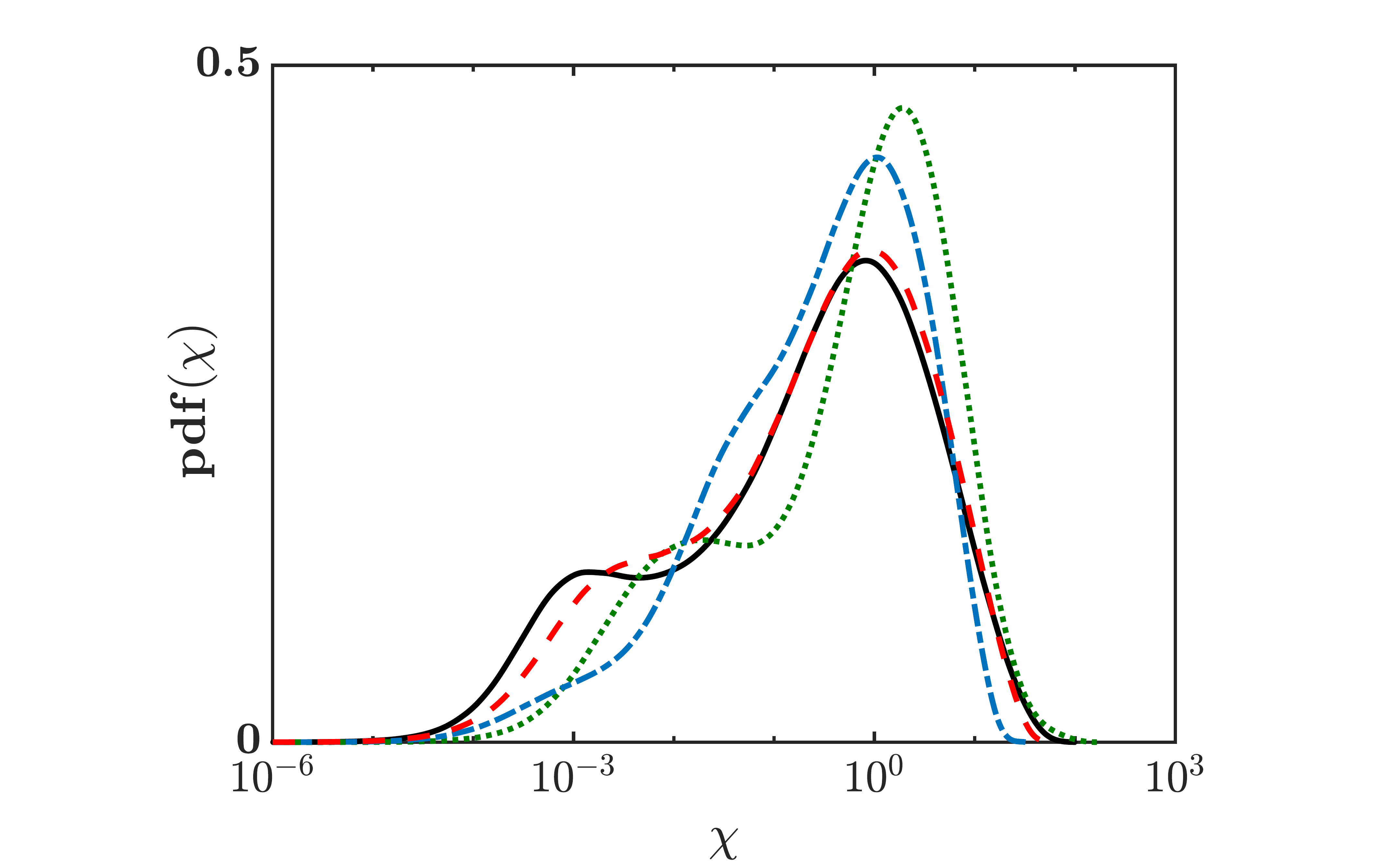}
\end{subfigure}%
\begin{subfigure}{0.33\textwidth}
\centering
\includegraphics[width=\linewidth]{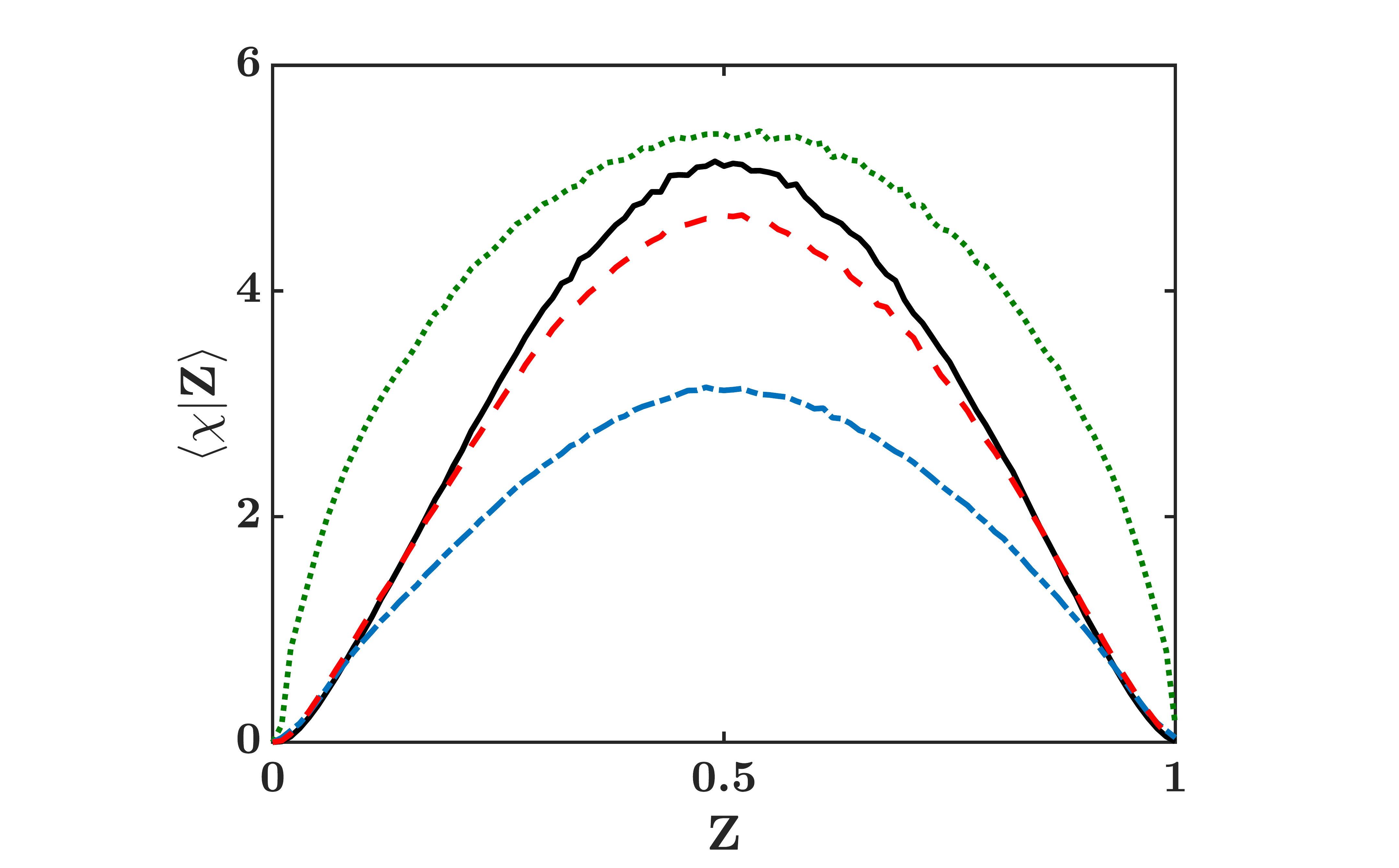}
\end{subfigure}%
\begin{subfigure}{0.33\textwidth}
\centering
\includegraphics[width=\linewidth]{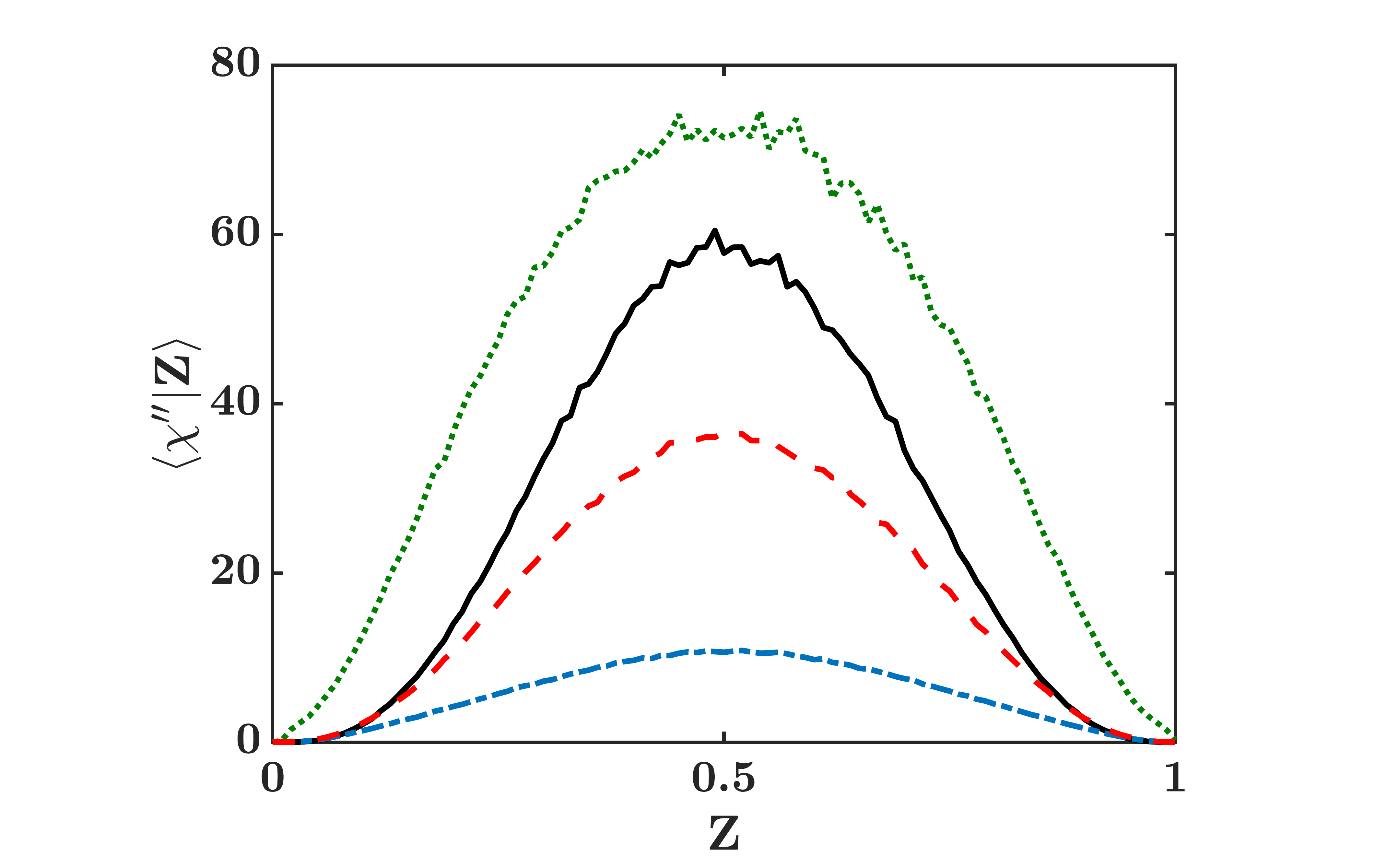}
\end{subfigure}%
\caption{Convergence of AIM-ROM in modeling mixture fraction dissipation rate ($\chi = 2D (\nabla Z)^2$) at $t/\tau \approx 0.25$ when maximum straining effect occurs. Left: probability density function of $\chi$, middle: conditional expectation of $\chi$, and right: conditional variance of $\chi$. DNS:{\mythickline{black}}, AIM with $k_m = 16$: {\mydot{ao(english)}}{\mydot{ao(english)}}{\mydot{ao(english)}}, AIM with $k_m = 32$: {\mythickdasheddottedline{blue}}, AIM with $k_m = 64$ \mythickdashedline{bostonuniversityred}.}
\label{fig:AIM2-X-Properties}
\end{figure}

AIM model performance in modeling scalar dissipation rate is evaluated further in Fig.~\ref{fig:AIM2-X-Properties}. While probability distribution and conditional distribution of $\chi$ are captured well at higher resolutions, conditional standard deviation of the dissipation field are underestimated considerably. The results shown here are when the turbulence effects are most prominent. In all cases, as the system moves towards equilibrium, the AIM performance improves. 

\begin{figure}[H]
\begin{subfigure}{0.5\textwidth}
\centering
\includegraphics[width=\linewidth]{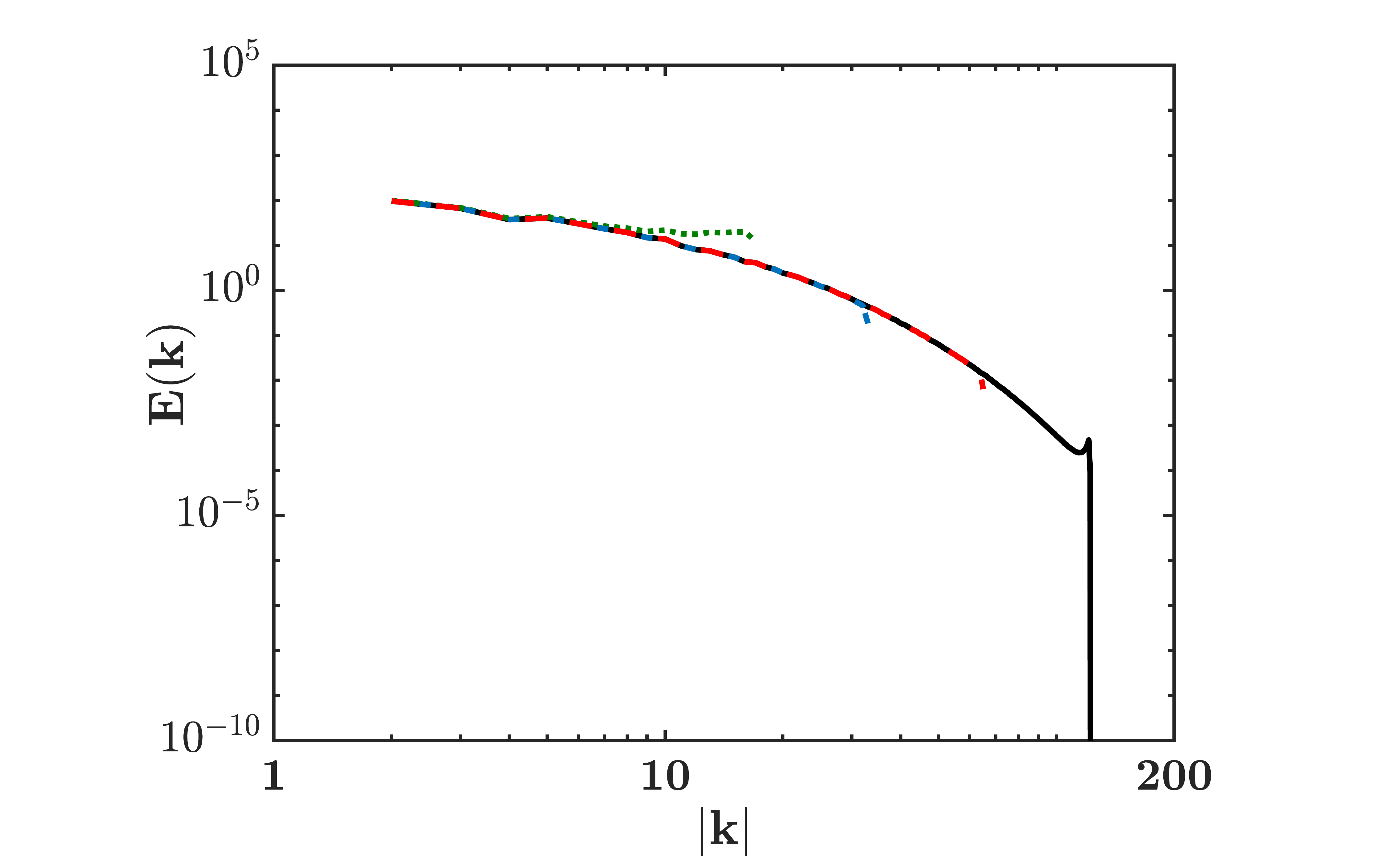}
\end{subfigure}%
\begin{subfigure}{0.5\textwidth}
\centering
\includegraphics[width=\linewidth]{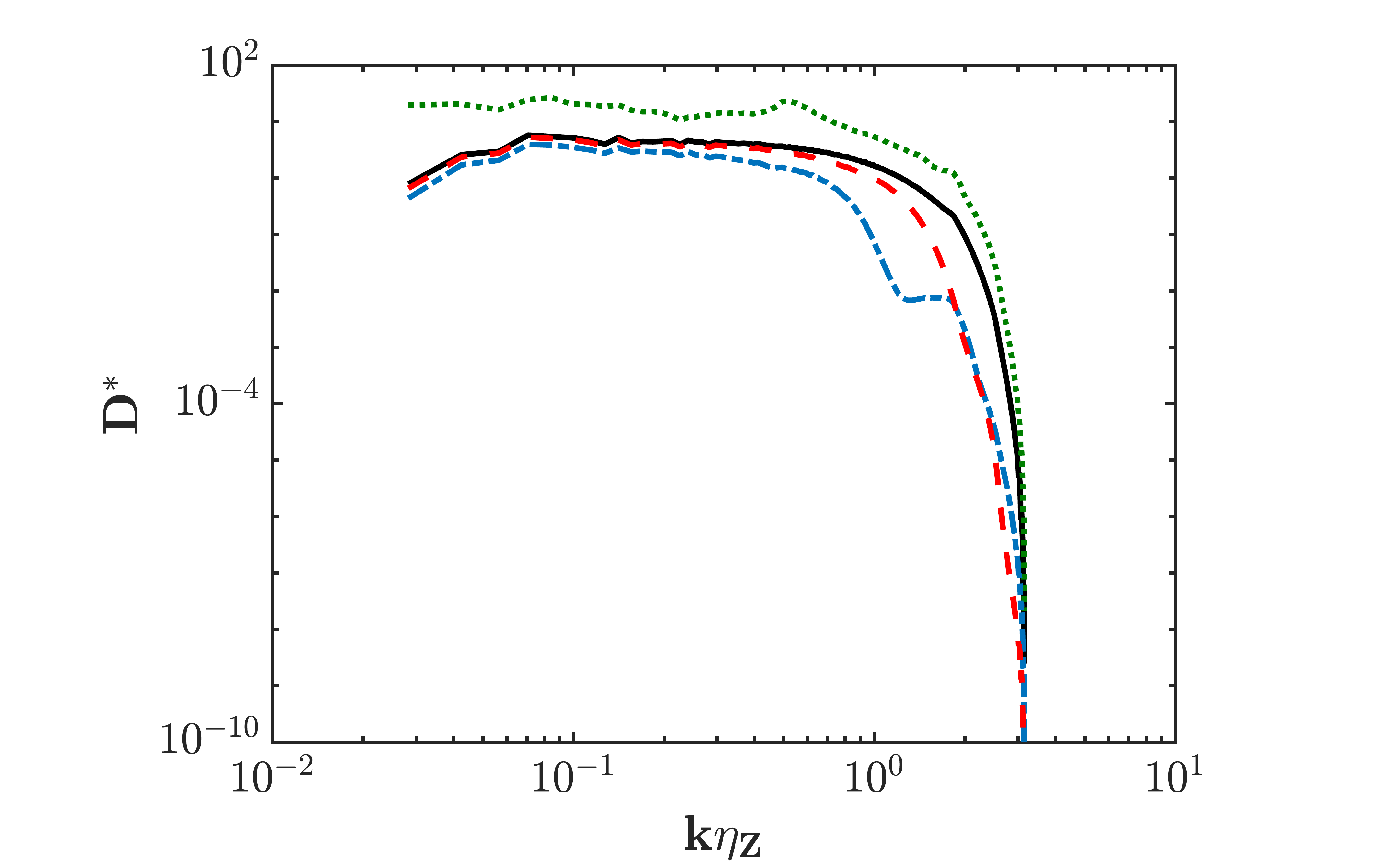}
\end{subfigure}%
\caption{Convergence of AIM-ROM in modeling energy spectrum of the velocity field (left) and mixture fraction dissipation rate (right) at $t/\tau \approx 0.25$ when maximum straining effect occurs. DNS:{\mythickline{black}}, AIM with $k_m = 16$: {\mydot{ao(english)}}{\mydot{ao(english)}}{\mydot{ao(english)}}, AIM with $k_m = 32$: {\mythickdasheddottedline{blue}}, AIM with $k_m = 64$ \mythickdashedline{bostonuniversityred}.}
\label{fig:AIM2-Spectrums}
\end{figure}

In Fig.~\ref{fig:AIM2-Spectrums}, the AIM modeled field is compared against the DNS field in the spectral space by comparing energy spectrum of the velocity field (left) and scalar dissipation rate (right) at different resolutions. The modeled turbulent energy spectrum contains only the modeled (resolved) scales at each AIM resolution. Different AIM resolutions are able to capture the large energy-containing and inertial range scales accurately. At the smallest AIM dimension ($k_m = 16$), energy spectrum of the smaller resolved scales (close to the cut-off wavenumber) is overestimated, as there is not enough dissipation at this resolution. This behavior has led to a more turbulent field at this resolution, which enhances mixing of the scalars and results in higher dissipation rate energy of scalars (Fig.~\ref{fig:AIM2-Spectrums}, right). For higher dimensional AIM, energy spectrum of turbulent field and mixture fraction dissipation rate is captured more accurately but still underestimated at the smallest scales. 

\begin{figure}[H]
\begin{subfigure}{0.5\textwidth}
\centering
\includegraphics[width=\linewidth]{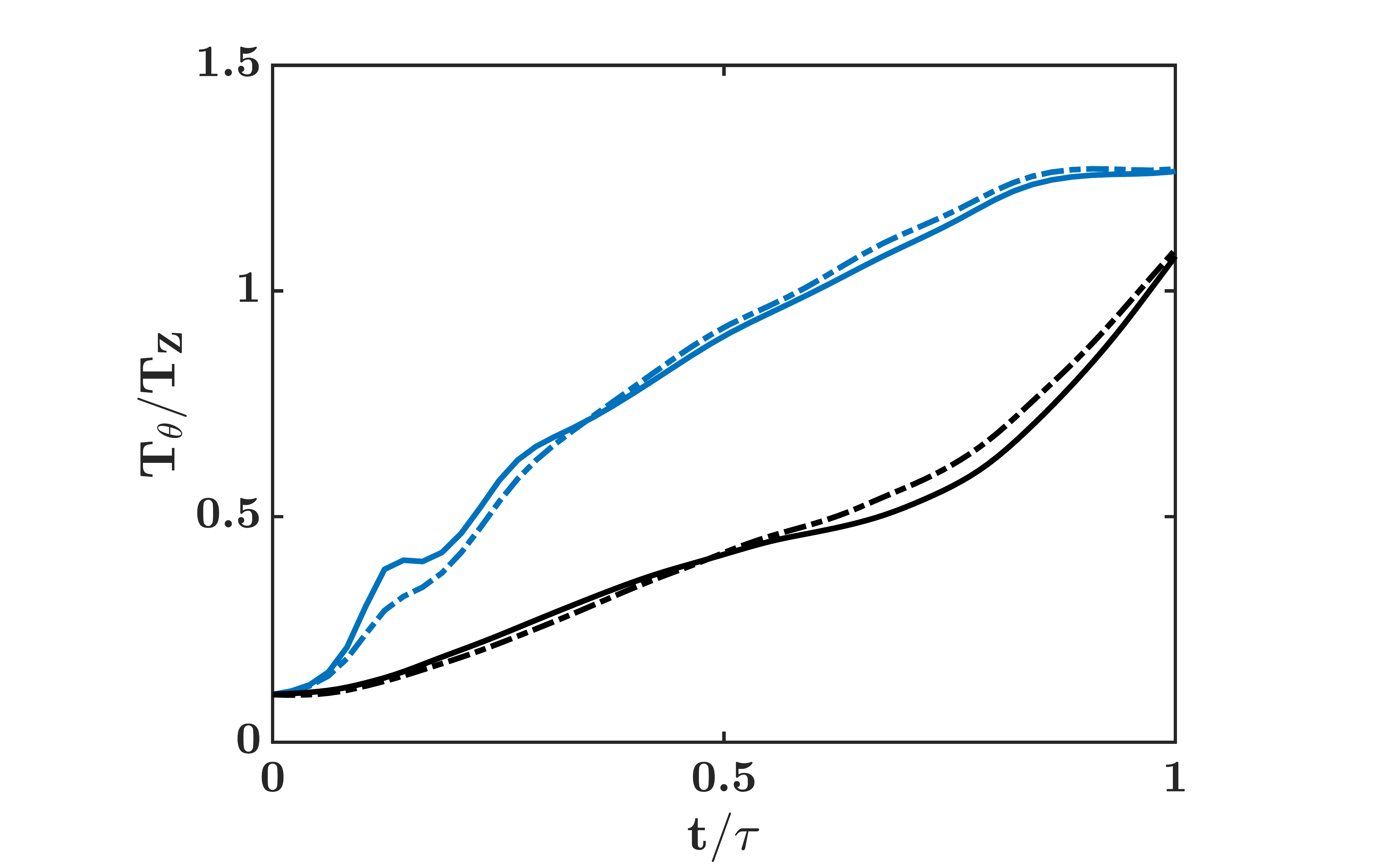}
\end{subfigure}%
\begin{subfigure}{0.5\textwidth}
\centering
\includegraphics[width=\linewidth]{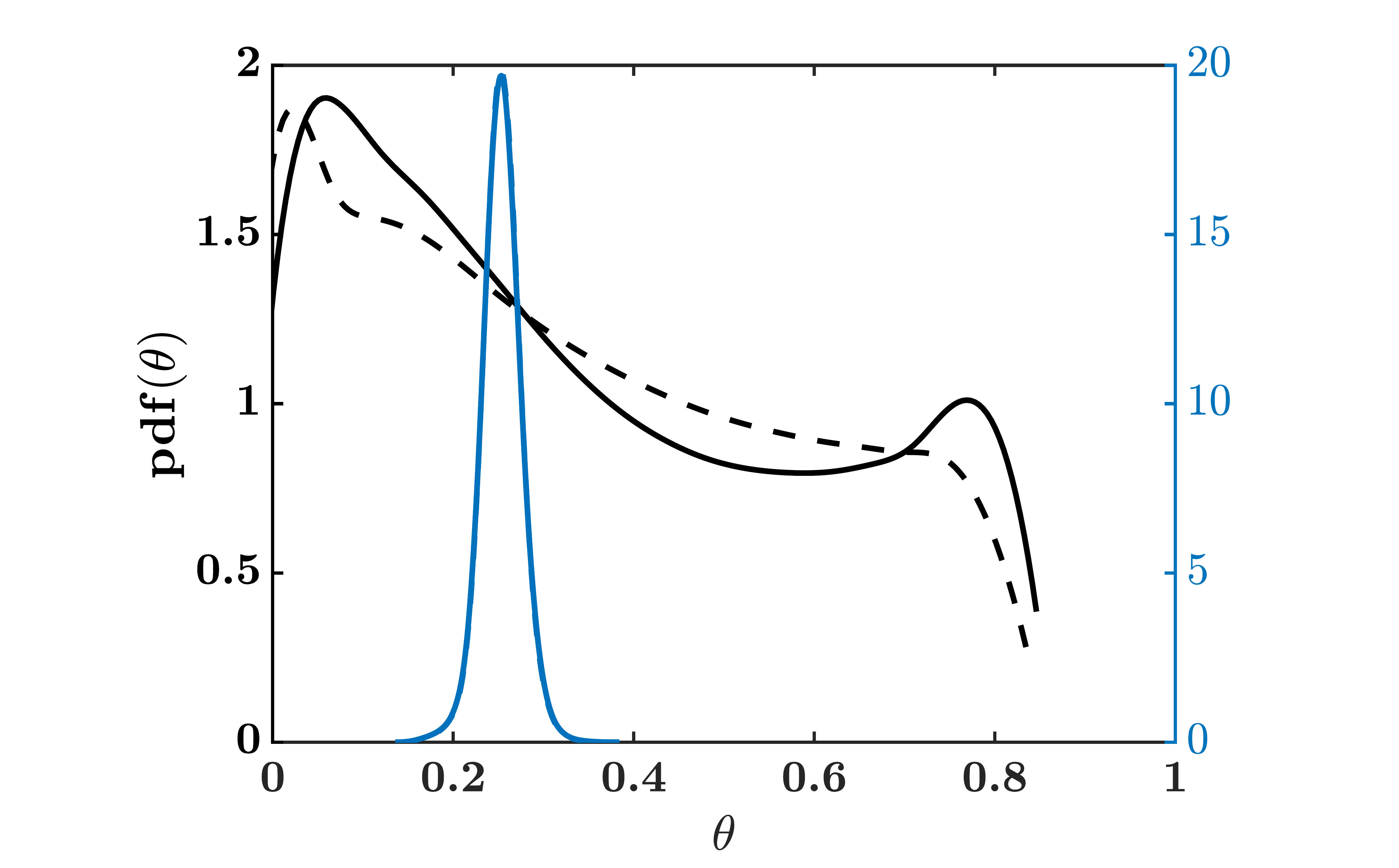}
\end{subfigure}%
\caption{Left: time evolution of mixing time scale ratio of progress variable and mixture fraction ($T_{\theta}/T_Z$), right: probability distribution of temperature at $t/\tau \approx 0.5$. DNS of flame $II$: {\mythickline{blue(ryb)}}, and flame $IV$: {\mythickline{black}}, AIM with $k_m = 32$ for flame $II$: {\mythickdashedline{blue(ryb)}} and flame $IV$: {\mythickdashedline{black}}.}
\label{fig:AIM2-mixingTimes}
\end{figure}

Figure~\ref{fig:AIM2-mixingTimes} (top left) compares $T_{\theta}/T_Z$ for flames $II$ and $IV$ up to $t/\tau = 1$, when both flames experience the highest scalar dissipation, causing flame $II$ to globally quench, but only partial extinction in flame $IV$. Conditional distribution of temperature of these flames when they experience their maximum extinction is compared in Fig.~\ref{fig:AIM2-mixingTimes}. As explained in the {\it a priori} analysis (Sec.~\ref{sec:apriori}), the faster reactions in flame $IV$ leads to smaller time scales, which is captured by the AIM approach. The global extinction reduces the support of the temperature distribution function, with near-Gaussian spread for flame $II$. On the other hand, there is a broader distribution for flame $IV$, with significant regions of low temperature consistent with local extinction-reignition process. 

\begin{figure}[H]
\begin{subfigure}{0.33\textwidth}
\centering
\includegraphics[width=\linewidth]{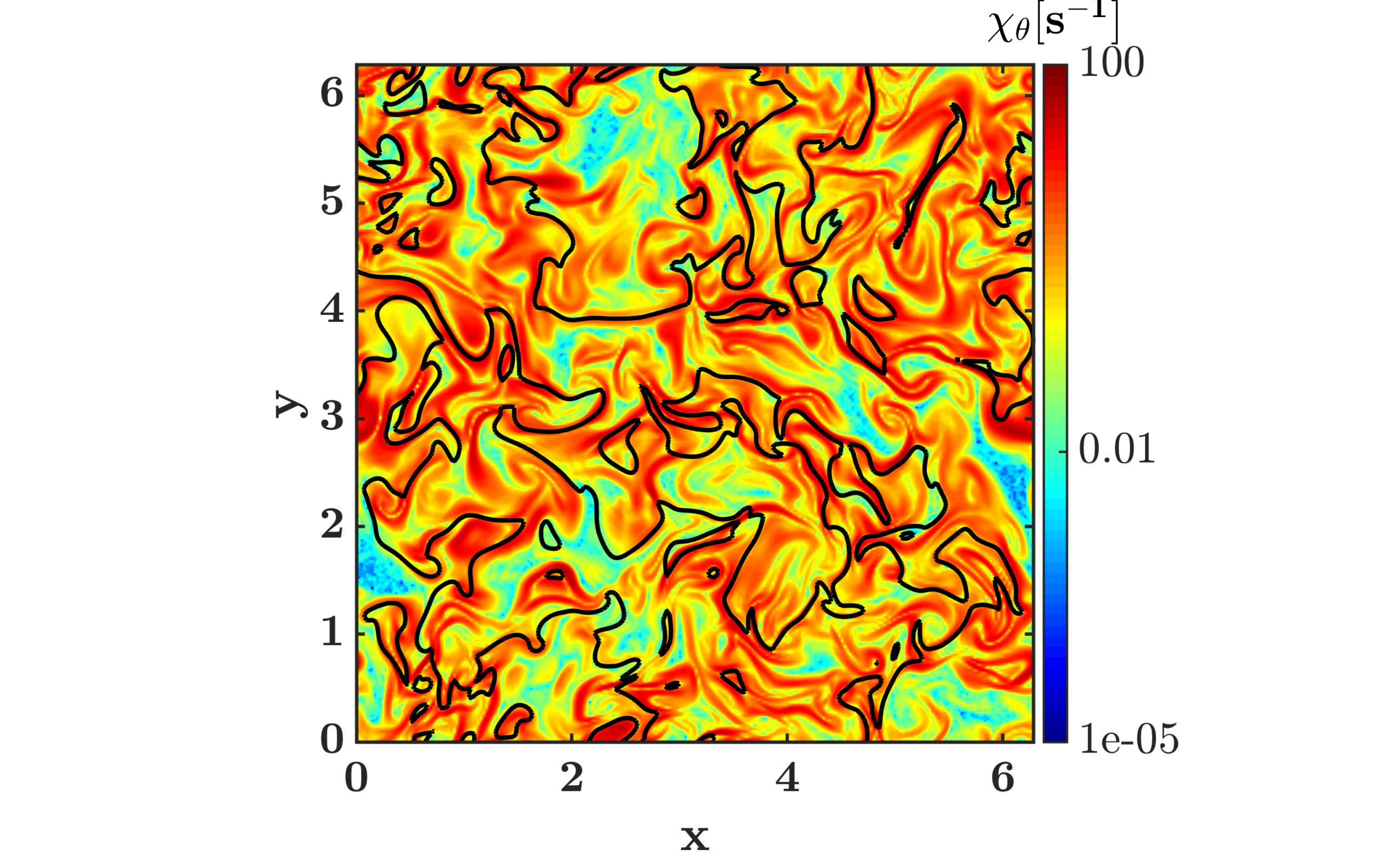}
\end{subfigure}%
\begin{subfigure}{0.33\textwidth}
\centering
\includegraphics[width=\linewidth]{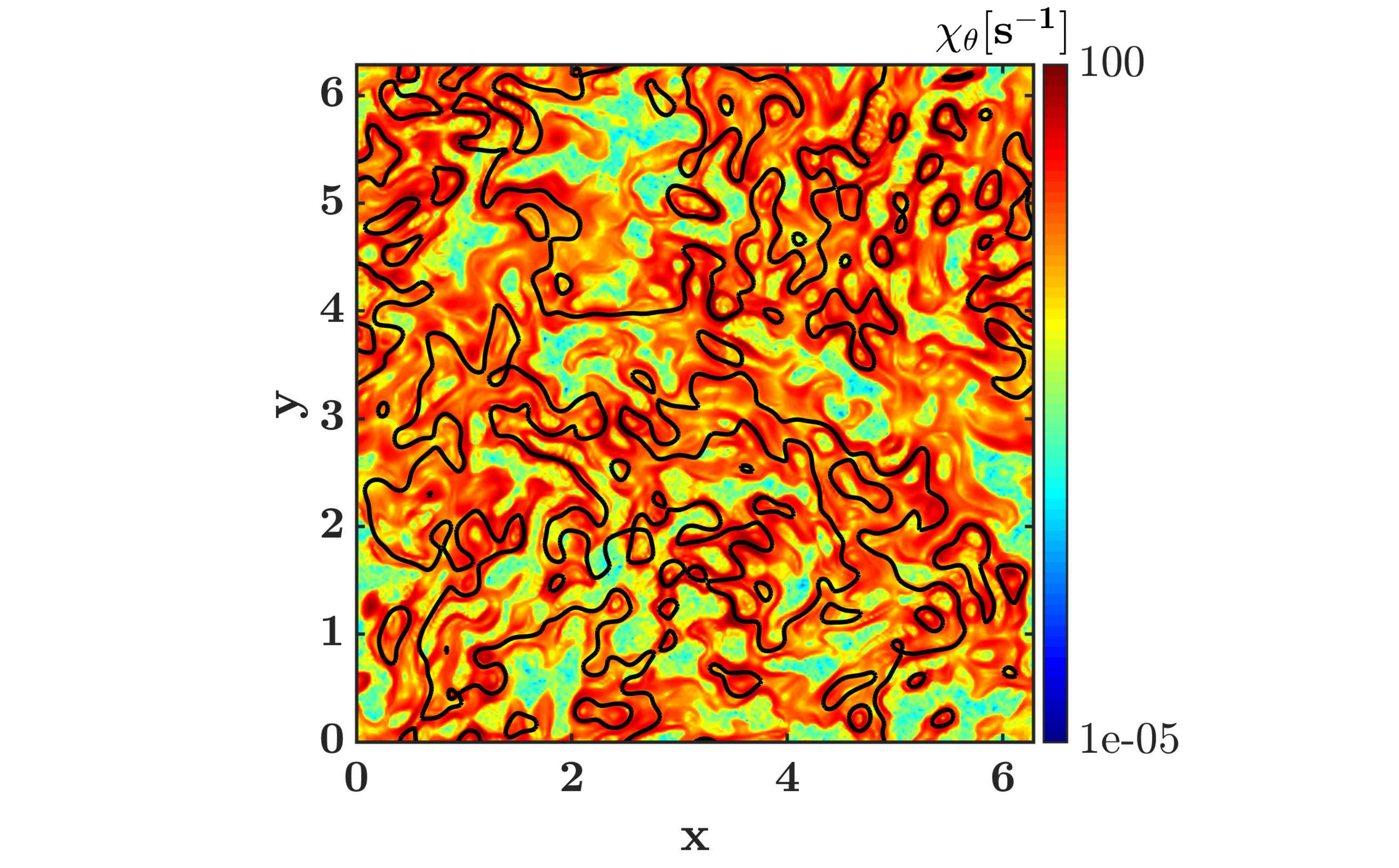}
\end{subfigure}%
\begin{subfigure}{0.33\textwidth}
\centering
\includegraphics[width=\linewidth]{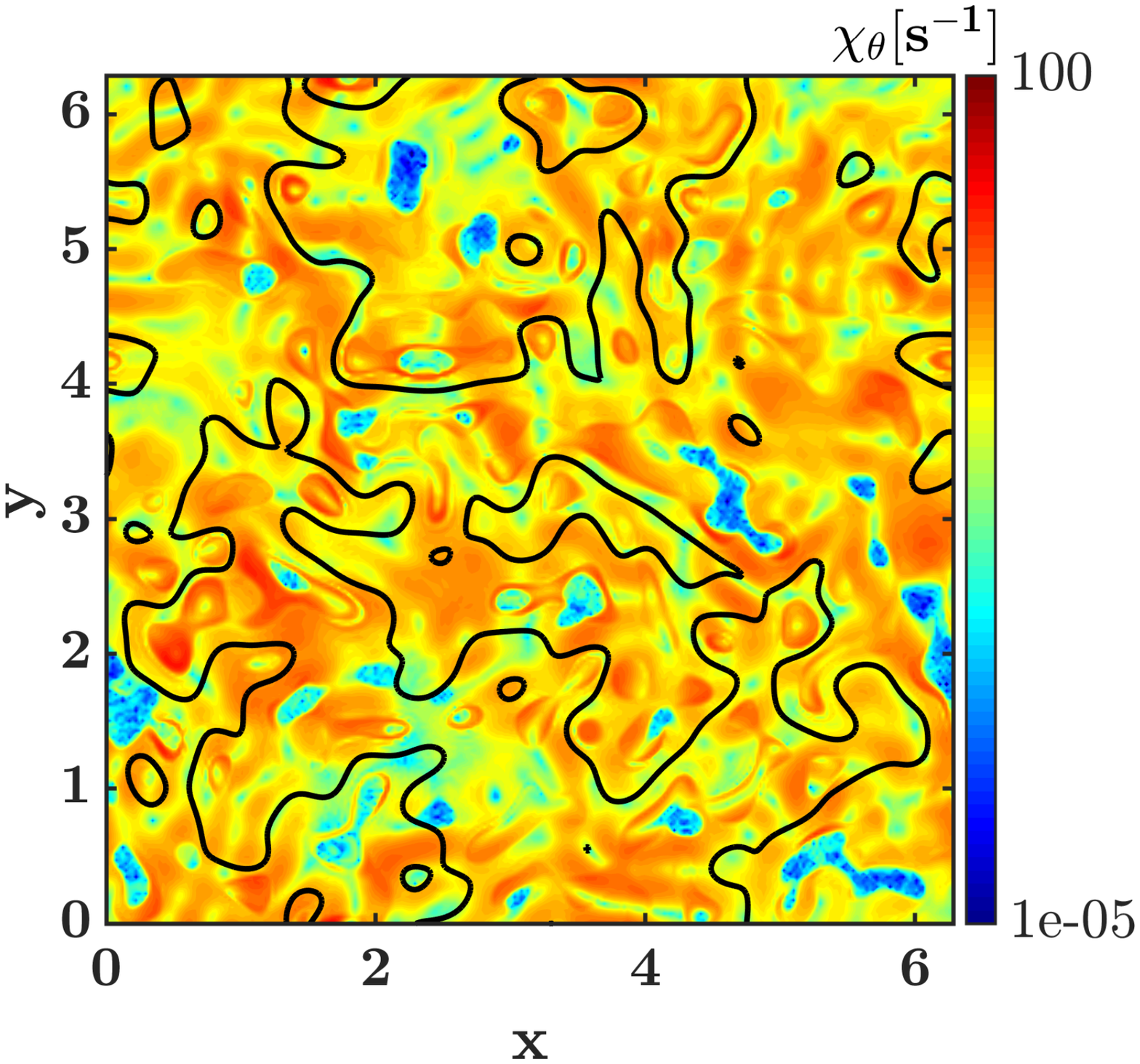}
\end{subfigure}%
\caption{Dissipation rate of progress variable ($\chi_{\theta} = 2D(\nabla \theta)^2$) for flame $IV$ at $t/\tau = 0.5$ when flame is locally extinguished. Left: DNS, middle: AIM, right: no model prediction. Projection operator is obtained with $k_m = 16$. Black lines represent stoichiometric mixture.}
\label{fig:AIM2-Flame4-X_theta}
\end{figure}

Dissipation rate of the progress variable (temperature) of flame $IV$ when local extinction is maximum (lowest conditional temperature in Fig.~\ref{fig:Stoi-AIM-conv}, middle). Near the stoichiometric mixture fraction surface (shown by black lines), higher reaction rates impose steeper temperature gradients, which emphasizes the reaction effects on the small scale mixing. Modeling of $\chi_{\theta}$ with the lowest dimensional AIM considered here is able to reproduce most features of the field, however, the dissipation rate is overestimated in general. Compared to the reduced order evolution of dynamics without any form of reconstruction of the unresolved modes (Fig.~\ref{fig:AIM2-Flame4-X_theta}, right), AIM has managed to recover small-scale features of the field. 

\begin{figure}[H]
\begin{subfigure}{0.5\textwidth}
\centering
\includegraphics[width=\linewidth]{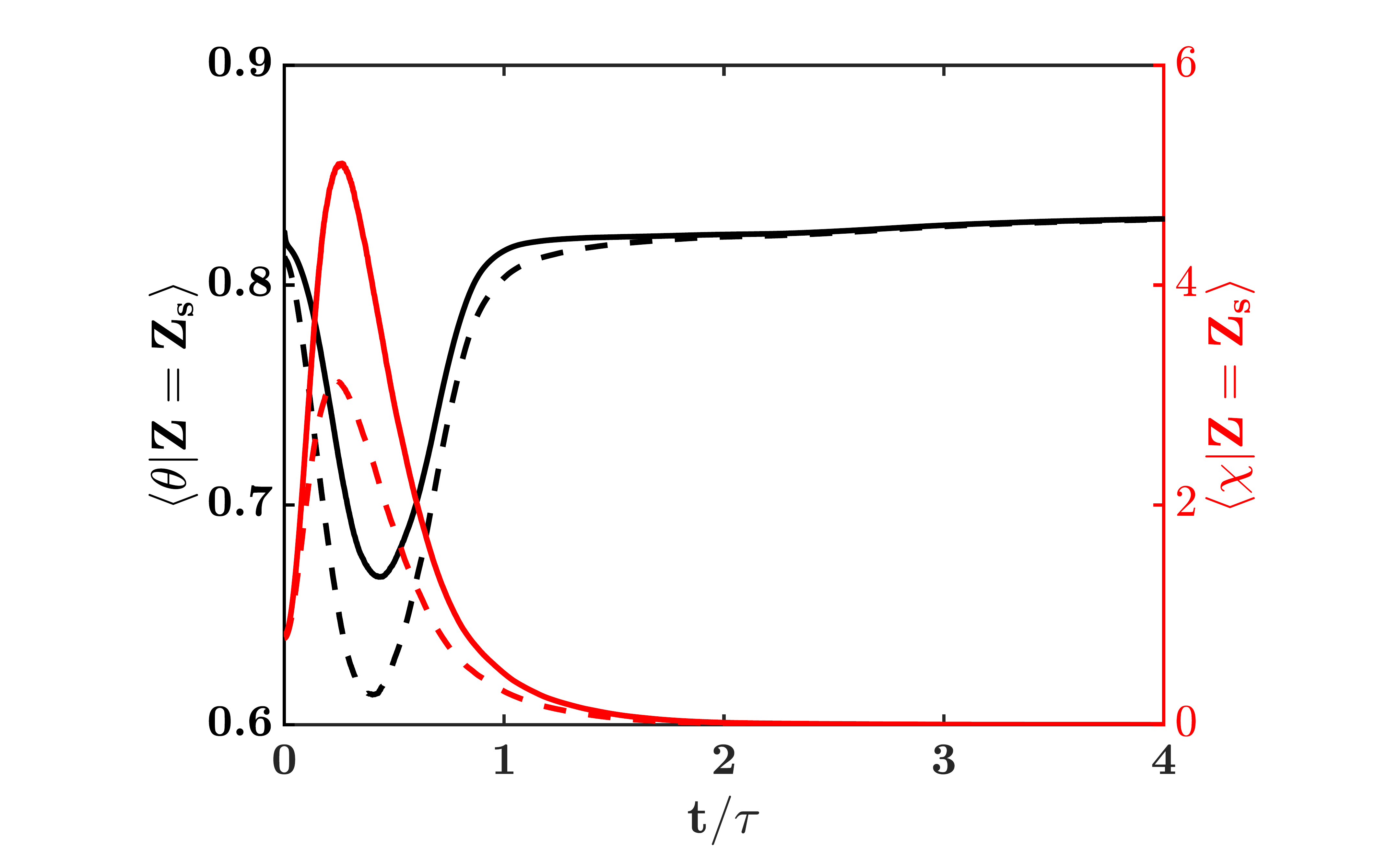}
\end{subfigure}%
\begin{subfigure}{0.5\textwidth}
\centering
\includegraphics[width=\linewidth]{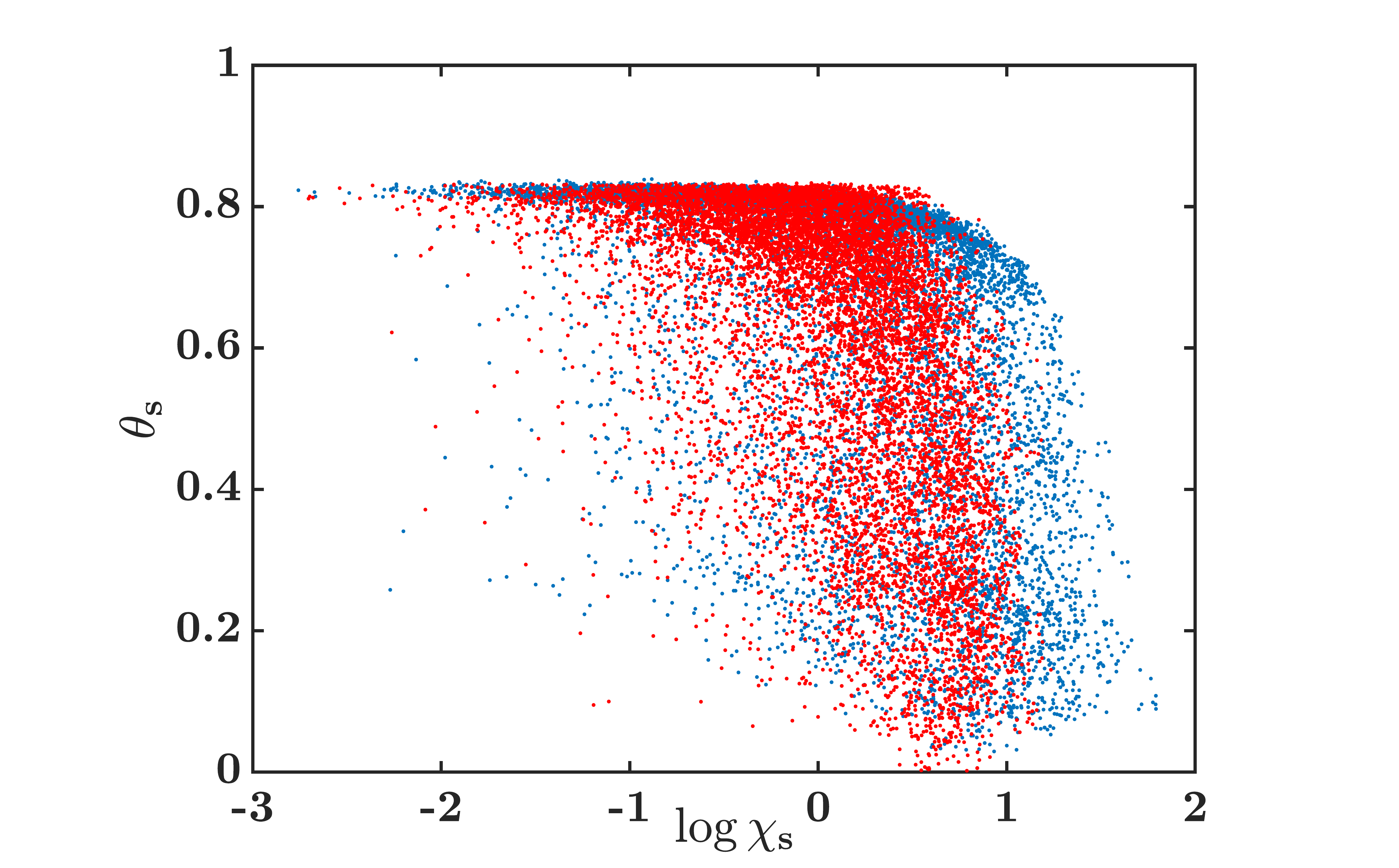}
\end{subfigure}%

\begin{subfigure}{0.5\textwidth}
\centering
\includegraphics[width=\linewidth]{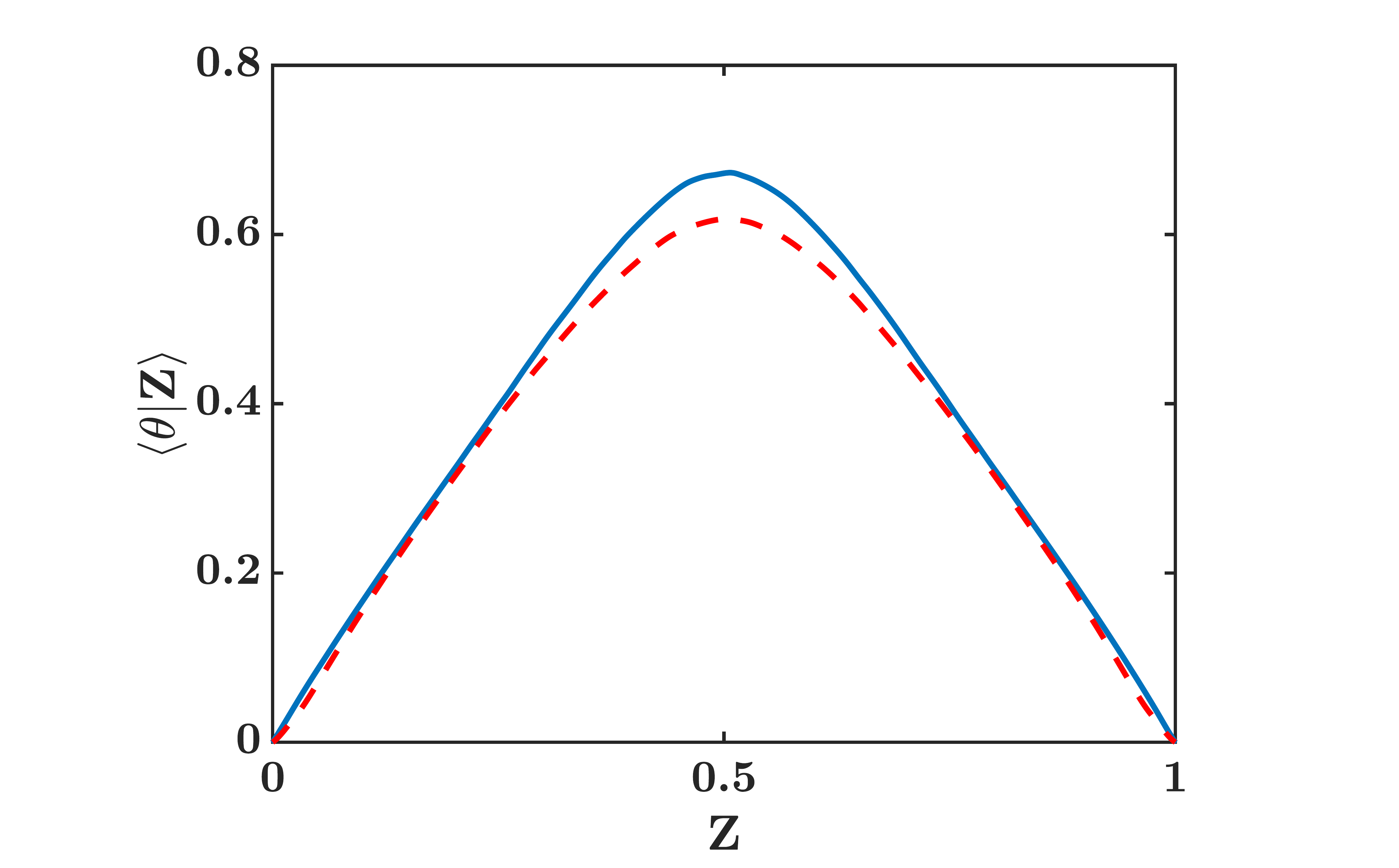}
\end{subfigure}%
\begin{subfigure}{0.5\textwidth}
\centering
\includegraphics[width=\linewidth]{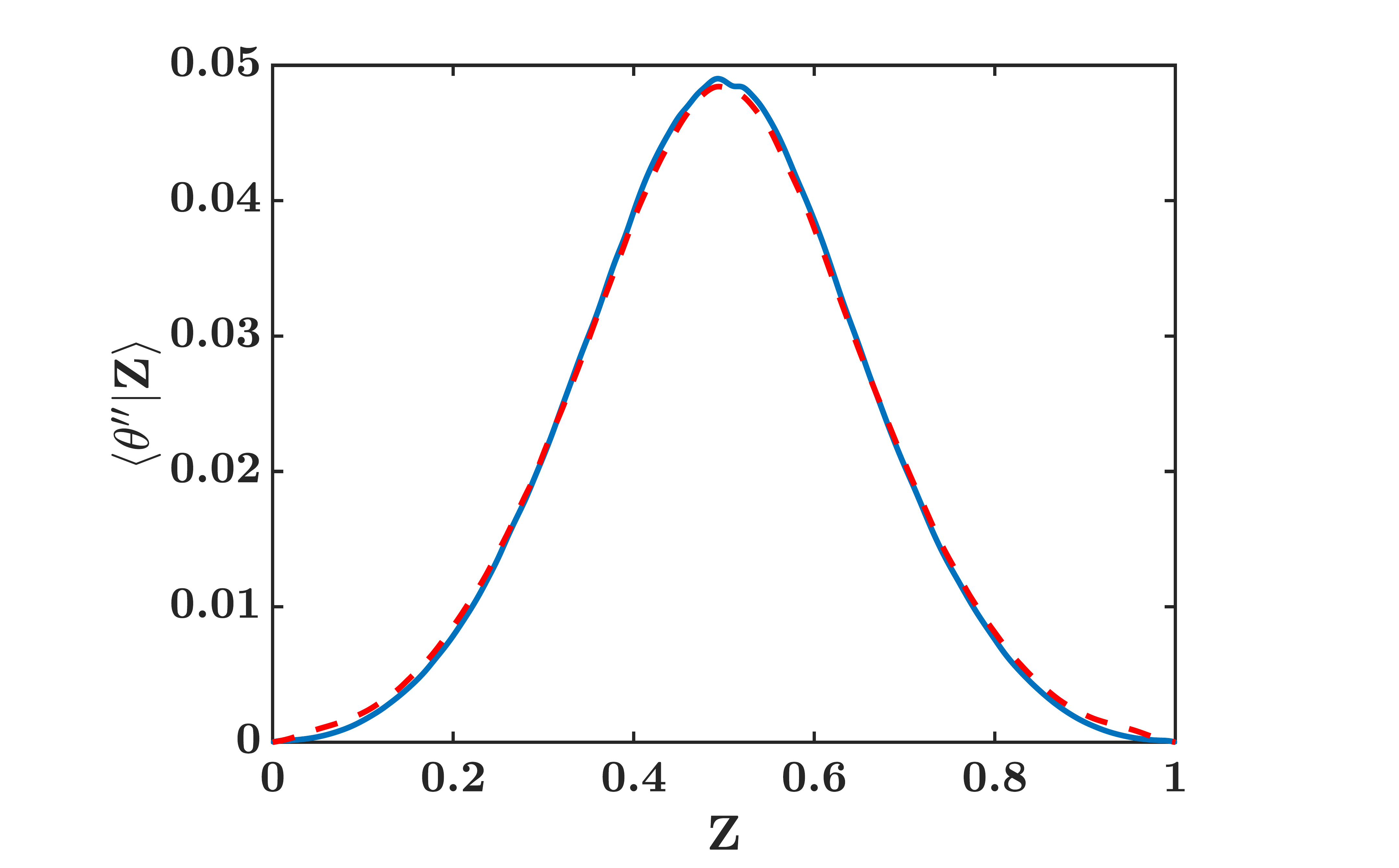}
\end{subfigure}%
\caption{Statistical properties of flame $IV$; top left: time evolution of conditionally averaged temperature and mixture fraction dissipation rate. Solid lines are DNS results, and dashed lines are AIM modeling results with $k_m = 32$. Top right: Distribution of temperature and mixture fraction dissipation rate at stoichiometric surface at $t/\tau = 0.5$. Bottom: conditional mean (left) and variance (right) of temperature at $t/\tau = 0.5$; DNS: {\mythickline{blue(ryb)}}, AIM: {\mythickdashedline{red}}. Cut-off wavenumber for AIM projection is $k_m = 32$.} 
\label{fig:AIM2-Flame4-temperature-prop}
\end{figure}

Figure~\ref{fig:AIM2-Flame4-temperature-prop} compares some statistics of temperature field for flame $IV$ predicted by {\it a posteriori} AIM against the DNS data. Global behavior of the flame is compared in Fig.~\ref{fig:AIM2-Flame4-temperature-prop} (top left) showing time evolution of conditioned temperature and mixture fraction dissipation rate. Overall, AIM captures the local extinction and subsequent reignition, but it opverpredicts the extent of low extinction. Given that the dissipation rate is underpredicted at the flame surface, this result shows that AIM suppresses reactions away from the stochiometric surface. Distribution of temperature and mixture fraction dissipation rate at stoichiometric mixture at $t/\tau \approx 0.5$ illustrates more details of extinction (Fig.~\ref{fig:AIM2-Flame4-temperature-prop},top right). While part of the stoichiometric mixture is burning close to the steady flamelet solution, localized extinction has created considerable cold and non-burning regions. Bi-modal distribution of the temperature shown in Fig.~\ref{fig:AIM2-mixingTimes} (right) is an evidence of this locally extinguished flame. It can be seen that the range of scalar dissipation values is underestimated by AIM. In particular, burning flamelets under higher strain rates are not captured. Additionally, AIM predicts lower reaction rates and temperatures under medium strain rates. These behaviors result in overprediction of the strength of extinction. Conditional distribution of temperature and its variance are compared in Fig.~\ref{fig:AIM2-Flame4-temperature-prop} (bottom). Overall, AIM is able to predict flame structure reasonably well, and handles local extinction and reignition without any additional modeling.  

\section{Conclusions}
\label{sec:conclusion}

The AIM approach has been formulated for turbulent combustion and tested using a canonical turbulent flow configuration. The main advantage of the AIM approach is that it models the small-scale features using a self-consistent approximation. In other words, no external modeling information regarding the structure of the chemical manifold or energy spectrum is needed. The underlying notion is that the slow modes of the dynamical system will traverse a stable low-dimensional manifold, and the fast variables are at a steady state that is controlled only by the slow variables. The AIM-based reduction requires that the dimension of the reduced order model should be higher than the dimension of the attractor. Since exactly obtaining this dimension is not feasible for most practical problems, the AIM reduction for different dimensions is used to understand the validity of this approximation. 

First, {\it a priori} analyses are carried out for a series of DNS configurations of forced isotropic turbulence with one-step chemical reactions. The parameters of the reaction rate are varied to introduce local extinction and reignition. For all the cases studied, the AIM reconstruction captures the variations in the small-scales. The model is particularly accurate when the large scales evolve slowly in time, for instance through a gradual reignition process rather than sudden return to equilibrium. Moreover, the AIM approach is able to provide scalar dissipation rate, mixing time for reactive scalars, and closures for nonlinear terms without any additional modeling. In this context, it overcomes some of the limitations of manifold-based techniques such as the flamelet-generated or flamelet based approaches. 

Second, {\it a posteriori} analyses are carried out for the same set of DNS configurations. Here, the AIM approximation of the small-scales is used to obtain the closure terms, which are then used in the resolved scale equations to evolve the system. It is observed that this feedback to the simulation does not cause accumulation of errors, and the solutions capture the dynamics of the flames. In particular, extinction and re-ignition processes are well-captured. However, the strength of extinction is overpredicted for small $m$ AIM simulations, with the accuracy improving as $m$ is increased.

Overall, the results of this study are promising. It shows for the first time that dynamical systems based modeling can be used to close turbulent reacting flow systems. While the concept of manifolds is well-known in the combustion community, this term generally refers to laminar flame or other non-turbulent configuration based representation. In this work, a fully turbulent manifold has been constructed through the AIM procedure. The next steps will involve extensions to non-homogeneous systems, for which the AIM approach will be cast in physical space \cite{foias1991determining}. More realistic combustion processes will also be considered.

\section*{Acknowledgements}
This work was financially supported through a grant from NASA (NNX16AP90A) with Dr. Jeff Moder as the program manager. 

\section*{Data Availability Statement}
The data that support the findings of this study are available from the corresponding author upon reasonable request.

\bibliography{Refs}
\end{document}